%% file: Buckling_arXiv_v3.tex
\documentclass[aps,pre,twocolumn,groupedaddress,noeprint,12p]{revtex4-1}
\usepackage[hidelinks]{hyperref}
\hypersetup{breaklinks=true}
\usepackage[utf8]{inputenc}
\usepackage{graphicx}
\usepackage{amsmath}
\usepackage{amsfonts} 
\usepackage{amssymb}
\usepackage{subcaption} 
\usepackage{psfrag}
\usepackage[pagewise,displaymath,mathlines]{lineno}     
\usepackage[normalem]{ulem}
\usepackage{xcolor}
\usepackage{xparse}
\ExplSyntaxOn
\NewDocumentCommand{\mref}{m}{\quinn_mref:n {#1}}
\seq_new:N \l_quinn_mref_seq
\cs_new:Npn \quinn_mref:n #1
 {
  \seq_set_split:Nnn \l_quinn_mref_seq { , } { #1 }
  \seq_pop_right:NN \l_quinn_mref_seq \l_tmpa_tl
  ( 
  \seq_map_inline:Nn \l_quinn_mref_seq
    { \ref{##1},\nobreakspace } 
  \exp_args:NV \ref \l_tmpa_tl 
  ) 
 }
\ExplSyntaxOff

\newcommand{\Ai}{\operatorname{Ai}}
\newcommand{\Bi}{\operatorname{Bi}}

\begin{document}
\title{Capillary control of collapse in soft composite columns}
\author{Marc Su\~n\'e}
\email{marc.sune.simon@su.se}
\affiliation{Nordita, Royal Institute of Technology and Stockholm University, Roslagstullbacken 23, SE-106 91 Stockholm, Sweden}
\author{John S. Wettlaufer}
\email{john.wettlaufer@su.se, john.wettlaufer@yale.edu}
\affiliation{Yale University, New Haven, CT 06520, USA}
\affiliation{Nordita, Royal Institute of Technology and Stockholm University, Roslagstullbacken 23, SE-106 91 Stockholm, Sweden}

\date{\today}

\begin{abstract}
Euler buckling is the elastic instability of a column subjected to longitudinal compression forces at its ends. The buckling instability occurs when the compressing load reaches a critical value and an infinitesimal fluctuation leads to a large amplitude deflection.  Since Euler's original study, this process has been extensively studied in homogeneous, isotropic, linear--elastic solids.  Here, we examine the nature of the buckling in {\em inhomogeneous} soft composite materials.  In particular, we consider a soft host with liquid inclusions both large and small relative to the {\em elastocapillarity length}, which lead to softening and stiffening of a homogeneous composite respectively.  However, by imposing a gradient of the inclusion volume fraction or by varying the inclusion size we can deliberately manipulate the spatial structure of the composite properties of a column and thereby control the nature of Euler buckling. 
\end{abstract}

\maketitle

\section{Introduction\label{sec:intro}}
An elastic beam under a sufficiently large compressible axial load collapses, or buckles, when an infinitesimal deflection destroys the equilibrium.
The critical load for the buckling of homogeneous, isotropic, linear--elastic rods with constant cross--section was derived by Euler in 1744~\cite{Euler1744,Euler1759}, and Lagrange analyzed the higher order modes in 1770~\cite{Lagrange1770}.
From the mechanical failure of structural elements in civil engineering to the storage of information through controlled buckling of nanoscale beams for future nanomechanical computing~\cite{ErbilHatipoglu20}, the buckling of slender structures has been a focus of studies in engineering, biology and physics for nearly 300 years~\cite{TimoshenkoGere63}.  

The macroscopic response of a solid body to an external force lies at the heart of buckling, the details of which depend on the body shape, material composition, and internal structure.
A vast range of distinct responses is displayed in materials with geometric inclusions of different elastic moduli~\cite{NezambadiYvonnet09}, foams modeled by anisotropic Kelvin cells~\cite{GongKyriakides05}, porous and particle--reinforced hyperelastic solids with circular inclusions of variable stiffness~\cite{TriantafyllidisNestorovic05}, fiber--reinforced elastomers with incompressible Neo--Hookean phases~\cite{deBottonHariton06}, long cylindrical shells with localized imperfections~\cite{JamalMidani99}, finitely strained porous elastomers~\cite{MichelLopez-Pamies07}, and hyperelastic cylindrical shells~\cite{GorielyVandiver08}, to mention but a few.
Attempts to describe buckling have lead to, among other things, the celebrated theory of elasticity~\cite{TimoshenkoGere63,LandauLifshitz86} and to finite--element simulation methods~\cite{PonteCastanedaSuquet97}.

Kirchhoff~\cite{Kirchhoff59,Kirchhoff76} and Clebsch~\cite{Clebsch62,Clebsch83} described the basic theoretical analysis of elastic rods by replacing the stress acting inside a volume element with a resultant force and the moment vectors attached to a body defining curve.  These ``Kirchhoff equations'' relate the averaged forces and moments to the curve's strains~\cite[e.g.,][]{ColemanDill93}.

Recent work shows how the elastic response of soft materials with liquid inclusions is governed by interfacial stresses~\cite{MoraPhou10,DuclouePitois14,StyleBoltyanskiy15,StyleWettlaufer15} and suggests the possibility that capillarity may play an important role in buckling instabilities.  To that end, we reformulate the study of compressed rods viewed from the perspective of the theory of elasticity~\cite{TimoshenkoGere63,LandauLifshitz86} to account for the surface tension effects of the inclusions.  We incorporate the physics of capillarity into the Kirchhoff equations through the elastic moduli as given by a generalization of Eshelby's theory of inclusions~\cite{StyleBoltyanskiy15,StyleWettlaufer15,MancarellaStyle16}.
Eshelby's theory describes how an inclusion of one elastic material deforms when it is embedded in an elastic host matrix~\cite{Eshelby57}.
However, it has recently been discovered that Eshelby's inclusion theory breaks down when the inclusion size $R$ approaches the {\em elastocapillary length}, $L\equiv \gamma/E$, where $\gamma$ is the inclusion/host surface tension and $E$ is the host Young's modulus \cite{StyleBoltyanskiy15,StyleWettlaufer15,MancarellaStyle16}.  Importantly, when $R > L$ ($R < L$) the {\em composite} softens (stiffens).  This basic physical process, wherein the inclusion size controls the Young's modulus of the the composite, $E_c$, reveals the possibility of controlling the buckling process through the properties and distribution of the inclusions. 

A quantitative treatment of how the inclusion size, $R$, and volume fraction, $\phi$, in soft composites influences their bulk mechanical properties underlies our understanding of their response under loads.  In particular, by determining how the {\em spatial variation} of $R$ and $\phi$ modify Euler buckling we provide a framework of either tailoring a material response or explaining observations in naturally occurring soft composites.  Canonical examples of  the latter include slender composite structures such as insect extremities~\cite{PeiskerMichels13,SchmittBuescher18}, plant stems~\cite{RuggebergSpeck08,SpeckBurgert11}, bones~\cite{FritschHellmich07}, bacterial biofilaments~\cite{Mendelson82} and plant tendrils~\cite{JaffeGalston68}. Indeed, these latter systems \cite{Mendelson82, JaffeGalston68} very often grow into axisymmetric elongated structures by adding new material at a small active growing zone located near the tip, creating a varying composition along the growth axis.

Although our analysis is confined to static elastic Kirchhoff rods, our results are clearly of use in interpreting the buckling instabilities prompted by tip growth~\cite{Goriely2017-4,Goriely2017-5} as well as the phenomenon of morphoelasticity induced by time--dependent compression~\cite{GoldsteinGoriely06,Goriely2017-6}.

The paper is organized as follows. In \S~\ref{sec:preliminaries}, we (i) review the general concepts of the Kirchhoff rod theory and the small deflection approximation;
(ii) introduce some concepts of static stability; and (iii) describe the models of composite mechanics that are used throughout the paper.
In \S~\ref{sec:homogeneous}, we outline classical Euler buckling of homogeneous rods and then in
\S~\ref{sec:inhomogeneous}, we examine in detail the buckling of \emph{inhomogeneous} composite materials. In particular, we (i) describe the stability analysis; (ii) study stiffened and softened composites; (iii) quantify the effect of \emph{inhomogeneity} on the critical compression; and (iv) consider the case with a ``polar'' inclusion configuration.
Conclusions and implications for experimental implementation of our results are presented in \S~\ref{sec:conclusions}.

\section{Preliminaries\label{sec:preliminaries}}
In this section, we formulate the equilibrium configurations of \emph{inhomogeneous} compressed rods in terms of planar elastica, we derive the corresponding approximation of small deflections, we introduce the key concepts of static stability of elastic rods adapted to the particular case at hand, we review generalized Eshelby theories for composite elastic materials with capillary effects, and we describe the non-dimensionalization of the problem.

\subsection{The planar equilibrium\label{sec:elastica}}
\begin{figure}[t]
	\psfrag{S}[ct][ct][1.]{$\mathcal{A}$}
\psfrag{ds}[ct][ct][1.]{$ds$}
	\psfrag{e1}[ct][ct][1.]{${\bf e}_1$}
	\psfrag{e2}[ct][ct][1.]{${\bf e}_2$}
	\psfrag{e3}[ct][ct][1.]{${\bf e}_3$}
	\psfrag{d1}[ct][ct][1.]{${\bf d}_1$}
	\psfrag{d2}[ct][ct][1.]{${\bf d}_2$}
	\psfrag{d3}[ct][ct][1.]{${\bf d}_3$}
	\psfrag{F}[ct][ct][1.]{${\bf T}$}
	\psfrag{M}[ct][ct][1.]{${\bf m}$}
	\psfrag{r}[ct][ct][1.]{${\bf r}(s)$}
	\psfrag{sig}[ct][ct][1.]{$\sigma_{zz}$}
	\psfrag{K}[ct][ct][1.]{${\bf K}$}
\includegraphics[scale=0.65]{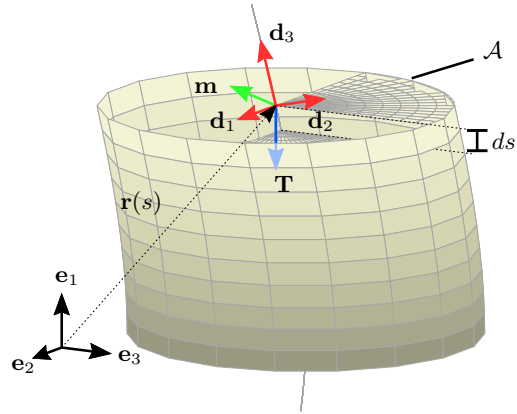}
\caption{Schematic representation of a Kirchhoff rod, with the quantities described in the main text.}
\label{fig_cartoon}
\end{figure}

Consider a straight, isotropic, inextensible and unshearable rod with constant circular cross--section of area $\mathcal{A}$, that can deform under end loading.
The absence of shear deformation and stretching are the fundamental assumptions of the Kirchhoff rod theory~\cite{Kirchhoff59,Kirchhoff76,Clebsch62,Clebsch83}, wherein the stress acting inside a volume element is replaced by the corresponding resultant force ${\bf T}$ and moment ${\bf m}$ vectors attached to the centerline, as shown in Fig.~\ref{fig_cartoon}.
The centerline is the space curve ${\bf r}(s)$ defining the rod axis, in which $s\in[0,1]$ is the dimensionless arc--length.

When the couples and forces exerted at either end of the rod are the only loads, the balance of the total forces and the total couple on a reference segment $ds$ are
\begin{eqnarray}
{\bf T}^{\prime}(s) & = & 0 \qquad \textrm{and},\label{eq_equilibrium_force}\\
{\bf m}^{\prime}(s)+{\bf r}^{\prime}(s)\times{\bf T}(s) & = & 0, \label{eq_equilibrium_moment}
\end{eqnarray}
where the primes denote differentiation with respect to arc--length.

The orientation of the normal cross--section of the rod at $s$ is specified by two material unit vector fields ${\bf d}_1(s)$ and ${\bf d}_2(s)$ that lie in the cross--section (Fig.~\ref{fig_cartoon}).
Adopting the same conventions as in~\cite{CaflischMaddocks84,Maddocks84}, we define a right--handed orthonormal basis $\{{\bf d}_1, {\bf d}_2, {\bf d}_3\}$ so that
\begin{align}
        {\bf d}_3(s)={\bf r}^{\prime}(s).
        \label{eq_tangent_vector}
\end{align}
The orthonormal directors $\{{\bf d}_i(s)\}$ change their orientation relative to a Cartesian fixed basis $\{{\bf e}_i\}$ smoothly and this change can be expressed as
\begin{align}
        {\bf d}_i^{\prime}={\boldsymbol\kappa}\times{\bf d}_i,~i=1,2,3;
\end{align}
where the components of the strain vector ${\boldsymbol\kappa}$ with respect to $\{{\bf d}_i(s)\}$ are
\begin{align}
        {\boldsymbol\kappa}=(\chi_1,\chi_2,\tau).
\end{align}
The components $\chi_1$ and $\chi_2$ are associated with bending, that is, associated with the centerline curvature.  The twisting, or rotation of the local basis around the ${\bf d}_3$ vector, is described by
$\tau$. 
It describes the torsion of the centerline (a measure of the curve nonplanarity) and the rotation of the cross section as the arc length increases.

The rod is assumed to be hyperelastic, and hence there is a strain energy density function, $W(\chi_1,\chi_2,\tau,s),\nonumber$ such that
\begin{align}
        m_1=\frac{\partial W}{\partial\chi_1},~m_2=\frac{\partial W}{\partial\chi_2},~m_3=\frac{\partial W}{\partial\tau},
\label{eq_moment_components}
\end{align}
where the $m_i(s)$ denote the components of ${\bf m}(s)$ with respect to the local basis $\{{\bf d}_i(s)\}$.
Recalling that we assume an isotropic rod with constant circular cross--section, and hence the case of linear constitutive stress--strain relations gives the strain energy density function as
\begin{align}
	W(\chi_1,\chi_2,\tau,s)=\frac{1}{2}E_c(s)I(\chi_1^2+\chi_2^2)+\frac{1}{2}C(s)\tau^2,
\label{eq_quadratic_energy}
\end{align}
where $E_c(s)$ is the composite Young's modulus, $I$ is the second moment of area about either ${\bf d}_1$ or ${\bf d}_2$, and $C(s)$ is the torsional rigidity of the cross--section at $s$.

The orientation of the local basis $\{{\bf d}_i(s)\}$ with respect to the fixed basis $\{{\bf e}_i\}$ is described by the set of Euler angles $\Theta(s)=\{\theta(s),\phi(s),\psi(s)\}$.
Explicit expressions for the directors in terms of the Euler angles can be found in~\cite{Maddocks84}, and the strain components are
\begin{eqnarray}
	\chi_1&=&-\phi^{\prime}\sin\theta\cos\psi+\theta^{\prime}\sin\psi,\label{eq_chi1}\\
	\chi_2&=&\phi^{\prime}\sin\theta\sin\psi+\theta^{\prime}\cos\psi ~\textrm{and}\label{eq_chi2}\\
	\tau&=&\phi^{\prime}\cos\theta+\psi^{\prime}.\label{eq_tau}
\end{eqnarray}

We consider the undeformed rod to be aligned along the ${\bf e}_1$ axis and subject to an axial compressional external force ${\bf T}$ at $s=1$, and hence
\begin{align}
        {\bf T}=-T\,{\bf e}_1,
        \label{eq_compression}
\end{align}
with $T>0$.
We treat the boundary value problem wherein both ends lie on the undeformed configuration axis, giving two isoperimetric constraints
\begin{align}
	\int_0^1\sin\theta(s)\,\sin\phi(s)\,ds&=0 ~\textrm{and} \label{eq_iso_I}\\
	\int_0^1\cos\theta(s)\,ds&=0.  \label{eq_iso_II}
\end{align}
Moreover, the ends are held in a ``ball--and--socket joint'' in that they can freely rotate and hence no moment is applied at either end; 
\begin{align}
        {\bf m}(0)={\bf m}(1)=0.
        \label{eq_ball-and-socket}
\end{align}
By virtue of the linear constitutive strain--stress relations, these boundary conditions can be rewritten in terms of the Euler angles as
\begin{align}
        \theta^{\prime}(0)=\theta^{\prime}(1)&=0, \nonumber \\~\phi^{\prime}(0)=\phi^{\prime}(1)&=0~\textrm{and} \nonumber \\~\psi^{\prime}(s)=\psi^{\prime}(s)&=0.
        \label{eq_bc}
\end{align}
Because the compression force is constant, Eq.~\eqref{eq_equilibrium_force} and the moment nullification, Eq.~\eqref{eq_ball-and-socket}, allow the equilibrium condition Eq.~\eqref{eq_equilibrium_moment} to
be integrated to yield 
\begin{align}
        {\bf m}(s)+{\bf r}(s)\times{\bf T}=0,
        \label{eq_int_eq_m}
\end{align}
after using the condition that both ends lie on the undeformed configuration; ${\bf r}(0)=0$.

By computing the scalar product of ${\bf d}_3$ with Eq.~\eqref{eq_equilibrium_moment} and integrating with boundary conditions~\eqref{eq_bc} we have $\tau(s)=0$. Therefore, all physical twist vanishes and hence the rod undergoes planar buckling.

We assume that any planar equilibrium lies in the $({\bf e}_1,{\bf e}_2)$--plane (or $\theta(s)=\pi/2$) so that planar solutions can be described by
\begin{eqnarray}
	\theta(s)=\frac{\pi}{2},~\psi(s)=\alpha,~\phi(s)=\Phi(s),\label{eq_planar_eq}\\
	\int_0^1\sin\Phi(s)\,ds=0.\label{eq_planar_isoperimetric}
\end{eqnarray}
We will further consider throughout that the director ${\bf d}_2$ lies in the $({\bf e}_1,{\bf e}_2)$--plane, and thus $\alpha=0$ and
\begin{align}
{\boldsymbol\kappa}=(-\Phi^{\prime},0,0).\label{eq_planar_kappa}
\end{align}

Thus any equilibrium configuration will be characterized by the coordinates in the (${\bf e}_1,{\bf e}_2$)--plane, $(x,Y(x))$, or, alternatively, by the pair $(s,\Phi(s))$, in which $\Phi(s)$ is the angle between the deformed rod tangent vector, ${\bf d}_3$, and the undeformed rod axis, ${\bf e}_1$.

Given the linear constitutive stress--strain relations Eqs.~\eqref{eq_moment_components} and~\eqref{eq_quadratic_energy}, the total angular momentum is
\begin{align}
        {\bf m}(s)=-\frac{\partial W}{\partial\Phi^{\prime}}{\bf d}_1=E_c(s)I\Phi^{\prime}(s)\,{\bf e}_3,
        \label{eq_total_m}
\end{align}
where we have used the set of Euler angles to characterize the orientation of the local basis $\{{\bf d}_i\}$ with respect to the fixed Cartesian basis $\{{\bf e}_i\}$ as per the criterion in~\cite{Maddocks84}.

Finally the balance of the total couple Eq.~\eqref{eq_equilibrium_moment} of the planar rod under consideration is
\begin{align}
	\frac{d}{ds}(E_c(s)\,I\,\Phi^{\prime}(s))+T\sin\Phi(s)=0, 
        \label{eq_Euler_Bernoulli}
\end{align}
which is also known as the Euler--Bernoulli equation of planar elastica. 
Its integral, corresponding to Eq.~\eqref{eq_int_eq_m} for planar elastica, is
\begin{align}
	E_c(s)\,I\,\Phi^{\prime}(s)+T\int_0^{s}\sin\Phi(u)\,du=0. 
        \label{eq_Euler_Bernoulli_int}
\end{align}

\subsection{The small deflection approximation\label{sec:small}}

We now further simplify Eq.~\eqref{eq_Euler_Bernoulli} by considering the limit of small deflections.

Given the definition of $\Phi(s)$ following Eq.~\eqref{eq_planar_kappa}, we note that its derivative with respect to the arc length is related to the turning rate of the tangent along the centerline.
Hence, using Eq.~\eqref{eq_tangent_vector} we have 
\begin{align}
        \Phi^{\prime}=|{\bf r}^{\prime\prime}|.
        \label{eq_deriv_Phi}
\end{align}

When deformations are small, so too is $\Phi$, and hence we can approximate the arc length derivative by the derivative along the undeformed axis as
\begin{align}
        |{\bf r}^{\prime\prime}|=\left|\frac{d^2{\bf r}}{ds^2}\right|\sim\left|\frac{d^2{\bf r}}{dx^2}\right|=\frac{d^2Y(x)}{dx^2},
        \label{eq_second_deriv_r}
\end{align}
where $|{\bf r}^{\prime\prime}|$ is the local curvature.

Now, upon substitution of Eqs.~\eqref{eq_deriv_Phi} and~\eqref{eq_second_deriv_r} into 
Eq.~\eqref{eq_Euler_Bernoulli_int}, and recalling that $Y(s)\equiv{\bf r}(s)\,{\bf e}_2=\int_0^{s}\sin\Phi(u)\,du$ as per Eq.~\eqref{eq_tangent_vector} and the planar equilibrium constraints Eq.~\eqref{eq_planar_eq}, the small deflection approximation of planar elastica Eq.~\eqref{eq_Euler_Bernoulli_int} is
\begin{align}
        \frac{d^2Y(x)}{dx^2}+\frac{T}{E_{c}(x)\,I} Y(x)=0.
        \label{eq_eq_m_small_deform}
\end{align}

The corresponding isoperimetric constraint~\eqref{eq_planar_isoperimetric} 
for the small displacements $Y(x)$ of the elastic line from the straight configuration is:
\begin{align}
	Y(0)=Y(1)=0.
\end{align}

\subsection{Remarks on stability\label{sec:stability}}
Standard calculus of variations asserts that an equilibrium configuration $\tilde\Phi(s)$, namely an extremal of the potential energy of the rod~\eqref{eq_energy}, is stable if the second variation of the potential energy at $\tilde\Phi(s)$ is positive definite.
\citet{CaflischMaddocks84} showed that stable equilibria according to this static  criterion are also stable in the dynamic sense due to Liapounov.

The potential energy, as given by \citet{Maddocks84}, of a Kirchhoff rod whose planar configurations are described by Eqs.~\eqref{eq_planar_eq} and~\eqref{eq_planar_isoperimetric}, is
\begin{align}
	V[\Phi,T]=\int_0^1\left\{\frac{1}{2}E_c(s)\,I\,(\Phi^{\prime})^2+T\cos\Phi(s)\right\}ds,
        \label{eq_energy}
\end{align}
where the first term in the integrand is the stored energy density of the rod and the second term is the work done by the external force.

The corresponding Euler--Lagrange equation for Eq.~\eqref{eq_energy} with respect to the variable $\Phi$ is the Euler--Bernoulli equation~\eqref{eq_Euler_Bernoulli} derived in \S\ref{sec:elastica}. This equation was exhaustively analyzed by~\citet{AntmanRosenfeld78}.  The solution set consists of the unbuckled solution $\Phi(s)=0$, for any compressing load $T$, and an infinite number of buckled modes bifurcating symmetrically from the trivial solution--pitchfork bifurcations.

Stability of the trivial solution $\Phi(s)=0$ is assessed by considering perturbations of the form
\begin{align}
        \Phi_{\epsilon}(s)=\epsilon\eta(s), 
\end{align}
wherein only variations satisfying 
\begin{align}
        \eta^{\prime}(0)=\eta^{\prime}(1)=0,
\label{eq_bc_perturbation}
\end{align}
need be considered because of the boundary conditions \eqref{eq_bc}, and
\begin{align}
        \int_0^1\eta(s)\,ds=0,
\end{align}
which correspond to the linearization of constraint \eqref{eq_planar_isoperimetric} about $\Phi(s)=0$.
Note that whilst estimating the second variation, linearization of the isoperimetric constraint suffices because nonlinear terms can be absorbed into the dominant linear term for small perturbations~\cite{Maddocks84}.

The second variation of $V[\Phi,T]$ about $\Phi(s)=0$ is given by
\begin{align}
	\delta^2V[0,T]=\int_0^1\{E_c(s)\,I\left(\frac{d\eta}{ds}\right)^2-T\,\eta^2(s)\}ds.
\label{eq_second_variation}
\end{align}
Integration by parts and imposing the boundary conditions~\eqref{eq_bc_perturbation} yields 
\begin{align}
	\int_0^1\{E_c(s)\,I\left(\frac{d\eta}{ds}\right)^2-T\,\eta^2(s)\}ds\equiv\langle\eta,P(\Phi=0)\eta\rangle,
\end{align}
where $P(\Phi=0) \equiv P(0)$ is a second order Sturm--Liouville operator defined by
\begin{align}
	P(0)\equiv-\frac{d}{ds}\left(E_c(s)\,I\frac{d}{ds}\right)-T,\label{eq_operator}
\end{align}
and $\langle\,,\rangle$ is the $L^2$--inner product.

The question of whether the second variation, Eq.~\eqref{eq_second_variation}, is positive--definite is then reduced to the study of the following eigenvalue problem; 
\begin{align}
        P(0)\eta=\mu\eta,\label{eq_eigenval}\\
        \eta^{\prime}(0)=\eta^{\prime}(1)=0,\label{eq_Neumann}\\
        \langle\eta,1\rangle=0.\label{eq_orthogonal_eigenfunc}
\end{align}

\subsection{Composite mechanics\label{sec:composite}}
The theory of effective elastic moduli of solid composites is generally ascribed to Eshelby~\cite{Eshelby57}. Initially conceived to treat composites of host materials such as glass or steel, with $E=\mathcal{O}(\text{GPa})$, containing dilute inclusions, Eshelby's theory has been extended to non--dilute composites~\cite{Hashin62,HashinShtrikman63,ChristensenLo79}.  
However, Eshelby's approach does not account for the energy between the inclusion and the host. Although this is quantitatively valid when the inclusion size, $R$, is much larger than the elastocapillary length, $L$, as defined above, such is not the case otherwise, which can be particularly important for a soft host.

Recently the other limit, where $R \lesssim L$ and surface--tension effects in soft solids are important, has been a major focus of research~\cite[e.g.,][and Refs. therein]{MoraMaurini13,StyleBoltyanskiy13,NadermannHui13,XuJagota14}.
In this limit a counterintuitive situation can occur wherein a soft composite is {\em stiffened} when the inclusions are {\em liquid}.  This effect is operative when, for example, liquid droplets of size $R = \mathcal{O}(100\mu\text{m})$ are embedded in soft materials, such as gels, with $E=\mathcal{O}(\text{kPa})$, whereas host materials with $E=\mathcal{O}(\text{MPa})$, like elastomers, may only exhibit composite stiffening when $R = \mathcal{O}(0.1 \mu\text{m})$~\cite{StyleBoltyanskiy15,StyleWettlaufer15}.

Following Eshelby's original approach, \citet{StyleBoltyanskiy15,StyleWettlaufer15} calculated the effective elastic modulus of composites consisting of a dilute dispersion of liquid droplets embedded in a homogeneous isotropic elastic solid matrix when the elastic stress at the surface of the droplets satisfies a generalized Young--Laplace equation \cite[e.g.,][]{MoraAbkarian11, StyleDufresne12}.
\citet{StyleBoltyanskiy15,StyleWettlaufer15} showed that, in terms of the elastic moduli of the host material (Young's modulus $E$ and Poisson's ratio $\nu$), the dimensionless number $\gamma\prime\equiv L/R$, and the inclusion volume fraction $\phi$, the effective elastic modulus of soft composite solids in the dilute limit is
\begin{equation}
 E_c(\phi,\gamma\prime)=E\,\frac{1+\frac{5}{2}\gamma\prime}{\frac{5}{2}\gamma\prime(1-\phi)+\left(1+\frac{5}{3}\phi\right)},\label{eq_Ec}
\end{equation}
where the solid is assumed to be incompressible; $\nu=1/2$.  We denote the dilute theory result Eq.~\eqref{eq_Ec} as DT.  

In the limit that surface tension vanishes,  $\gamma\prime\to 0$, Eshelby's result $E_c/E=(1+\frac{5}{3}\phi)^{-1}$~\cite{Eshelby57}  of a softening composite as $\phi$ increases is recovered from 
Eq.~\eqref{eq_Ec}.  In the surface tension dominated limit, $\gamma\prime\gg1$, we have $E_c/E=(1-\phi)^{-1}$, the inclusions maintain sphericity and hence the composite stiffens as $\phi$ increases.  The delineation between these two different behaviors is $\gamma\prime=2/3$, when  $E_c/E=1$, and the surface tension effect leads to elastic cloaking, wherein the inclusions are mechanically invisible.

Two approaches have been used to treat the non--dilute limit.  In the first, \citet{MancarellaStyle16b,MW2017} used a three-phase generalized self-consistent (GSC) theory, which replaces the actual inclusions by composite spheres.  In the second, \citet{MancarellaStyle16} (MSW) extended the multiphase scheme of~\citet{MoriTanaka73} to treat the fluid inclusions in the solid matrix---with isotropic interfacial tension---as elastic inclusions with no interfacial tension.  Although both the GSC and MSW approaches recover Eq.~\eqref{eq_Ec} in the dilute limit, the results of the former are too cumbersome for incorporation into the buckling of composite rods we study here.  Therefore, we use the MSW approach for which the effective Young's modulus of the composite is
\begin{align} 
	E_c(\phi,\gamma\prime)=E\,\frac{2-2\phi+\gamma\prime(5+3\phi)}{2+(4/3)\,\phi+\gamma\prime(5-2\phi)},  
 \label{eq_Ec_Mori-Tanaka} 
\end{align}
in which the transition between stiffening and softening at $\gamma\prime=2/3$, is the same value as for the DT expression Eq.~\eqref{eq_Ec}.  Therefore, we will use the effective composite Young's moduli in Eqs.~\eqref{eq_Ec} and \eqref{eq_Ec_Mori-Tanaka} in our analysis of the collapse of composite columns.

\subsection{Scaling\label{sec:scaling}}
We note that, because of the inextensibility of the rod, from the outset the arc length $s$ has been dimensionless.  In order to avoid clutter in notation, the other independent variables, mass (${m^*}$), space (${x^*}$) and time (${t^*}$) were not labeled in the usual manner (e.g., with a superscript *) to distinguish that they carried dimensions.  We now render them dimensionless as follows, 
\begin{align} 
m = \frac{m^*}{ \rho\sqrt{\mathcal{A} I}}, ~ x = \frac{x^*}{ \sqrt{I/\mathcal{A}}} ~ \textrm{and}~ t = \frac{t^*}{ \sqrt{\rho I/\mathcal{A} E}},  \label{eq_scaling} 
\end{align} 
where $\rho$ is the mass per unit reference volume, $\mathcal{A}$ denotes the rod cross--sectional area, and $E$ is the Young's modulus of the host matrix as above, so that the dimensionless composite Young's modulus is $E_{\text{rel}}(s)=E_c(s)/E$, and hence the dimensionless force is $\Gamma=T/(E\,\mathcal{A})$.

Under this rescaling, the small--deflections equilibrium equation~\eqref{eq_eq_m_small_deform} is given by
\begin{align}
	\frac{d^2Y(x)}{dx^2}+\frac{\Gamma}{E_{\text{rel}}(x)} Y(x)=0. 
        \label{eq_eq_scaled}
\end{align}

\section{Buckling of homogeneous composite rods\label{sec:homogeneous}}
The classical Euler buckling problem \cite{Euler1744} treats a compressed homogeneous rod ($E_{\text{rel}}$ = constant) as a boundary--value problem for the small--deflection Euler--Bernoulli equation, Eq.~\eqref{eq_eq_scaled}, with Dirichlet boundary conditions.
The trivial solution $Y(x)=0$ corresponds to the undeformed rod.
However, such a configuration is only stable if the compressing force $\Gamma$ is less than the critical value $\Gamma_{\text{cr}}$~\cite{Euler1744},

\begin{align}
	\Gamma_{\text{cr}}=\pi^2E_{\text{rel}},\label{eq_gamma_crit}
\end{align}
which corresponds to the smallest non--zero eigenvalue of the Dirichlet problem. At the critical compression a first bifurcation of the solution is encountered: the unbuckled solution becomes unstable and two buckled stable symmetric configurations appear;
\begin{equation}
	Y(x)\propto \pm\sin\left(\pi\,x\right).\label{eq_reference_1st_mode}
\end{equation}

Beyond the bifurcation, that is for $\Gamma>\Gamma_{\text{cr}}$, the solution can be computed explicitly because the Euler--Bernoulli equation~\eqref{eq_Euler_Bernoulli} is integrable. The solutions of the non--linear Kirchhoff equations are beyond the scope of this paper.
Here we treat the isoperimetric linearized planar elastica, Eq.~\eqref{eq_eq_scaled}, whose eigenvalues,
\begin{align}
	\Gamma_{\text{cr}}^{(n)}=\left(n\pi\right)^2E_{\text{rel}},~n=1,2,\dots,\label{eq_gamma_crit_n}
\end{align}
describe the bifurcation points. The associated buckled configurations are described by the corresponding eigenfunctions,
\begin{equation}
	Y_n(x)\propto \pm\sin\left(n \pi\,x\right).\label{eq_reference_modes}
\end{equation}

\section{Buckling of \emph{inhomogeneous} composites\label{sec:inhomogeneous}}
Given the expressions for the effective Young's modulus of a soft composite, Eqs.~\eqref{eq_Ec} and \eqref{eq_Ec_Mori-Tanaka}, an axially \emph{inhomogeneous} elastic modulus can be constructed by varying {\em either} the inclusion volume fraction, $\phi(x)$, or the ratio of the elastocapillary length to the inclusion radius, $\gamma\prime(x)\equiv L/R(x)$, both of which we discuss presently. 

A linear inclusion volume fraction profile,
\begin{equation}
 \phi(x)=\phi_0 \left(1-x\right),
 \label{eq_volume_fraction}
\end{equation}
where $\phi(x=0) \equiv \phi_0$, leads to the bulk modulus decreasing ($\gamma\prime>2/3$) or increasing ($\gamma\prime<2/3$) with the distance along the column.

A linear variation in $\gamma\prime(x)$,
\begin{equation}
	\gamma\prime(x)=\gamma_0\prime-\left(\gamma_0\prime-\gamma_1\prime\right)x\,,
	\label{eq_gamma}
\end{equation}
where $\gamma_0\prime$ and $\gamma_1\prime$ are the values of $\gamma\prime$ at $x=0$ and $x=1$ respectively, leads to the possibility of a ``polar'' configuration in which one side of a column will be stiffer than the bulk host and the other will be softer.

Equipped with Eqs. \eqref{eq_volume_fraction} and \eqref{eq_gamma}, in what follows we examine the nature of the buckling in \emph{inhomogeneous} soft--composite rods by studying the dimensionless small deviation equilibrium equation~\ref{eq_eq_scaled}, with Dirichlet boundary conditions $Y(0)=Y(1)=0$.
The operator associated with this second order differential equation is Hermitian for the functionals of $E_{\text{rel}}(x)$ given by the DT, Eq.~\eqref{eq_Ec}, and MSW, Eq.~\eqref{eq_Ec_Mori-Tanaka}, theories with spatial inhomogeneity introduced through either Eq.~\eqref{eq_volume_fraction} or Eq.~\eqref{eq_gamma}.

\subsection{Stability analysis for \emph{inhomogeneous} elastic rods\label{sec:stability_inhom}}
We follow the framework described in \S\ref{sec:stability}, and analyze the stability of the trivial solution, $Y(x)=0$, of Eq.~\eqref{eq_eq_scaled} via the eigenvalue problem given by 
Eqs.~\mref{eq_eigenval,eq_Neumann,eq_orthogonal_eigenfunc}, in which the dimensionless version of the operator $P(0)$, Eq.~\eqref{eq_operator}, is
\begin{align}
	P(0)=-\frac{d}{dx}\left(E_{\text{rel}}(x)\frac{d}{dx}\right)-\Gamma ,
	\label{eq_operator_scaled}
\end{align}
which is a Hermitian Sturm--Liouville operator on $x\in[0,1]$ with Neumann boundary conditions~\eqref{eq_Neumann}.
Note that the small deflection approximation of Eq.~\eqref{eq_second_deriv_r}, lead to approximating $s$ with $x$ in \eqref{eq_operator_scaled}.

The eigenvalue problem in Eqs.~\mref{eq_eigenval,eq_Neumann,eq_orthogonal_eigenfunc} corresponds to the first order approximation in $\Phi$ (i.e., $\Phi$ small) of the Euler--Bernoulli equation~\eqref{eq_Euler_Bernoulli} when $\mu(\Gamma)=0$.
Thus, the eigenfunction $\eta(x,\Gamma)$ associated with $\mu(\Gamma)=0$ constitutes a first order solution of the equilibrium Euler--Bernoulli equation~\eqref{eq_Euler_Bernoulli} and hence gives an equilibrium configuration of the rod.

On the other hand the equilibrium configurations for small deformations, also to first order in $\Phi$, correspond to the eigenfunctions of the eigenvalue problem~\eqref{eq_eq_scaled}, which are determined by the sequence of critical tensions, $\Gamma_{\text{cr}}^{(n)}$, or loads at which the rod is at equilibrium.

Therefore, because equilibrium configurations require both $\Gamma=\Gamma_{\text{cr}}^{(n)}$ and $\mu(\Gamma)=0$, we have
\begin{align}
	P\left(\Phi=0,\Gamma=\Gamma_{\text{cr}}^{(n)}\right)\,\eta(x,\Gamma_{\text{cr}}^{(n)})=0,
\end{align}
and hence Eq.~\eqref{eq_eigenval} can be rewritten as
\begin{align}
	\left(\Gamma_{\text{cr}}^{(n)}-\Gamma\right)\eta=\mu\eta.
\end{align}
Namely, eigenvalues $\mu^{(n)}(\Gamma)$ of operator~\eqref{eq_operator_scaled} that each correspond to a certain $\Gamma_{\text{cr}}^{(n)}$ are
\begin{align}
        \mu^{(n)}(\Gamma)=\Gamma_{\text{cr}}^{(n)}-\Gamma,
        \label{eq_mu_Gamma}
\end{align}
and hence they are positive definite when $\Gamma<\Gamma_{\text{cr}}^{(n)}$.

As implied by Eq.~\eqref{eq_compression}, compression occurs for positive values of $\Gamma$ and hence $\Gamma_{\text{cr}}^{(n)}> 0,~\forall~n>0$.
In the case of no compression, the lowest eigenvalue of Eq.~\eqref{eq_eq_scaled} is $\Gamma_{\text{cr}}^{(0)}=0$, whose eigenfunction is the undeformed trivial solution $Y(x)=0$.  Therefore, 
the eigenvalue problem~\eqref{eq_eigenval}, with Neumann boundary conditions~\eqref{eq_Neumann}, can be integrated to give the lowest eigenvalue $\mu^{(0)}(\Gamma)=-\Gamma$ and the associated constant eigenfunction $\eta^{(0)}(x,\Gamma)=1$.  However, we note that this solution does not satisfy the isoperimetric constraint~\eqref{eq_orthogonal_eigenfunc}.  Therefore, the second variation will be positive--definite when
\begin{align}
        \mu^{(1)}(\Gamma)>0,
\end{align}
and hence
\begin{align}
        \Gamma<\Gamma_{\text{cr}}^{(1)},
\end{align}
where $\Gamma_{\text{cr}}^{(1)}$ denotes the first buckling critical compression (provided that the corresponding eigenfunction $\eta^{(1)}(x,\Gamma)$ is orthogonal to the first eigenfunction $\eta^{(0)}(x,\Gamma)$).  Importantly, although $\Gamma_{\text{cr}}^{(1)}$ is the first buckling load, it is the force at which the second eigenvalue of Eq.~\eqref{eq_eigenval} crosses zero.
In consequence, we recover Euler's classical buckling result for a homogeneous column~\cite{Euler1744}: namely, the undeformed equilibrium configuration $\Phi=0$ will be stable when the compressive force does not exceed $\Gamma_{\text{cr}}^{(1)}$, and unstable otherwise.

\subsection{Stiffened inhomogeneous composites\label{sec:stiff}}
\begin{figure}[h]
\begin{subfigure}[b]{0.45\textwidth}
	\psfrag{Y}[ct][ct][1.]{$Y(x)$}
	\psfrag{Erel}[ct][ct][1.]{$E_{\text{rel}}(x)$}
        \psfrag{x}[cb][cb][1.]{$x$}
\includegraphics[width=\linewidth]{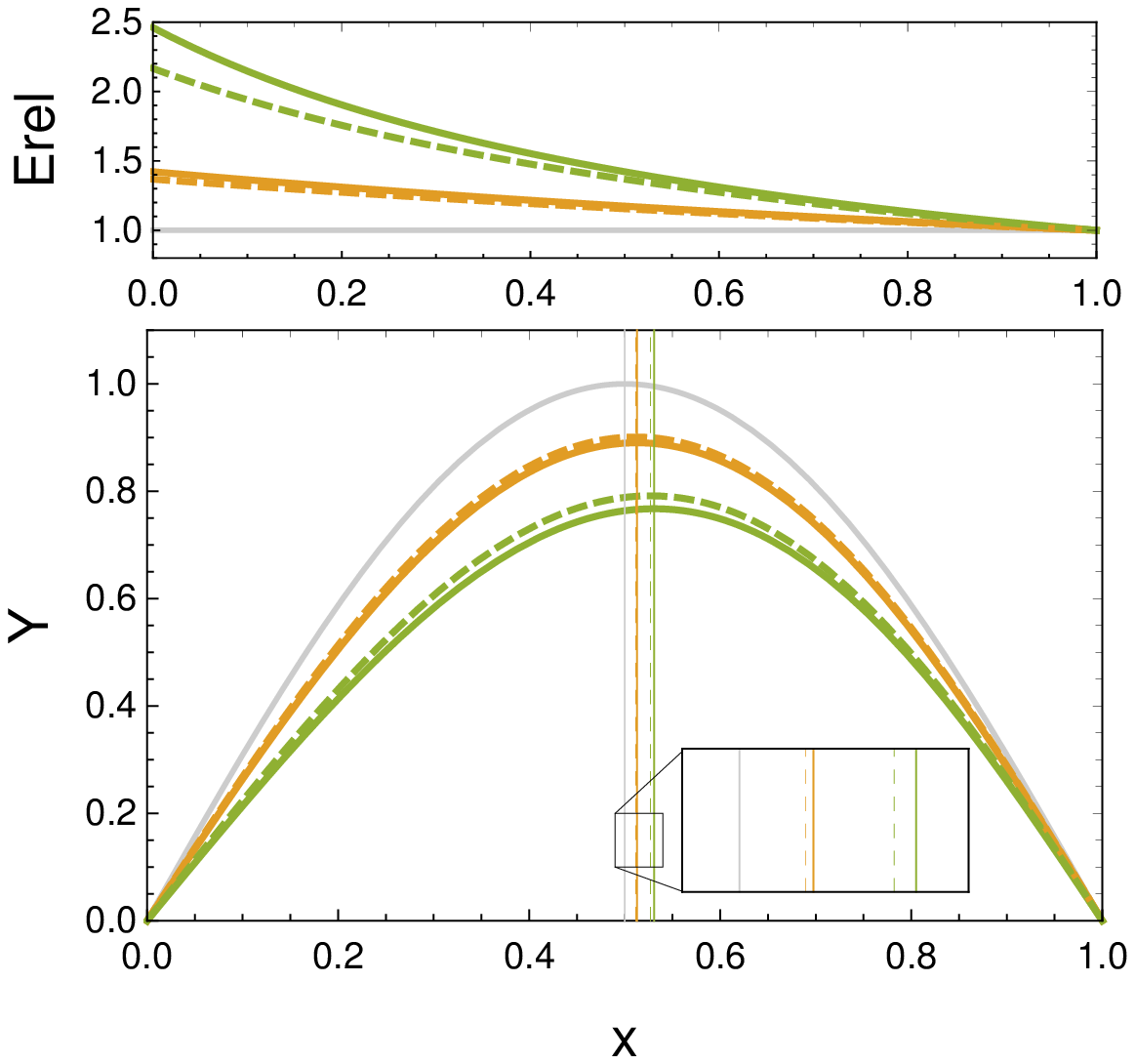}
	\caption{First buckling mode.}
\label{fig_1st_buckling_DT}
\end{subfigure}
\begin{subfigure}[b]{0.45\textwidth}
	\psfrag{Y}[ct][ct][1.]{$Y(x)$}
	\psfrag{Erel}[ct][ct][1.]{$E_{\text{rel}}(x)$}
        \psfrag{x}[cb][cb][1.]{$x$}
	\psfrag{u}[l][l][1]{$E_{\text{rel}}=1$}
        \psfrag{d}[l][l][1]{$\gamma\prime=100$, $\phi_0=0.3$}
        \psfrag{t}[l][l][1]{$\gamma\prime=100$, $\phi_0=0.6$}
        \psfrag{q}[l][l][1]{$\gamma\prime=10$, $\phi_0=0.3$}
        \psfrag{cinc}[l][l][1]{$\gamma\prime=10$, $\phi_0=0.6$}
\includegraphics[width=\linewidth]{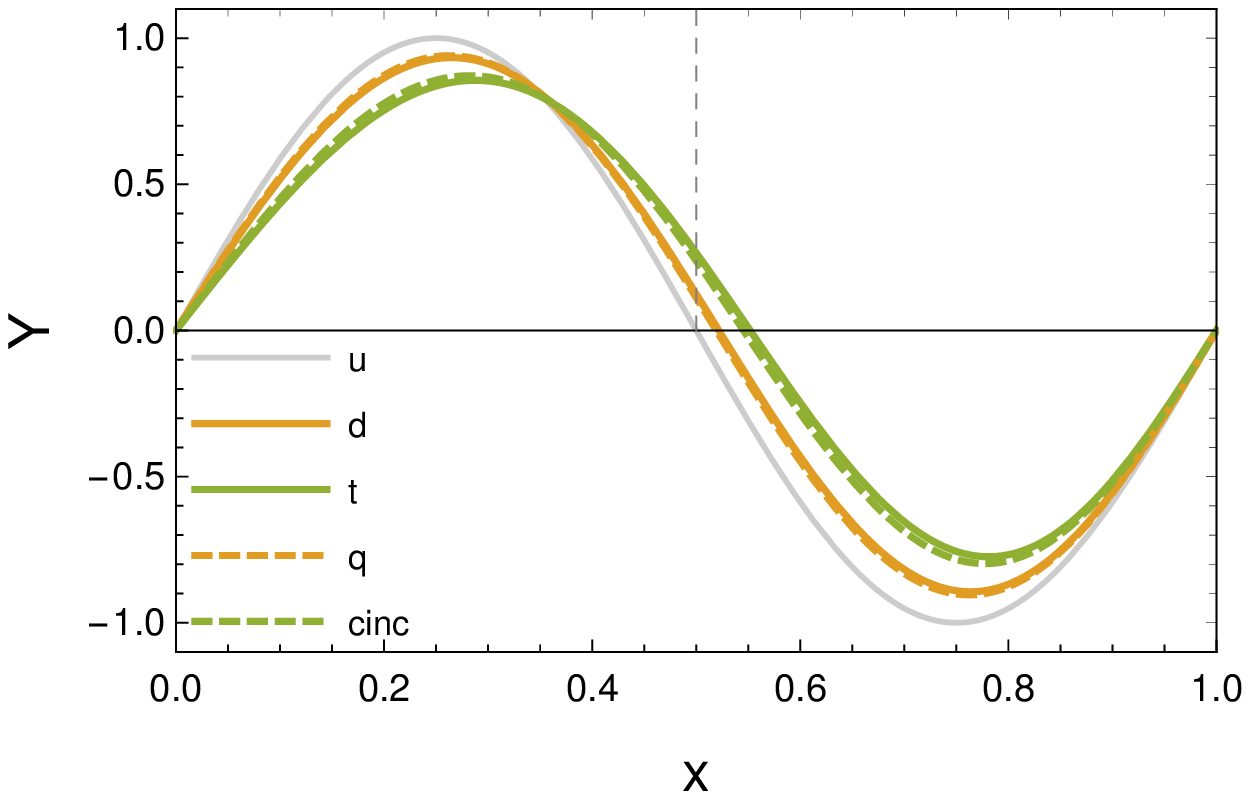}
	\caption{Second buckling mode.}
\label{fig_2nd_buckling_DT}
\end{subfigure}
\caption{Effective Young's Moduli and Buckling Modes.  The upper panel of (a) is the relative effective Young's modulus Eq.~(\ref{eq_Ec}) as a function of $\gamma\prime$ and $\phi$.  
The first (a) (lower panel) and second (b) buckling modes of homogeneous and linearly stiffened soft composite columns with the same potential energy.
	The vertical lines in (a) denote the coordinate $x$ at which the color corresponding column reaches the maximum deflection.
	The dashed vertical line in (b) denotes the position of the middle of the undeformed rod; $x=1/2$.  }
\label{fig_buckling_DT}
\end{figure}

In this section we examine a stiffened composite ($\gamma\prime>2/3$) rod with a linear axial gradient of the liquid content using Eq.~\eqref{eq_volume_fraction} and the DT expression for the effective Young's modulus, Eq.~\eqref{eq_Ec}.   In this manner we can impose a linearly decreasing stiffness along the rod.  We show in the upper panel of Fig.~\ref{fig_1st_buckling_DT} the effective Young's modulus for this class of composite rods as we vary the parameters: $\gamma\prime=10,100$; $\phi_0=0.3,0.6$.
Clearly, the stiffness of the column, $E_{\text{rel}}(x)$, decreases with $x$, as the volume fraction, $\phi(x)$, decreases from $\phi_0$ to 0.

The small deviation Euler buckling boundary value problem for Eq.~\eqref{eq_eq_scaled}, with $Y(0)=Y(1)=0$, takes the form of the Airy equation, 
 \begin{equation}
	Y^{\prime\prime}(x)+(a+b\,x)\,\Gamma\,Y(x)=0\,,
	\label{eq_equilibrium_DT}
\end{equation}
where $a=1/E_{\text{rel}}(\phi=\phi_0)$ and $b=(\phi_0/l)\,(5/2\, \gamma\prime-5/3)/(1+5/2\,\gamma\prime)$, the solutions to which can be written in terms of Airy functions $\Ai,\Bi$ as
\begin{align}
\label{eq_sol_bc}
	Y(x)  &= {\frac{C}{\Bi\left(-\Gamma_{\text{cr}}\,a/|-\Gamma_{\text{cr}}\,b|^{2/3}\right)}}\\
	&\times{\Bigg[}\Ai\left(\frac{-\Gamma_{\text{cr}}\,a-\Gamma_{\text{cr}}\,b\,x}{|-\Gamma_{\text{cr}}\,b|^{2/3}}\right) \Bi\left(\frac{-\Gamma_{\text{cr}}\,a}{|-\Gamma_{\text{cr}}\,b|^{2/3}}\right)\nonumber\\
	&-\Ai\left(\frac{-\Gamma_{\text{cr}}\,a}{|-\Gamma_{\text{cr}}\,b|^{2/3}}\right) \Bi\left(\frac{-\Gamma_{\text{cr}}\,a-\Gamma_{\text{cr}}\,b\,x}{|-\Gamma_{\text{cr}}\,b|^{2/3}}\right) {\Bigg]}\nonumber, 
\end{align}
where $C$ is a constant.  When the compression exerted on the ends of the inhomogeneous column exceeds a critical value, $\Gamma_{\text{cr}}$, the column deflections are given by Eq.~\eqref{eq_sol_bc} as a function of the strength of the gradient in inclusion volume fraction, $|{\phi_0}|$.
The values of $\Gamma_{\text{cr}}$ are now given by the non--trivial solutions of Eq.~\eqref{eq_equilibrium_DT}, and are the roots of the transcendental equation 
\begin{align}
	\Ai\left(\frac{-\Gamma}{|-\Gamma b|^{2/3}}\right)\Bi\left(\frac{-\Gamma a}{|-\Gamma b|^{2/3}}\right)=\nonumber\\ 
	\Ai\left(\frac{-\Gamma a}{|-\Gamma b|^{2/3}}\right)\Bi\left(\frac{-\Gamma}{|-\Gamma b|^{2/3}}\right), 
 \label{eq_condition_bc}
\end{align}
which we solve numerically to determine the failure modes.
Table~\ref{tab:table_gamma_cr} shows the results for the two lowest critical loads (i.e., the first two non--zero roots of Eq.~\ref{eq_condition_bc}) of inhomogeneous stiffened rods for a range of $\gamma\prime$ and $\phi_0$.

\begin{table}[t]
\begin{ruledtabular}
\begin{tabular}{ l c }
$\gamma\prime,~\phi_0$&
$\Gamma_{\text{cr}}^{(1)},~\Gamma_{\text{cr}}^{(2)}$\\
\colrule
	$\gamma\prime=100,~\phi_0=0.3$ & 11.574(4), 46.430(4)\\
	$\gamma\prime=100,~\phi_0=0.6$ & 13.929(0), 56.631(0)\\
	$\gamma\prime=10,~\phi_0=0.3$ & 11.392(9), 45.676(1)\\
	$\gamma\prime=10,~\phi_0=0.6$ & 13.427(4), 54.386(4)\\
\end{tabular}
\end{ruledtabular}
\caption{\label{tab:table_gamma_cr}%
	Critical loads of \emph{inhomogeneous} stiffened rods with a linear gradient of liquid inclusions,  Eq.~\eqref{eq_volume_fraction}, as a function of $\gamma\prime$ and $\phi_0$.
}
\end{table}

\begin{figure}[h]
	\psfrag{Y}[ct][ct][1.]{$Y^{\prime\prime}(x)$}
        \psfrag{x}[cb][cb][1.]{$x$}
\includegraphics[width=\linewidth]{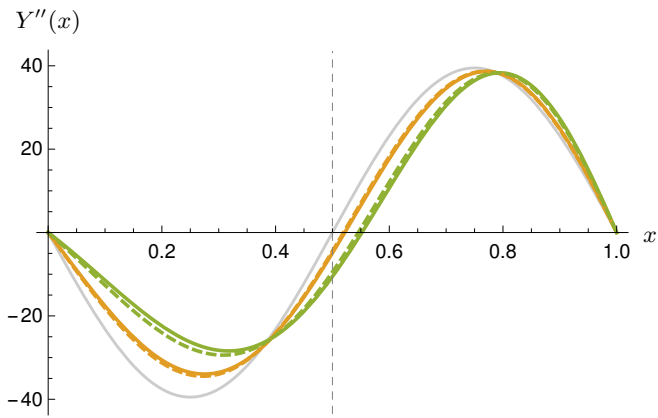}
\caption{Buckling curvature. The second derivative of the second buckling modes of homogeneous and linearly stiffened soft composite columns with the same potential energy.
The dashed vertical line denotes the position of the middle of the naturally straight rod; $x=1/2$.  
The same color legend as in Fig.~\ref{fig_buckling_DT} applies.}
\label{fig_2nd_buckling_deriv_DT}
\end{figure}

Substituting the critical loads of Table~\ref{tab:table_gamma_cr} into Eq.~\eqref{eq_sol_bc}, we obtain the first two buckling modes $(n=1,2)$, which we compare to the corresponding failure configurations of a column with a constant elasticity given by Eq.~\eqref{eq_reference_modes}
~\footnote{In order to benchmark our results for heterogeneous elastic rods against the constant Young's modulus case, we have to assume an additional constraint associated with the fact that the bending shapes are defined up to a constant. Here we impose the condition that the reference case Eqs.~\eqref{eq_reference_modes}, and the heterogeneous elastic rod Eqs.~\eqref{eq_sol_bc},\eqref{eq_sol_bc-Mori-Tanaka} have the same potential energy as given by Eq.~\eqref{eq_energy}.}. 
Fig.~\ref{fig_buckling_DT} shows how the axially varying Young's modulus, associated with a linear gradient of the liquid volume fraction, breaks the buckling symmetry associated with the Kirchhoff rod.
The distinction between the classical homogeneous and the heterogeneous column is seen in the first buckling mode, through the shift in the apex of the deflection towards the compliant end, $x=1$, as shown in the magnified inset of Fig.~\ref{fig_1st_buckling_DT}.  The distinction between the stiffened and compliant ends is more striking for the second buckling mode, shown in Fig.~\ref{fig_2nd_buckling_DT}, and enhanced when we plot the curvature of the profile as done in Fig.~\ref{fig_2nd_buckling_deriv_DT}.

Although we can tailor the response of the stiffened inhomogeneous column by changing $\gamma\prime$ and $\phi_0$, their effect is the same: the more we increase the composite Young's modulus at the stiffened end, either by reducing the inclusion size (increasing $\gamma\prime$), or by increasing the liquid volume fraction and gradient $\phi_0$, the more asymmetric the response.
Moreover, because we are comparing the buckling modes at the same potential energy, a larger stiffness implies a smaller deflection from the straight configuration.

\begin{figure}[t]
\begin{subfigure}[t]{0.5\textwidth}
	\psfrag{g}[ct][ct][1.]{$\Gamma$}
        \psfrag{e}[cb][cb][1.]{$\mu^{(i)}(\Gamma)$}
        \psfrag{u}[l][l][1]{$\mu^{(0)}(\Gamma)$}
        \psfrag{uu}[l][l][1]{$\mu^{(1)}(\Gamma)$}
        \psfrag{d}[l][l][1]{$\mu^{(2)}(\Gamma)$}
	\includegraphics[width=\linewidth]{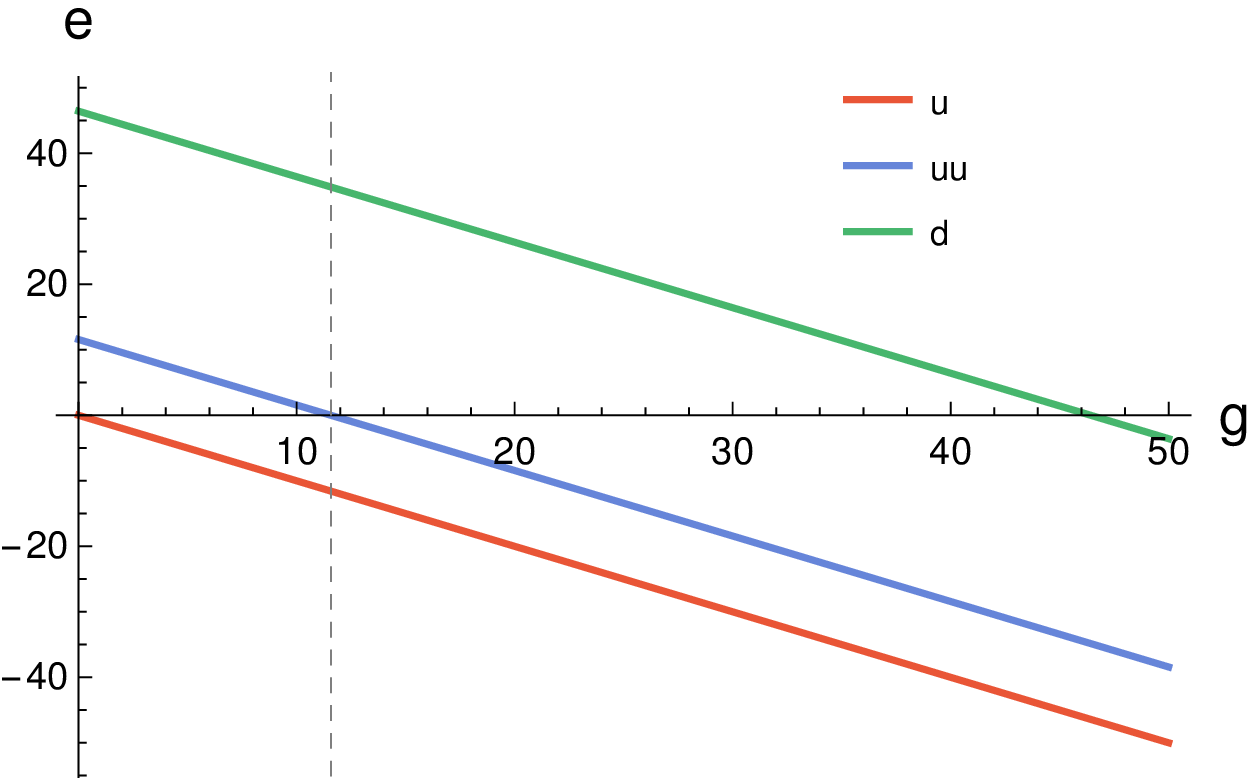}
        \caption{The three lowest eigenvalues.}
\label{fig_eigenvals_DT}
\end{subfigure}
\begin{subfigure}[t]{0.5\textwidth}
        \psfrag{s}[ct][ct][1.]{$x$}
        \psfrag{eta}[cb][cb][1.]{$\eta^{(i)}(x,\Gamma)$}
	\includegraphics[width=\linewidth]{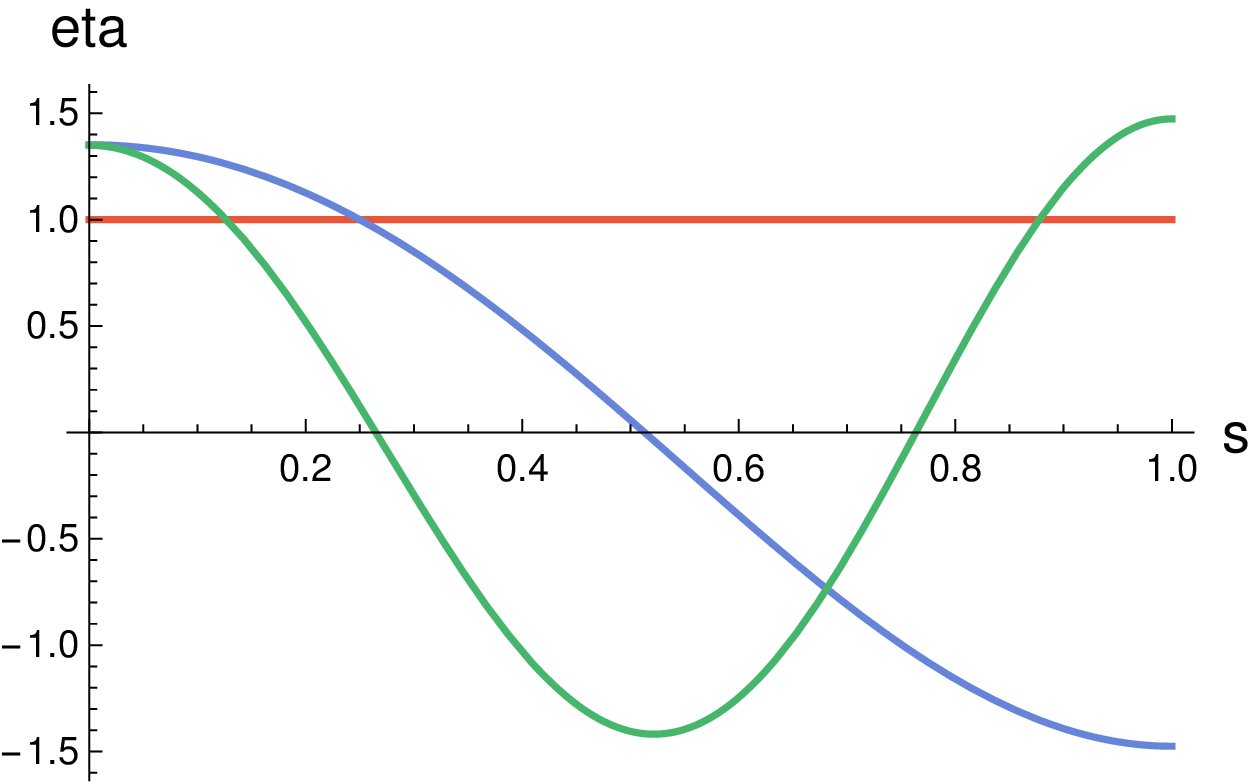}
        \caption{The three lowest eigenfunctions.}
\label{fig_eigenfuncs_1st_buckling_DT}
\end{subfigure}
	\caption{Stability analysis. Eigenvalue problem Eq.~\mref{eq_eigenval,eq_Neumann} for an \emph{inhomogeneous} stiffened rod characterized by the DT approach with a linear gradient of its liquid volume fraction, and parameters $\gamma\prime=100$ and $\phi_0=0.3$.
	Panel (a): The three lowest eigenvalues of the operator~\eqref{eq_operator_scaled}; the dashed vertical line pinpoints the critical compression $\Gamma_{\text{cr}}^{(1)}$.
	Panel (b): Corresponding eigenfunctions evaluated at the first buckling load $\Gamma=\Gamma_{\text{cr}}^{(1)}$.}
\label{fig_stability_DT}
\end{figure}

Finally, we compute the three smallest eigenvalues and eigenfunctions of operator~\eqref{eq_operator_scaled} under different loads $\Gamma$. This provides a numerical check of  Eq.~\eqref{eq_mu_Gamma}, which is essential in the stability analysis of inhomogeneous rods as described in \S~\ref{sec:stability_inhom}.  Fig.~\ref{fig_eigenvals_DT} shows that 
the eigenvalues decrease linearly with $\Gamma$, crossing the abscissa at $\Gamma=\Gamma_{\text{cr}}^{(i)}$ $\{\Gamma_{\text{cr}}^{(0)}=0,\Gamma_{\text{cr}}^{(1)}=11.574(4),\Gamma_{\text{cr}}^{(2)}=46.431(0)\}$ in agreement with Eq.~\eqref{eq_mu_Gamma} and the results in Table~\ref{tab:table_gamma_cr}.

Fig.~\ref{fig_eigenfuncs_1st_buckling_DT}, shows the corresponding eigenfunctions, $\nu^{(i)}(\Gamma)$, of the first buckling load $\Gamma=\Gamma_{\text{cr}}^{(1)}$.  Clearly, the eigenfunction $\eta^{(0)}(x,\Gamma)$ associated with the lowest eigenvalue is unity, and hence it is not a solution to the eigenvalue problem given by Eqs.~\mref{eq_eigenval,eq_Neumann,eq_orthogonal_eigenfunc}.

\subsection{Softened inhomogeneous composites \label{sec:soft}}

\begin{figure}[t]
\begin{subfigure}[b]{0.45\textwidth}
	\psfrag{Y}[ct][ct][1.]{$Y(x)$}
	\psfrag{Erel}[ct][ct][1.]{$E_{\text{rel}}(x)$}
        \psfrag{x}[cb][cb][1.]{$x$}
\includegraphics[width=\linewidth]{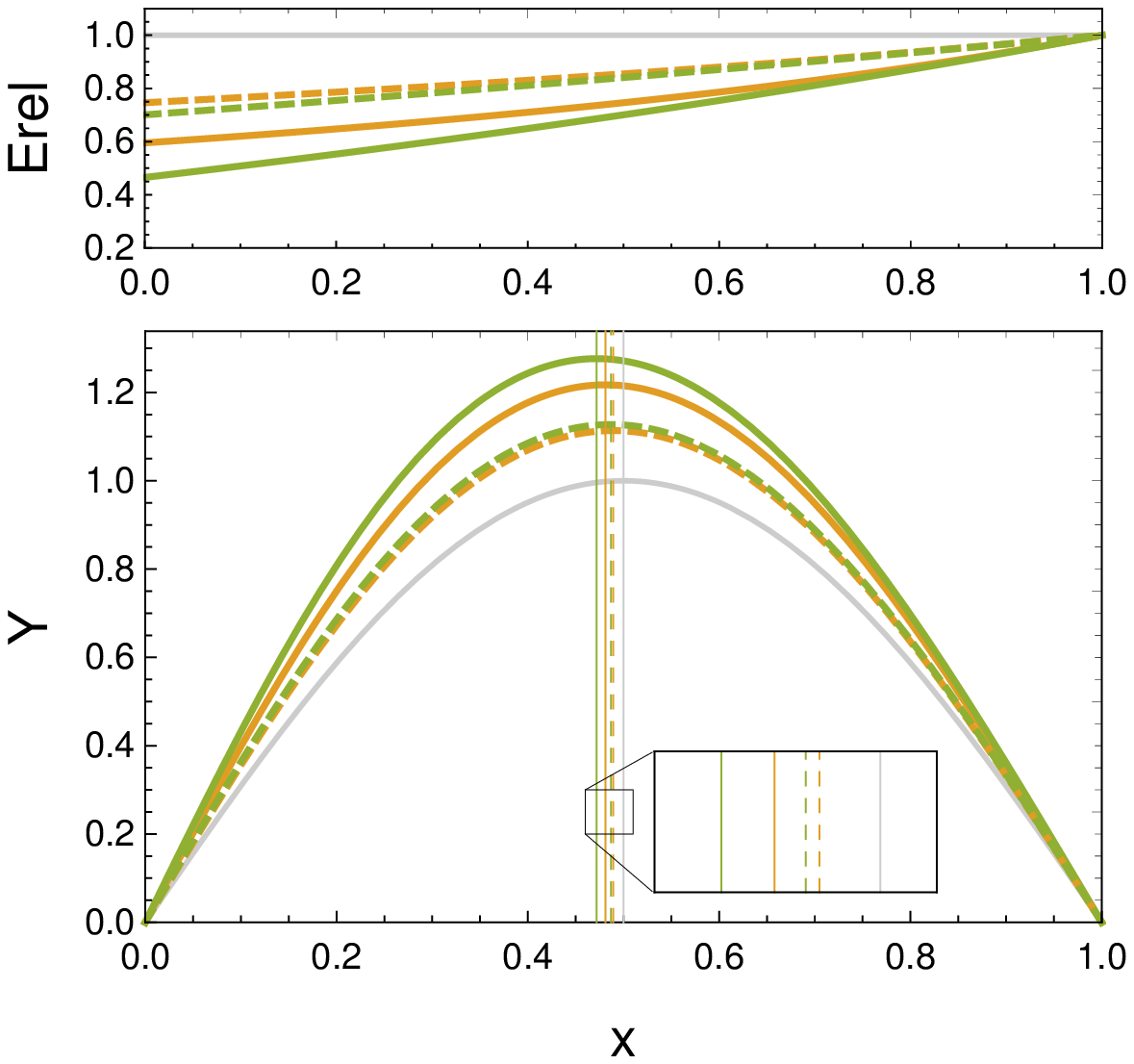}
	\caption{First buckling mode.}
\label{fig_1st_buckling_MSW_DT}
\end{subfigure}
\begin{subfigure}[b]{0.45\textwidth}
	\psfrag{Y}[ct][ct][1.]{$Y(x)$}
        \psfrag{x}[cb][cb][1.]{$x$}
	\psfrag{u}[l][l][1]{$E_{\text{rel}}=1$}
	\psfrag{d}[l][l][1]{$\phi_0=0.6$, (DT)}
	\psfrag{t}[l][l][1]{$\phi_0=0.6$, (MSW)}
	\psfrag{q}[l][l][1]{$\phi_0=0.3$, (DT)}
	\psfrag{c}[l][l][1]{$\phi_0=0.3$, (MSW)}
\includegraphics[width=\linewidth]{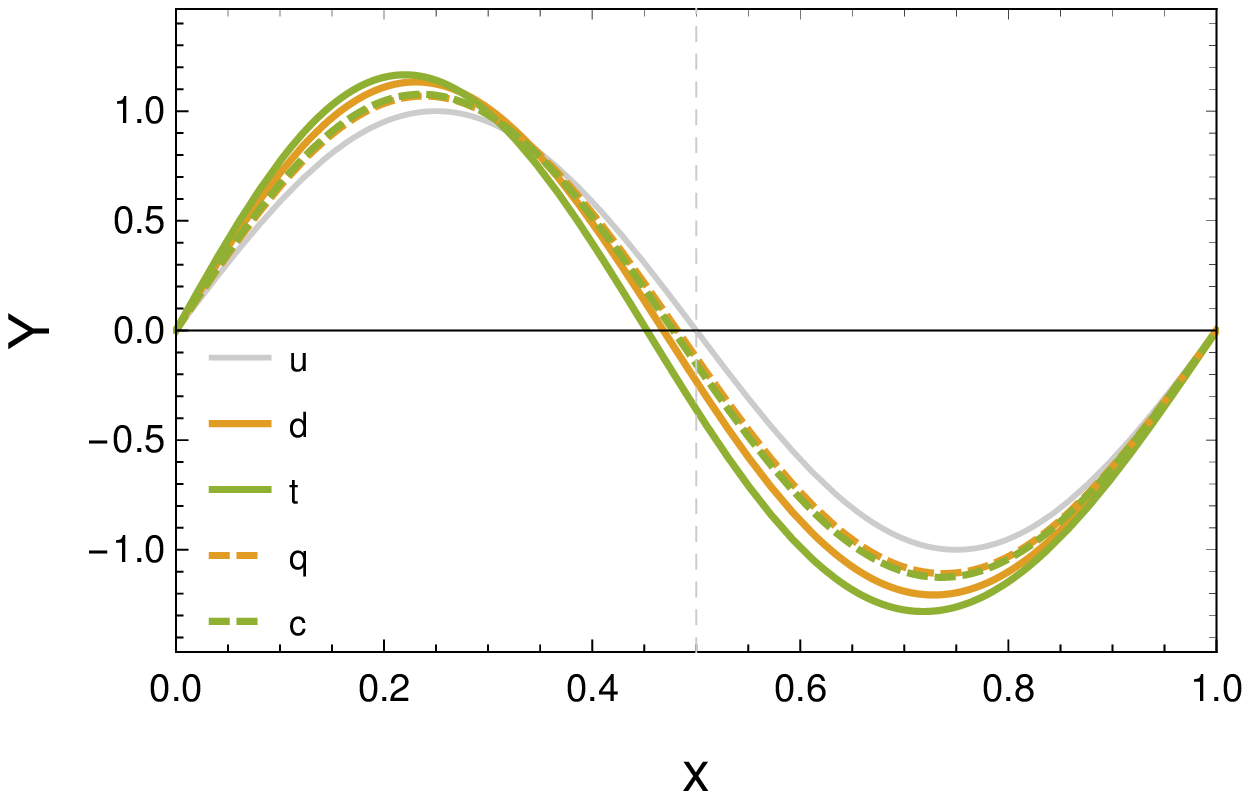}
	\caption{Second buckling mode.}
\label{fig_2nd_buckling_MSW_DT}
\end{subfigure}
\caption{Effective Young's Moduli and Buckling Modes.  The upper panel of (a) is the relative effective Young's moduli from Eqs.~\mref{eq_Ec,eq_Ec_Mori-Tanaka}
 as a function of $\gamma\prime$ and $\phi$.  
The first (a) (lower panel) and second (b) buckling modes of homogeneous and linearly softened soft composite columns with the same potential energy,
for $\gamma\prime=0.1$.
	The vertical lines in panel (a) denote the coordinate $x$ at which the color corresponding column reaches the maximum deflection.
	The dashed vertical line in panel (b) denotes the position of the middle of the undeformed rod; $x=1/2$.}
\label{fig_buckling_MSW_DT}
\end{figure}

In contrast to \S \ref{sec:stiff}, here we examine a {\em softened} composite ($\gamma\prime<2/3$) rod with a linear axial gradient of the liquid content using Eq.~\eqref{eq_volume_fraction}.  Moreover, we compare and contrast the DT expression for the effective Young's modulus, Eq.~\eqref{eq_Ec}, with the MSW theory given by Eq.~\eqref{eq_Ec_Mori-Tanaka}, which is valid in the non--dilute regime (see top panel in Fig.~\ref{fig_1st_buckling_MSW_DT}). 

In this case, the equilibrium equation \eqref{eq_eq_scaled} can be rewritten as
\begin{equation}
	\frac{d^2Y(x)}{dx^2}+\frac{a+b\,x}{c+d\,x}\,\Gamma\,Y(x)=0\,,
	\label{eq_equilibrium_MSW}
\end{equation}
where $a=2+(4/3)\phi_0+\gamma\prime(5-2\phi_0)$, $b=-(\phi_0/l)(4/3-2\gamma\prime)$, $c=2-2\phi_0+\gamma\prime(5+3\phi_0)$ and $d=-(\phi_0/l)(-2+3\gamma\prime)$.

The solution of the boundary value problem for Eq.~\eqref{eq_equilibrium_MSW} with $Y(0)=Y(1)=0$ is
\begin{widetext}
\begin{multline}
	Y(x)=\frac{C\,e^{-\frac{g(x,\Gamma_{\text{cr}})}{2}}\,e^{-\frac{g(0,\Gamma_{\text{cr}})}{2}}}{L_{-n(\Gamma_{\text{cr}})}^{(-1)}\left(g(0,\Gamma_{\text{cr}})\right)}\,\biggl[U(n(\Gamma_{\text{cr}}),0;g(x,\Gamma_{\text{cr}}))\,L_{-n(\Gamma_{\text{cr}})}^{(-1)}\left(g(0,\Gamma_{\text{cr}})\right)\\
	-U(n(\Gamma_{\text{cr}}),0;g(0,\Gamma_{\text{cr}}))\,L_{-n(\Gamma_{\text{cr}})}^{(-1)}\left(g(x,\Gamma_{\text{cr}})\right)\biggr]\,,\label{eq_sol_bc-Mori-Tanaka}
\end{multline}
	where $U(n(\Gamma),0;g(x,\Gamma))$ is the confluent hypergeometric function; $L_{n(\Gamma)}^{(-1)}(g(x,\Gamma))$ are the associated Laguerre polynomials;
$g(x,\Gamma)\equiv 2\rm{i}\frac{(c+d\cdot x)\sqrt{b\,\Gamma}}{d^{3/2}}$, $n(\Gamma)\equiv\rm{i}\frac{(-c\,b+a\,d)\sqrt{\Gamma}}{2\sqrt{b}\,d^{3/2}}$, and $C$ is a constant.
Because $b<0$ and $d>0$ when $\gamma\prime<2/3$, and $\Gamma>0$ under compression, then $g(x,\Gamma)$ and $n(\Gamma)$ are real--valued functions.
The critical buckling forces $\Gamma_{\text{cr}}$ are the roots of 
\begin{multline}
	\exp\left(-\frac{g(1,\Gamma)}{2}\right)\,U\left(n(\Gamma),0;g(1,\Gamma)\right)\, \exp\left(-\frac{g(0,\Gamma)}{2}\right)\,L_{-n(\Gamma)}^{(-1)}\left(g(0,\Gamma)\right)\\
	= \exp\left(-\frac{g(0,\Gamma)}{2}\right)\,U\left(n(\Gamma),0;g(0,\Gamma)\right)\, \exp\left(-\frac{g(1,\Gamma)}{2}\right)\,L_{-n(\Gamma)}^{(-1)}\left(g(1,\Gamma)\right)\,.
 \label{eq_condition_bc-Mori-Tanaka}
\end{multline}
\end{widetext}
Substituting the roots of Eq.~\eqref{eq_condition_bc-Mori-Tanaka}
\footnote{The critical loads for the buckling modes in Fig.~\ref{fig_buckling_MSW_DT}, $\gamma\prime=0.1$, are---MSW in first place, DT second: $\Gamma_{\text{cr}}^{(1)}=6.785(4),7.344(9)$; $\Gamma_{\text{cr}}^{(2)}=27.104(2),29.557(6)$, for $\phi_0=0.6$; and $\Gamma_{\text{cr}}^{(1)}=8.265(4),8.427(8)$; $\Gamma_{\text{cr}}^{(2)}=33.059(4),33.778(8)$, for $\phi_0=0.3$.}
into Eq.~\eqref{eq_sol_bc-Mori-Tanaka}, we find the corresponding buckling modes of the softened composite column.
They correspond to the equilibrium configurations of the rod when the axial load $\Gamma$ exceeds the first critical compression $\Gamma_{\text{cr}}^{(1)}$ and hence, the trivial solution $Y(x)=0$ becomes unstable.

In Fig.~\ref{fig_buckling_MSW_DT} we compare the first two modes with the homogeneous reference case of Eq.~\eqref{eq_reference_modes}, and with the DT result Eq.~\eqref{eq_sol_bc}.

We see in  Fig.~\ref{fig_buckling_MSW_DT} the anticipated asymmetric profile for the inhomogeneous column buckling modes.
Namely, the first buckling mode maximum deflection is shifted towards the compliant end to a degree that is proportional to the gradient of liquid content, $\phi_0$
(lower panel Fig.~\ref{fig_1st_buckling_MSW_DT}).

The curvature of the second buckling mode is a larger in the softer first half period than in the second, as seen in Fig.~\ref{fig_2nd_buckling_MSW_DT}.
Moreover, relative to the Kirchhoff case, a softened rod achieves a greater maximum deflection when the potential energy criterion is adopted.  We also see in Fig.~\ref{fig_buckling_MSW_DT} the expected difference between the MSW and the DT cases in the non--dilute regime ($\phi_0=0.6$) where the latter is invalid, but the buckling modes are indistinguishable in the dilute regime.

\subsection{Critical loads\label{sec:critical_loads}}

We have studied the basic geometry of the buckling modes through a comparison and contrast between homogeneous and inhomogeneous (stiffened and softened) composite columns.  Now we compare these cases in terms the magnitude of the critical collapse loads.  We assume a fixed inclusion size by holding $\gamma\prime$ constant, and introduce the average inclusion volume fraction $\bar\phi$.
We choose $\bar\phi = \phi_0/2$, which is the constant inclusion volume fraction for the Kirchhoff case and the average for the {inhomogeneous} rod with a linear gradient of liquid inclusions Eq.~\eqref{eq_volume_fraction}.  In this manner we can compare composite materials with the same liquid volume fraction and study the effect of spatial gradients on the collapse of a column.

\begin{figure}[t]
 \psfrag{y}[ct][ct][1.]{$\Gamma_{\text{cr}}$}
\psfrag{x}[ct][ct][1.]{$\bar\phi$}                                                                   
	\psfrag{q}[cl][cl][.8]{$\gamma\prime=0.01$ (MSW)}
        \psfrag{t}[cl][cl][.8]{$\gamma\prime=0.1$ (MSW)}
        \psfrag{d}[cl][cl][.8]{$\gamma\prime=10$ (DT)}
        \psfrag{u}[cl][cl][.8]{$\gamma\prime=100$ (DT)}
 \includegraphics[width=\linewidth]{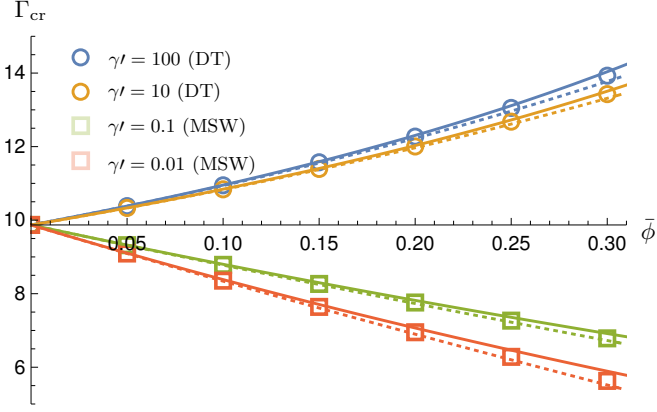}
	\caption{Critical forces for the first buckling mode as a function of the average inclusion volume fraction, obtained as the numerical roots of the transcendental equation considering the dilute theory (DT, circles), and of the non--dilute Mori--Tanaka approach (MSW, squares).
        Both stiffened ($\gamma\prime=100,10$) and softened ($\gamma\prime=0.1,0.01$) rods are considered.
	The analytic critical compression for the homogeneous rod, Eq.~\eqref{eq_gamma_crit}, is depicted with color lines as per the color legend for symbols.
	The color legend also applies for the dotted lines depicting the perturbation approximation Eq.~\eqref{eq_gamma_epsilon}, with Eqs.~\mref{eq_gamma_crit,eq_gamma2_epsilon_DT,eq_gamma2_epsilon_MSW}, to the critical compression.}
 \label{fig_critical_tension-DT-MSW}
\end{figure}

We plot in Fig.~\ref{fig_critical_tension-DT-MSW} the
critical loads as function of $\bar\phi$ for the inhomogeneous stiffened and softened columns, together with the critical compression force of a homogeneous rod Eq.~\eqref{eq_gamma_crit}, derived in \S \ref{sec:homogeneous}. 
Unlike the previous sections, in which the reference Kirchhoff rod had a unit Young's modulus $E_{\text{rel}}=1$ as if it were a single compound elastic material, we now consider the homogeneous rod stiffness to be given by either the DT Eq.~\eqref{eq_Ec} or the MSW Eq.~\eqref{eq_Ec_Mori-Tanaka} effective Young's modulus, but with constant values of $\gamma\prime$ and $\phi(x)=\bar\phi=\phi_0/2$.

The stiffening and softening regimes are in clear evidence in Fig.~\ref{fig_critical_tension-DT-MSW}.  
As the average fluid inclusion volume increases, the critical compression becomes larger (smaller) in stiffening (softening) conditions.  This qualitative behavior is found in both homogeneous and inhomogeneous columns.  Moreover, despite remarkably good numerical agreement between the analytical critical compression for the Kirchhoff rod, Eq.~\eqref{eq_gamma_crit}, and the numerical results for inhomogeneous columns, there is a systematic deviation between them in that the critical loads of the \emph{inhomogeneous} rod are \emph{smaller}.
This deviation increases with $\bar\phi$, and hence the spatial gradient of inclusions.

\begin{figure}[t]
\psfrag{x}[ct][ct][1.]{$x$}
\psfrag{Erel}[ct][ct][1.]{$E_{\text{rel}}$}
\psfrag{ez}[ct][ct][1.]{$\varepsilon=0$}
\psfrag{eu}[ct][ct][1.]{$\varepsilon=1$}
 \includegraphics[width=\linewidth]{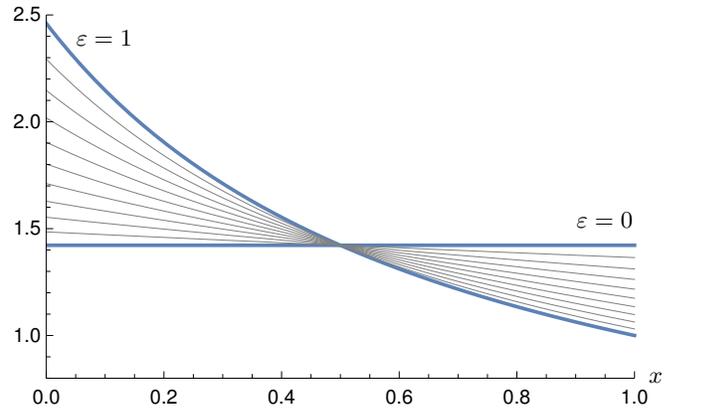}
        \caption{Effective Young's modulus Eq.~\eqref{eq_Ec} with inclusions volume fraction profile controlled by $\varepsilon$ varying between 0 and 1; $\phi_0=0.6$ and $\gamma\prime=100$.}
 \label{fig_Erel_epsilon_DT}
\end{figure}

We use perturbation methods~\cite{BenderOrszag} to explain this deviation as follows.  We approximate the inhomogeneous case from the analytic solution to the equilibrium equation~\eqref{eq_eq_scaled} for a homogeneous rod as a power series in $\varepsilon \ll 1$.  The spatial variation is modeled as
\begin{align}
        \phi(x,\varepsilon)=\frac{\phi_0}{2}+\phi_0\left(\frac{1}{2}-x\right)\,\varepsilon,\label{eq_volume_fraction_epsilon}
\end{align}
so that $\varepsilon=0$ corresponds to the homogeneous rod and the degree of inhomogeneity increases with $\varepsilon$.  See the Young's modulus in Fig.~\ref{fig_Erel_epsilon_DT}.

To leading order, the perturbation approximation for the critical compression of the  first mode
\begin{align}
        \Gamma_{\text{cr}}^{(1)}=\sum_{l=0}^{\infty}\Gamma_{\text{cr},l}^{(1)}\,\varepsilon^l,\label{eq_gamma_epsilon}
\end{align}
equals Eq.~\eqref{eq_gamma_crit}.
The first--order correction vanishes for both the DT and MSW models, and hence the second--order correction accounts for the principal deviations from leading order.
For the DT and the MSW approaches, this second--order corrections are:
\begin{widetext}
\begin{align}
	\Gamma_{\text{cr,DT},2}^{(1)}=-\frac{\pi^2}{2}\phi_0^2\frac{\left(1+\frac{5}{2}\gamma\prime\right)\left(\frac{5}{2}\gamma\prime-\frac{5}{3}\right)^2}{\left[1+\frac{5}{2}\gamma\prime-\frac{\phi_0}{2}\left(\frac{5}{2}\gamma\prime-\frac{5}{3}\right)\right]^3}\,\left(-\frac{1}{24}-\frac{1}{\pi}+\frac{1}{4\pi^2}+\frac{12}{\pi^3}\right)~\textrm{and}~\label{eq_gamma2_epsilon_DT}
\end{align}
	\begin{multline}
        \Gamma_{\text{cr,MSW},2}^{(1)}=-\pi^2\phi_0^2\frac{10}{3}\frac{(2-3\gamma\prime)^2(2+5\gamma\prime)}{\left(2+\frac{2}{3}\phi_0+\gamma\prime(5-\phi_0)\right)^2\left(2-\phi_0+\gamma\prime\left(5+\frac{3}{2}\right)\right)}\\
        \cdot\left[\frac{5}{12}\frac{(2+5\gamma\prime)}{2+\frac{2}{3}\phi_0+\gamma\prime(5-\phi_0)}\left(-\frac{1}{24}-\frac{1}{\pi}+\frac{1}{4\pi^2}+\frac{12}{\pi^3}\right)+\frac{1}{24}-\frac{1}{4\pi^2}\right],\label{eq_gamma2_epsilon_MSW}
\end{multline}
\end{widetext}
respectively.  Further details of the perturbative calculations can be found in the Supplemental Material in Appendix~\ref{SI}.

In Fig.~\ref{fig_critical_tension-DT-MSW} we plot the critical compression from Eqs. \mref{eq_gamma_crit,eq_gamma2_epsilon_DT,eq_gamma2_epsilon_MSW} along with 
perturbation expansion, Eq.~\eqref{eq_gamma_epsilon}. We note that the numerical results of the {inhomogeneous} rod---symbols in Fig.~\ref{fig_critical_tension-DT-MSW}---correspond to $\varepsilon\to 1$ in the perturbation theory, although the accuracy of the latter is restricted to $\varepsilon <<1$.  Thus, whereas we cannot reproduce the {inhomogeneous} case with the perturbation theory
the latter provides  valuable insight into the interpretation of the results.  In particular, the first non--zero correction to the homogeneous critical compression in Eq.~\eqref{eq_gamma_epsilon} is negative for any set of parameters in both the DT and MSW models.  Therefore, spatial gradients in the elastic properties of a column promotes buckling under smaller loads.

Physically,  the fact that an {inhomogeneous} elastic rod buckles more easily can be seen in terms of the perturbed Young's modulus plotted in Fig.~\ref{fig_Erel_epsilon_DT}.
Namely, a gradient in the elasticity gives one half of the rod with $E_{\text{rel}}(x)<E_{\text{rel}}(\phi=\phi_0/2)$, which is softened relative to the homogeneous counterpart.  This weaker region is sufficient to lead to collapse under lower loads despite the fact that the other half of the rod has an increased stiffness compared to the homogeneous reference case.

\subsection{``Polar'' elasticity\label{sec:polar}}

In \S \ref{sec:stiff} and \ref{sec:soft} we either linearly stiffened or softened a column using Eq.~\eqref{eq_volume_fraction} for a given ratio of elastocapillary length to inclusion radius, $\gamma\prime$.  In consequence, the bulk modulus either decreased or increased with distance along the column.
Here we consider a constant inclusion volume fraction $\phi$ and a linear variation in $\gamma\prime(x)$ as in Eq.~\eqref{eq_gamma}, to create a ``polar'' rod that changes from softened to stiffened along its axis.

As described in \S \ref{sec:composite} the DT and MSW extensions of Eshelby's theory are equivalent in the dilute limit.  However, in the softening (stiffening) regime in the non-dilute limit the effective Young's modulus estimates of \citet{StyleBoltyanskiy15,StyleWettlaufer15} deviate from the three--phase model of \citet{MancarellaStyle16b,MW2017} (MSW method \cite{MancarellaStyle16}).  
Here, we solve the equilibrium equation Eq.~\eqref{eq_eq_scaled} for $\gamma\prime(x)$ as in Eq.~\eqref{eq_gamma}, $\phi$ constant, and use the DT and MSW theories in the dilute limit. 

The resulting effective Young's modulus, for both the DT and the MSW models, can be expressed as a ratio of two linear polynomials.  The equilibrium equation corresponding to Eq.~\eqref{eq_eq_scaled} is thus mathematically analogous to Eq.~\eqref{eq_equilibrium_MSW}, namely the MSW theory with $\phi(x)$ obeying Eq.~\eqref{eq_volume_fraction} with constant $\gamma\prime$ treated in \S \ref{sec:soft}.
Thus, as in Eq.~\eqref{eq_sol_bc-Mori-Tanaka}, we can write the deflections of a composite polar rod with hinged ends in terms of the confluent hypergeometric function and the associated Laguerre polynomials.

The coefficients in Eq.~\eqref{eq_sol_bc-Mori-Tanaka} when $\phi$ is constant and $\gamma\prime$ obeys the linear gradient  of Eq.~\eqref{eq_gamma}, are: $a=(1+5/3\,\phi)+5/2\,(1-\phi)\gamma_0\prime$, $b=-5/2\,(1-\phi)(\gamma_0\prime-\gamma_1\prime)$, $c=1+5/2\, \gamma_0\prime$, and $d=-5/2\, (\gamma_0\prime-\gamma_1\prime)$, for the DT approach; $a=2+4/3\,\phi+(5-2\phi)\gamma_0\prime$, $b=-(5-2\phi)(\gamma_0\prime-\gamma_1\prime)$, $c=2-2\phi+(5+3\phi)\gamma_0\prime$, and $d=-(5+3\phi)(\gamma_0\prime-\gamma_1\prime)$, for the MSW approach.
Parameters $b$ and $d$ for both models are negative when $\gamma_0\prime>\gamma_1\prime$ ($0\leq\phi\leq 1$) yielding complex roots of Eqs.~\eqref{eq_sol_bc-Mori-Tanaka} and~\eqref{eq_condition_bc-Mori-Tanaka}. 
However, we note that in both of these equations $y_1(x)\equiv\exp(-g(x,\Gamma)/2)\,U(n(\Gamma),0;g(x,\Gamma))$ and $y_2(x)\equiv\exp(-g(x,\Gamma)/2)\,L_{-n(\Gamma)}^{(-1)}(g(x,\Gamma))$ 
are eigenfunctions of the linear differential operator acting on $Y(x)$ and have eigenvalue 0, corresponding to the equilibrium equation~\eqref{eq_equilibrium_MSW}.
We noted at the outset of \S \ref{sec:inhomogeneous} that, for either the DT or MSW effective Young's moduli, 
the second--order linear operator of Eq.~\eqref{eq_eq_scaled}, with Dirichlet boundary conditions, is Hermitian.
Therefore, Eqs.~\eqref{eq_sol_bc-Mori-Tanaka} and~\eqref{eq_condition_bc-Mori-Tanaka} can be rewritten in terms of linear combinations of the real-valued eigenfunctions $y_1(x)$ and $y_2(x)$.  Hence, since the deflection of a polar compressed rod given by Eq.~\eqref{eq_sol_bc-Mori-Tanaka} is defined up to an undetermined constant, without loss of generality we consider the real part of Eq.~\eqref{eq_sol_bc-Mori-Tanaka} for the remainder of this subsection. 

\begin{figure}
\psfrag{ve}[ct][ct][1.]{$E_{\text{rel}}(\gamma\prime)$}
\psfrag{ho}[ct][ct][1.]{$\gamma\prime$}
\psfrag{so}[ct][ct][1.]{Soft}
\psfrag{st}[ct][ct][1.]{Stiff}
	\psfrag{DT}[cr][cr][1.]{DT}
	\psfrag{MT}[cr][cr][1.]{MSW}
 \includegraphics[width=\linewidth]{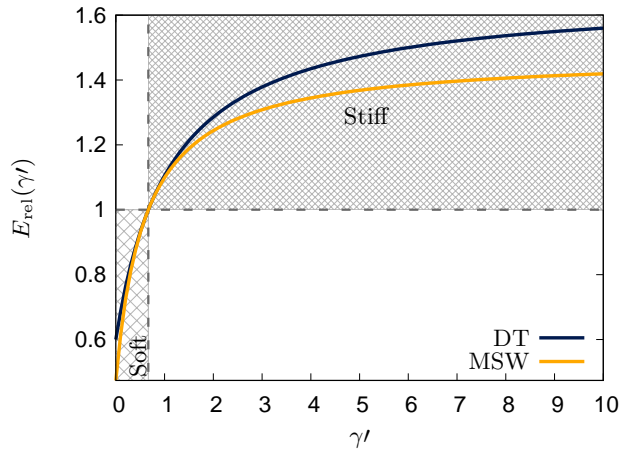}
	\caption{Effective Young's modulus as function of $\gamma\prime$, when $\phi=0.4$, both for the DT, Eq.~\eqref{eq_Ec}, and the MSW, Eq.~\eqref{eq_Ec_Mori-Tanaka} models. The vertical dashed line  delineates the softening $\gamma\prime<2/3$, and stiffening $\gamma\prime>2/3$ regimes.}
\label{fig_Ec_gammap}
\end{figure}

\begin{figure}[h]
\begin{subfigure}[b]{0.45\textwidth}
	\psfrag{Y}[ct][ct][1.]{$Y(x)$}
	\psfrag{Erel}[ct][ct][1.]{$E_{\text{rel}}(x)$}
        \psfrag{x}[cb][cb][1.]{$x$}
\includegraphics[width=\linewidth]{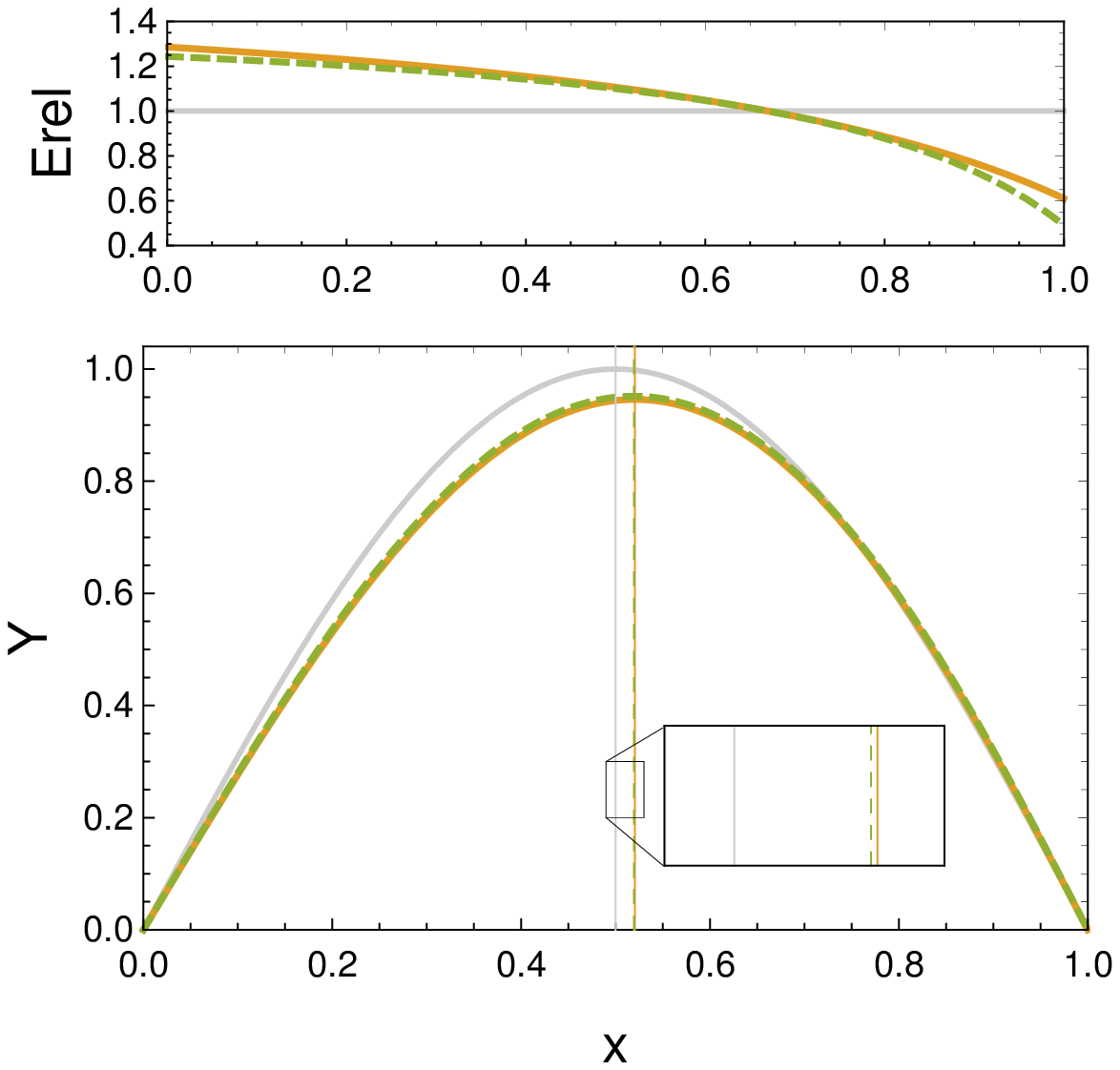}
	\caption{First buckling mode.}
\label{fig_1st_buckling_polar}
\end{subfigure}
\begin{subfigure}[b]{0.45\textwidth}
	\psfrag{Y}[ct][ct][1.]{$Y(x)$}
        \psfrag{x}[cb][cb][1.]{$x$}
	\psfrag{u}[l][l][1]{$E_{\text{rel}}=1$}
	\psfrag{d}[l][l][1]{Linear $\gamma\prime(x)$, (DT)}
	\psfrag{t}[l][l][1]{Linear $\gamma\prime(x)$, (MSW)}
\includegraphics[width=\linewidth]{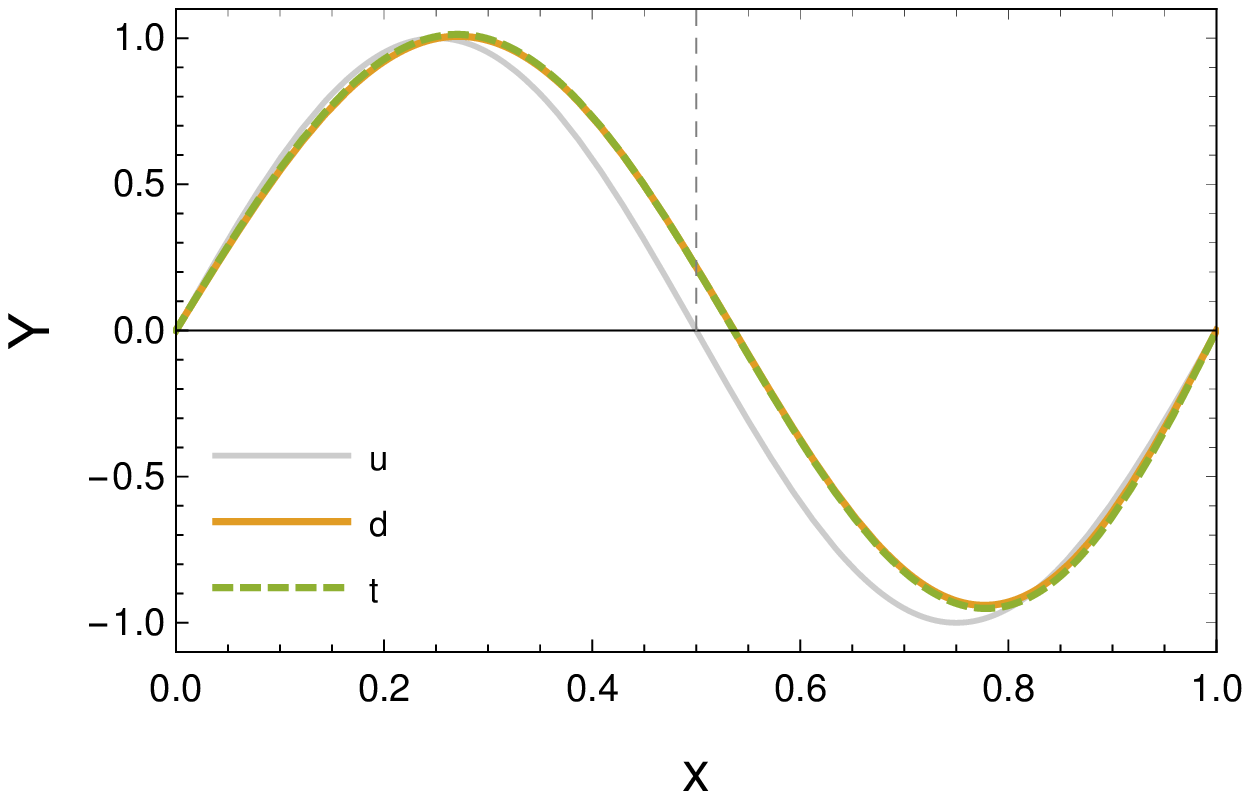}
\caption{Second buckling mode.}
\label{fig_2nd_buckling_polar}
\end{subfigure}
	\caption{The buckling modes of polar elastic rods, from the real part of Eq.~\eqref{eq_sol_bc-Mori-Tanaka}, compared to the homogeneous rod. The DT and the MSW models are used to compute the composite Young's modulus, Eqs.~\mref{eq_Ec,eq_Ec_Mori-Tanaka}, plotted in the upper panel of (a), with $\gamma\prime_0=2,\,\gamma\prime_1=0.01,\,\phi=0.4$.
	}
	\label{fig_buckling_polar}
\end{figure}

As shown in Fig.~\ref{fig_Ec_gammap}, the effective Young's moduli from Eqs.~\eqref{eq_Ec} and \eqref{eq_Ec_Mori-Tanaka} exhibit a nonlinear dependence on $\gamma\prime$.  Therefore, the range of Young's modulus is limited in the dilute limit with a linear model such as that in Eq.~\eqref{eq_gamma}.  The buckling modes of a compressed polar elastic compressed rod using Eq.~\eqref{eq_gamma} are shown in Fig.~\ref{fig_buckling_polar} 
	\footnote{The numerical results for the critical forces corresponding to the first two buckling modes of a polar rod with \{$\gamma\prime_0=2,\,\gamma\prime_1=0.01,\,\phi=0.4$\} are $\Gamma_{\text{cr}}^{(1)}=10.601(6)$ and $\Gamma_{\text{cr}}^{(2)}=41.265(1)$ (DT), and $\Gamma_{\text{cr}}^{(1)}=10.502(6)$ and $\Gamma_{\text{cr}}^{(2)}=40.549(2)$ (MSW)\label{foot_polar}}, 
wherein as in previous sections these modes have the same potential energy, given by Eq.~\eqref{eq_energy}.

Without going too deeply into the non--dilute regime, an intermediate volume fraction ($\phi=0.4$) was chosen to tailor the transition of the Young's modulus midway between the rod's ends (see Fig.~\ref{fig_1st_buckling_polar}--upper panel).  Indeed, although strictly speaking $\phi=0.4$ is non--dilute, we note that there is near perfect agreement between the DT and MSW models.

On the one hand, the ``polar'' elasticity drives an asymmetrical response to a critical compression, qualitatively like the stiffening (\S \ref{sec:stiff}) and softening (\S \ref{sec:soft}) cases in that the maximum of the first buckling mode is shifted towards the compliant end (Fig.~\ref{fig_1st_buckling_polar}), and the curvature of the second buckling mode is reduced (increased) in the stiff (compliant) half--period, as shown in Fig.~\ref{fig_2nd_buckling_polar}.  On the other hand, the extreme values of the profiles of the ``polar'' configuration are closer to the reference Kirchhoff rod of the same energy.  Namely, whereas a ``polar'' rod buckles unevenly, it does so within the same deflection range as the Kirchhoff rod.

\section{Conclusions}\label{sec:conclusions}
We have studied the buckling of \emph{inhomogeneous} soft composite columns, or rods, with axially varying elasticity.
Their spatial structure is tailored either by changing the volume fraction  of inclusions ($\phi$) or the ratio of the elastocapillarity length to the inclusion size ($\gamma\prime$) along the column axis.
We have extended the classical theoretical description of compressed elastic rods by incorporating these inclusion/host surface tension effects on the effective elastic modulus of the mixture.
The resulting equilibrium equation that accounts for the rod's response to a compressing force has analytical solutions for a linear model for the variation of either $\phi$ or $\gamma\prime$. 
This provides a framework of broad relevance to soft composite materials and could be tested by considering a distribution of sizes of the liquid droplets embedded in a soft solid host, which might be possible by suitable variations of the experiment described by \citet{StyleBoltyanskiy15}. 

We studied three different problems of heterogeneous rods: stiffening (softening) by increasing the volume fraction of small (large) inclusions (i.e., of radius $R<(3/2)L$ ($R>(3/2)L$)) as a function of distance along the axis as described by Eq.~\eqref{eq_volume_fraction}, and ``polar'' elasticity by introducing a gradient of the inclusion size by Eq.~\eqref{eq_gamma} at a constant volume fraction.  Their principal common feature is the intuitive result that a compressed column of variable elasticity bends most easily where it is softer.
Accordingly, the symmetric buckling characteristic of a rod with homogeneous properties is broken whenever a rod with {inhomogeneous} stiffness is considered. In both the dilute and non--dilute regimes, these three general cases, {\em mutatis mutandis}, include the overall behavior of the of the buckling instabilities in linearly heterogeneous soft composite rods.

When comparing the critical compression forces of homogeneous and {inhomogeneous} rods with the same average inclusion volume fraction, we find that the latter exhibit smaller values.  
Perturbation theory provided both a physical and mathematical  rationale for why {inhomogeneous} columns fail under smaller loads.

We have studied these rather simple models with the hope that considering an elementary architecture might motivate experimental testing of these ideas as well as their potential for biological relevance \cite{Goriely2017-4,Goriely2017-5}. 
Indeed, recent advances in understanding multi--scale morphomechanical effects in biological systems \cite{Hofhuis2016} and experimental realizations of soft--solid cavitation~\cite{KimLiu20} and of stiffness--modulated elastic solids~\cite{RivaRosa20},  provide useful techniques and settings to implement and explore our models.

\begin{acknowledgements}
The authors thank Crist\'obal Arratia, Anthony Bonfils and Alain Goriely for helpful comments. 
The authors acknowledge the support of Swedish Research Council Grant No. 638-2013-9243.
\end{acknowledgements}

\bibliography{bibliography}

\appendix
\renewcommand\thefigure{\thesection\arabic{figure}} 
\setcounter{figure}{0}
\input{Buckling_SI.tex} 

\end{document}

%% file: Buckling_SI.tex
\title{Supplemental Material: Controlling collapse of soft composites with capillarity}
\author{Marc Su\~n\'e}
\email{marc.sune.simon@su.se}
\affiliation{Nordita, Royal Institute of Technology and Stockholm University, Roslagstullbacken 23, SE-106 91 Stockholm, Sweden}
\author{John S. Wettlaufer}
\affiliation{Yale University, New Haven, CT 06520, USA}
\affiliation{Nordita, Royal Institute of Technology and Stockholm University, Roslagstullbacken 23, SE-106 91 Stockholm, Sweden}

\section{Supplemental Material}
\label{SI}
We develop a theory that approximates the critical compression of \emph{inhomogeneous} rods in the limit of small deflections, so that the equilibrium condition is 
\begin{align}
	\frac{d^2 Y(x)}{dx^2}+\frac{\Gamma}{E_{\text{rel}}} Y(x)=0, \label{SI:eq_eq_scaled}
\end{align}
with boundary conditions, 
\begin{align}
	Y(0)=Y(1)=0.\label{SI:eq_bc}
\end{align}

As described in the main text, the composite Young's moduli are expressed in terms of the elastic moduli of the host material (Young's modulus $E$ and Poisson's ratio $\nu$),  the dimensionless ratio 
$\gamma\prime\equiv L/R$ of the elastocapillarity length $L$ to the inclusion radius $R$, and the inclusion volume fraction $\phi$, in two forms: 
(a) the dilute theory (DT) due to \citet{StyleBoltyanskiy15,StyleWettlaufer15}; 
\begin{align} 
	E_{\text{rel,DT}}(\phi,\gamma\prime)=\frac{1+\frac{5}{2}\gamma\prime}{\frac{5}{2}\gamma\prime(1-\phi)+\left(1+\frac{5}{3}\phi\right)},\label{SI:eq_Ec} 
\end{align}
or (b) the MSW method of \citet{MancarellaStyle16} wherein 
\begin{align}  
	E_{\text{rel,MSW}}(\phi,\gamma\prime)=\frac{2-2\phi+\gamma\prime(5+3\phi)}{2+(4/3)\,\phi+\gamma\prime(5-2\phi)}, 
 \label{SI:eq_Ec_Mori-Tanaka}  
\end{align} 
which exploits the equivalence between a droplet embedded in an elastic solid with an isotropic interfacial tension and a corresponding elastic inclusion with no interfacial tension. 
Both cases assume an incompressible host; $\nu=1/2$. 
The former will be of particular relevance for stiffening conditions, $\gamma\prime>2/3$, whereas the latter for softened composites, $\gamma\prime<2/3$.

\subsection{Perturbation theory}
The equilibrium equation~\eqref{SI:eq_eq_scaled} for a homogeneous rod with fixed inclusion volume fraction $\phi=\bar\phi$ and $\gamma\prime$, and hence a constant Young's modulus (either Eq.~\mref{SI:eq_Ec} or \mref{SI:eq_Ec_Mori-Tanaka}), can be solved analytically.
Thus, the critical compressing force and the buckling modes for the Kirchhoff rod are
\begin{eqnarray}
	\Gamma_{\text{cr},0}^{(n)}&=&(n\pi)^2 E_{\text{rel}}(\bar\phi),\qquad n=1,2,\dots\label{SI:eq_gamma_crit}\\
	Y_{0}^{(n)}(x)&\propto&\sin(n\pi x).\label{SI:eq_reference_modes}
\end{eqnarray}
We introduce two indices here: the superscript $n$ accounts for the buckling mode, whereas the subscript $l$ ($l=0$ until otherwise specified) labels each term in a perturbative expansion for the solution of Eq.~\eqref{SI:eq_eq_scaled}.

When we consider a linear inclusion volume fraction profile as
\begin{align}
	\phi(x)=\phi_0(1-x),\label{SI:eq_volume_fraction}
\end{align}
the rod's Young's modulus varies parallel to its axis and the eigenvalue problem in Eqs.~\mref{SI:eq_eq_scaled, SI:eq_bc} has no closed form solution, because the eigenvalues depend on the numerical roots of the transcendental equation.
We use perturbation theory~\cite[see e.g.,][]{BenderOrszag} to approximate the critical compression in the eigenvalue problem Eqs.~\mref{SI:eq_eq_scaled, SI:eq_bc} when $E_{\text{rel}}$ is given by Eqs.~\mref{SI:eq_Ec,SI:eq_Ec_Mori-Tanaka} with a constant $\gamma\prime$ and a linear inclusion volume fraction profile, viz., Eq.~\eqref{SI:eq_volume_fraction}.

\begin{figure}                                                                                        
\begin{subfigure}[b]{0.45\textwidth}
\psfrag{x}[ct][ct][1.]{$x$}
\psfrag{Erel}[ct][ct][1.]{$E_{\text{rel}}$}                                                                   
\psfrag{ez}[ct][ct][1.]{$\varepsilon=0$}                                                                   
\psfrag{eu}[ct][ct][1.]{$\varepsilon=1$}                                                                   
 \includegraphics[width=\linewidth]{Erel_epsilon_DT_100.eps}
 \caption{DT, $\gamma\prime=100$.}
 \label{SI:fig_Erel_epsilon_DT}
 \end{subfigure}
 \begin{subfigure}[b]{0.45\textwidth} 
\psfrag{x}[ct][ct][1.]{$x$}
\psfrag{Erel}[ct][ct][1.]{$E_{\text{rel}}$}                                                                   
\psfrag{ez}[ct][ct][1.]{$\varepsilon=0$}                                                                   
\psfrag{eu}[ct][ct][1.]{$\varepsilon=1$}                                                                   
 \includegraphics[width=\linewidth]{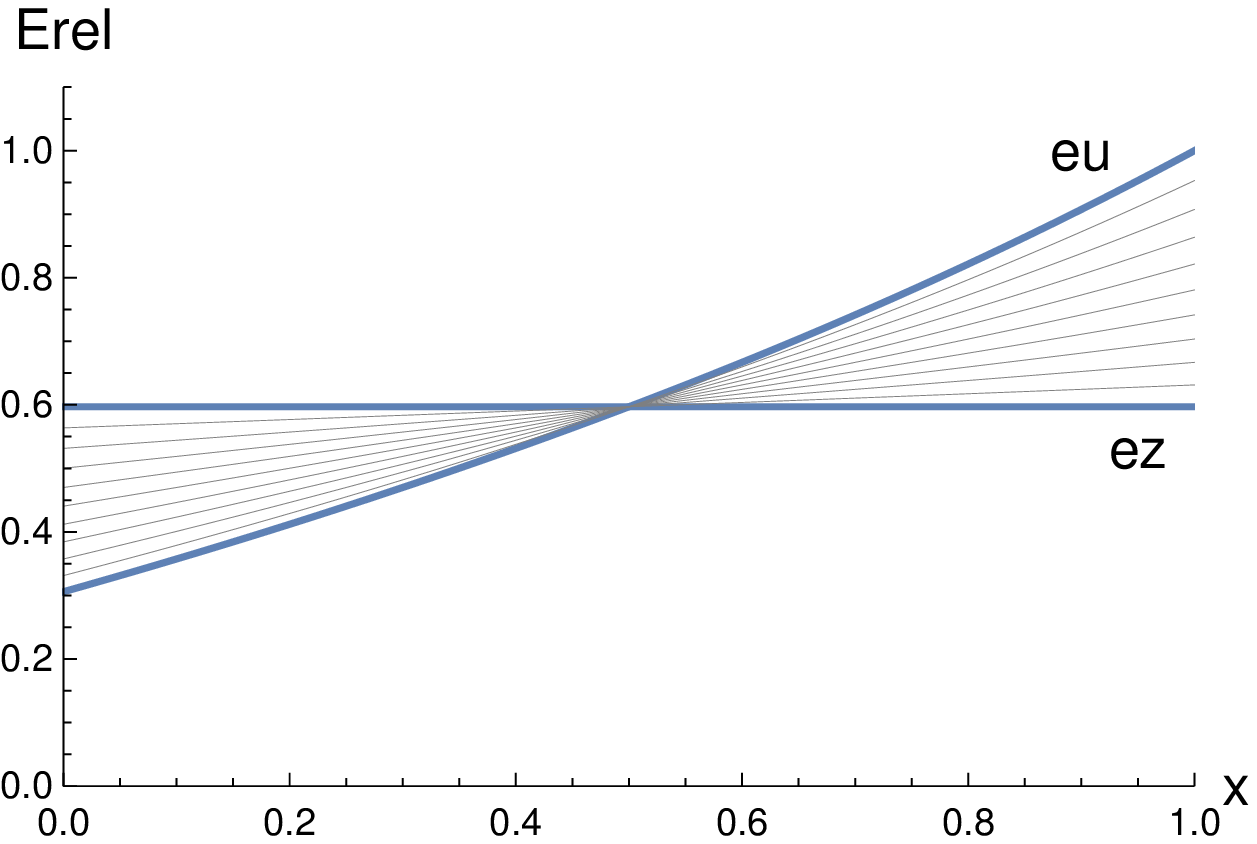}
 \caption{MSW, $\gamma\prime=0.01$}
 \label{SI:fig_Erel_epsilon_MSW}
 \end{subfigure}                                                                                      
	\caption{Effective Young's modulus Eqs.~\mref{SI:eq_Ec,SI:eq_Ec_Mori-Tanaka} with inclusions volume fraction as given by Eq.~\eqref{SI:eq_volume_fraction_epsilon} and $\varepsilon$ varying between 0 and 1; $\phi_0=0.6$.} 
 \label{SI:fig_Erel_epsilon}
\end{figure}

Because the average inclusion volume fraction for a linear profile Eq.~\eqref{SI:eq_volume_fraction} is $\bar\phi=\phi_0/2$, we can replace $\phi(x)$ by
\begin{align}
	\phi(x,\varepsilon)=\frac{\phi_0}{2}+\phi_0\left(\frac{1}{2}-x\right)\,\varepsilon,\label{SI:eq_volume_fraction_epsilon}
\end{align}
and treat the spatial inhomogeneity using the control parameter $\varepsilon$.  Therefore, as $\varepsilon$ varies from $0$ to $1$ a homogeneous rod develops a steeper volume fraction gradient and becomes increasingly \emph{inhomogeneous}.   This is reflected in the Young's modulus plotted in Fig.~\ref{SI:fig_Erel_epsilon}.

Consistent with the remarks above, the unperturbed problem
\begin{align}
	\frac{d^2 Y_{0}^{(n)}(x)}{dx^2}+\frac{\Gamma_{\text{cr},0}^{(n)}}{E_{\text{rel}}(\phi=\phi_0/2,\gamma\prime)} Y_{0}^{(n)}(x)=0,\label{SI:eq_eq_unperturb}
\end{align}
is exactly soluble, so we seek a perturbative solution to the equilibrium Eq.~\eqref{SI:eq_eq_scaled} of the form
\begin{eqnarray}
	Y^{(n)}(x)&=&\sum_{l=0}^{\infty}Y_l^{(n)}(x)\,\varepsilon^l,\label{SI:eq_Y_epsilon}\qquad \textrm{with}\\
	\Gamma_{\text{cr}}^{(n)}&=&\sum_{l=0}^{\infty}\Gamma_{\text{cr},l}^{(n)}\,\varepsilon^l;\label{SI:eq_gamma_epsilon}
\end{eqnarray}
and the inverse of the Young's modulus $E_{\text{rel}}^{-1}$ expanded as a power series in $\varepsilon$,
\begin{align}
	E_{\text{rel}}^{-1}=\sum_{l=0}^{\infty}E_{\text{rel},l}^{-1}\,\varepsilon^l.\label{SI:eq_Erel_epsilon}
\end{align}

Substituting Eqs.~\mref{SI:eq_Y_epsilon,SI:eq_gamma_epsilon,SI:eq_Erel_epsilon} into Eq.~\eqref{SI:eq_eq_scaled} gives the following equation at first order in $\varepsilon$:
\begin{multline}
	\frac{d^2Y^{(n)}_1(x)}{dx^2}+\Gamma_{\text{cr},1}^{(n)}E_{\text{rel},0}^{-1}(x)\,Y_0^{(n)}(x)\\
	+\Gamma_{\text{cr},0}^{(n)}E_{\text{rel},0}^{-1}(x)\,Y_1^{(n)}(x)\\
	+\Gamma_{\text{cr},0}^{(n)}E_{\text{rel},1}^{-1}(x)\,Y_0^{(n)}(x)=0,\label{SI:eq_perturb_1st_order}
\end{multline}
whose solutions must satisfy the boundary conditions
\begin{align}
	Y^{(n)}_1(0)=0, Y^{(n)}_1(1)=0.\label{SI:eq_bc_1st_order}
\end{align}
Equation~\eqref{SI:eq_perturb_1st_order} for $Y_1^{(n)}(x)$ is linear and inhomogeneous.
The associated homogeneous equation is of the same type as the unperturbed problem, Eq.~\eqref{SI:eq_eq_unperturb}, and can thus be solved.
Of the two linearly independent solutions, only that satisfying the boundary conditions is assumed known.
Therefore, we use reduction of order and substitute
\begin{align}
	Y_1^{(n)}(x)=Y_0^{(n)}(x)\,F_1^{(n)}(x)
\end{align}
into Eq.~\eqref{SI:eq_perturb_1st_order}.
Simplifying the result using Eq.~\eqref{SI:eq_eq_unperturb} and multiplying by $Y_0^{(n)}$ gives
\begin{multline}
	\frac{d}{dx}\left[\left(Y_0^{(n)}(x)\right)^2 \left(F_1^{(n)}(x)\right)^{\prime}\right]\\
	+\left(Y_0^{(n)}(x)\right)^2\left[\Gamma_{\text{cr},1}^{(n)}E_{\text{rel},0}^{-1}(x)+\Gamma_{\text{cr},0}^{(n)}E_{\text{rel},1}^{-1}(x)\right]=0.
\end{multline}
We integrate this equation from 0 to 1 and use the fact that 
\begin{multline}
\left(Y_0^{(n)}(x)\right)^2 \left(F_1^{(n)}(x)\right)^{\prime}\\=Y_0^{(n)}(x)\, \left(Y_1^{(n)}(x)\right)^{\prime}-\left(Y_0^{(n)}(x)\right)^{\prime} Y_1^{(n)}(x), \nonumber
\end{multline}
vanishes at $x=0,1$ because of the boundary conditions~\mref{SI:eq_bc,SI:eq_bc_1st_order}.  In this manner we obtain the first order correction to $\Gamma_{\text{cr}}$ as 
\begin{align}
	\Gamma_{\text{cr},1}^{(n)}=-\frac{\Gamma_{\text{cr},0}^{(n)}\int_0^{1}\left(Y_0^{(n)}(x)\right)^2 E_{\text{rel},1}^{-1}(x)\,dx}{\int_0^1\left(Y_0^{(n)}(x)\right)^2 E_{\text{rel},0}^{-1}(x)\,dx}.
	\label{SI:eq_1st_ord_Gamma_cr}
\end{align}

\subsubsection{DT and MSW Young's modulus expansion}
When we use the DT expression for the effective Young's modulus, Eq. ~\eqref{SI:eq_Ec}, with $\gamma\prime$ constant and the volume fraction profile in Eq. \eqref{SI:eq_volume_fraction_epsilon}, we find that Eq.~\eqref{SI:eq_Erel_epsilon} reduces to
\begin{widetext}
\begin{eqnarray}
	E_{\text{rel,DT}}^{-1}(x)&=&1-\frac{\phi_0}{2}\frac{\frac{5}{2}\gamma\prime-\frac{5}{3}}{1+\frac{5}{2}\gamma\prime}-\phi_0\frac{\frac{5}{2}\gamma\prime-\frac{5}{3}}{1+\frac{5}{2}\gamma\prime}\left(\frac{1}{2}-x\right)\,\varepsilon\nonumber\\
	&=&E_{\text{rel,DT}}^{-1}\left(\phi=\frac{\phi_0}{2},\gamma\prime\right)+2\left[E_{\text{rel,DT}}^{-1}\left(\phi=\frac{\phi_0}{2},\gamma\prime\right)-1\right]\left(\frac{1}{2}-x\right)\,\varepsilon.\label{SI:eq_Ec_DT_epsilon}
\end{eqnarray}
On the other hand, the MSW inverse Young's modulus, Eq.~\eqref{SI:eq_Ec_Mori-Tanaka}, with $\gamma\prime$ constant and inclusion profile given  by Eq.~\eqref{SI:eq_volume_fraction_epsilon}, can be written as
\begin{align}
	E_{\text{rel,MSW}}^{-1}(x)=\frac{A_1+\frac{2}{3}A_2\left(\frac{1}{2}-x\right)\varepsilon}{A_3-A_2\left(\frac{1}{2}-x\right)\varepsilon},\label{SI:eq_inverse_Erel_Mori-Tanaka}
\end{align}
where $A_1=2+\frac{2}{3}\phi_0+\gamma\prime(5-\phi_0)$, $A_2=\phi_0(2-3\gamma\prime)$, and $A_3=2-\phi_0+\gamma\prime\left(5+\frac{2}{3}\phi_0\right)$.
We note again that our boundary value problem is defined on $x\in[0,1]$ and hence the terms multiplying $\varepsilon$ in Eq.~\eqref{SI:eq_inverse_Erel_Mori-Tanaka} are finite ($\phi_0\in[0,1)$ and $\gamma\prime\in[0,2/3]$), so that a Taylor expansion for $\varepsilon$ small of $E_{\text{rel,MSW}}^{-1}$ gives
\begin{eqnarray}
	E_{\text{rel,MSW}}^{-1}(x)&\sim&\frac{A_1}{A_3}+\frac{A_2}{A_3^2}\left(\frac{2}{3}A_3+A_1\right)\left(\frac{1}{2}-x\right)\,\varepsilon+\frac{A_2^2}{A_3^3}\left(\frac{2}{3}A_3+A_1\right)\left(\frac{1}{2}-x\right)^2\,\varepsilon^2+\dots\nonumber\\
	&\sim&E_{\text{rel,MSW}}^{-1}\left(\phi=\frac{\phi_0}{2},\gamma\prime\right)+\frac{A_2}{A_3}\left[\frac{2}{3}+E_{\text{rel,MSW}}^{-1}\left(\phi=\frac{\phi_0}{2},\gamma\prime\right)\right]\left(\frac{1}{2}-x\right)\,\varepsilon\nonumber\\
	&+&\frac{A_2^2}{A_3^2}\left[\frac{2}{3}+E_{\text{rel,MSW}}^{-1}\left(\phi=\frac{\phi_0}{2},\gamma\prime\right)\right]\left(\frac{1}{2}-x\right)^2\,\varepsilon^2+\dots\,.\label{SI:eq_Ec_MSW_epsilon}
\end{eqnarray}

We observe the same functional dependence on $x$ in the expansion for both the DT, Eq.~\eqref{SI:eq_Ec_DT_epsilon}, and the MSW, Eq.~\eqref{SI:eq_Ec_MSW_epsilon}, expressions and thus simplify the notation
and rewrite these as 
\begin{eqnarray}
	E_{\text{rel,DT}}^{-1}(x)&=&\langle E_{\text{rel,DT}}\rangle^{-1}+E_{\text{rel,DT},1}^{-1}\cdot\left(\frac{1}{2}-x\right)\,\varepsilon ~~ \textrm{and}\label{SI:eq_Ec_DT_epsilon_re}\\
	E_{\text{rel,MSW}}^{-1}(x)&\sim&\langle E_{\text{rel,MSW}}\rangle^{-1}+E_{\text{rel,MSW},1}^{-1}\cdot\left(\frac{1}{2}-x\right)\,\varepsilon+E_{\text{rel,MSW},2}^{-1}\cdot\left(\frac{1}{2}-x\right)^2\,\varepsilon^2+\dots\,.\label{SI:eq_Ec_MSW_epsilon_re}
\end{eqnarray}
\end{widetext}
respectively, where $\langle E_{\text{rel,DT}}\rangle^{-1}\equiv E_{\text{rel,DT}}^{-1}(\phi=\phi_0/2,\gamma\prime)$ and $\langle E_{\text{rel,MSW}}\rangle^{-1}\equiv E_{\text{rel,MSW}}^{-1}(\phi=\phi_0/2,\gamma\prime)$.

\subsubsection{First order correction to the first buckling mode: DT and MSW}
Consider the first buckling mode $n=1$ for both the DT and the MSW effective Young's moduli.
Because for both cases the zeroth order term in the inverse Young's modulus expansions is a constant inverse Young's modulus with $\phi=\phi_0/2$, namely $\langle E_{\text{rel}}\rangle^{-1}$, by substituting Eq.~\eqref{SI:eq_reference_modes} into Eq.~\eqref{SI:eq_1st_ord_Gamma_cr} we can easily solve the integral in the denominator, giving the first order correction $\Gamma_{\text{cr},1}^{(1)}$ as
\begin{align}
	\Gamma_{\text{cr},1}^{(1)}=-2\pi^2 \langle E_{\text{rel}}\rangle^2\,E_{\text{rel},1}^{-1}\, \int_0^{1}\sin^2(\pi x) \left(\frac{1}{2}-x\right)\,dx,
	\label{SI:eq_1st_ord_1st_mode_Gamma_cr}
\end{align}
where we have also used Eq.~\eqref{SI:eq_gamma_crit}.

Note that he integral $\int_0^{1}\sin^2(\pi x) \left(\frac{1}{2}-x\right)\,dx$ vanishes because its integrand is the product of the even function $\sin^2(\pi x)$ and an odd function $(1/2 -x)$ over the domain $x\in[0,1]$. 
Therefore, for both the DT and MSW models the first order term in the expansion of the first critical load is zero
\begin{align}
	\Gamma_{\text{cr},1}^{(1)}=0.
	\label{SI:eq_1st_ord_1st_mode_Gamma_cr_0}
\end{align}

Substituting this result into the ODE for the first order correction, Eq.~\eqref{SI:eq_perturb_1st_order}, gives
\begin{align}
	\frac{d^2Y^{(1)}_1(x)}{dx^2}+\pi^2\,Y_1^{(1)}(x)=-\pi^2 E_{\text{rel},1}^{-1}\cdot\left(\frac{1}{2}-x\right)\,\sin(\pi x),\label{SI:eq_perturb_1st_order_1st_mode}
\end{align}
where we have again used the unperturbed expressions Eqs.~\mref{SI:eq_gamma_crit,SI:eq_reference_modes}.

The solution of the homogeneous part of Eq.~\eqref{SI:eq_perturb_1st_order_1st_mode} is $y_c(x)=c_1\sin(\pi x)+c_2\cos(\pi x)$, where $c_1,\,c_2$ are constants.
The particular solution can be computed by variation of parameters, and thus after some algebra we obtain the solution to Eq.~\eqref{SI:eq_perturb_1st_order_1st_mode} with boundary conditions~\eqref{SI:eq_bc_1st_order} as 
\begin{multline}
	Y_1^{(1)}(x)=\left[c_1+\frac{\langle E_{\text{rel}}\rangle\,E_{\text{rel},1}^{-1}}{4}(1-x)\right]\sin(\pi x)\\
	+\frac{\pi \langle E_{\text{rel}}\rangle\,E_{\text{rel},1}^{-1}}{4}x(x-1)\cos(\pi x). \label{SI:eq_1st_ord_term_Y1}
\end{multline}

\begin{figure*}                                                                                        
\begin{subfigure}[b]{0.45\textwidth}
\psfrag{x}[ct][ct][1.]{$x$}
	\psfrag{y}[ct][ct][1.]{$Y(x)$}                                                                   
\psfrag{ez}[ct][ct][1.]{$\varepsilon=0.001$}                                                                   
\psfrag{eu}[ct][ct][1.]{$\varepsilon=1$}                                                                   
 \includegraphics[width=\linewidth]{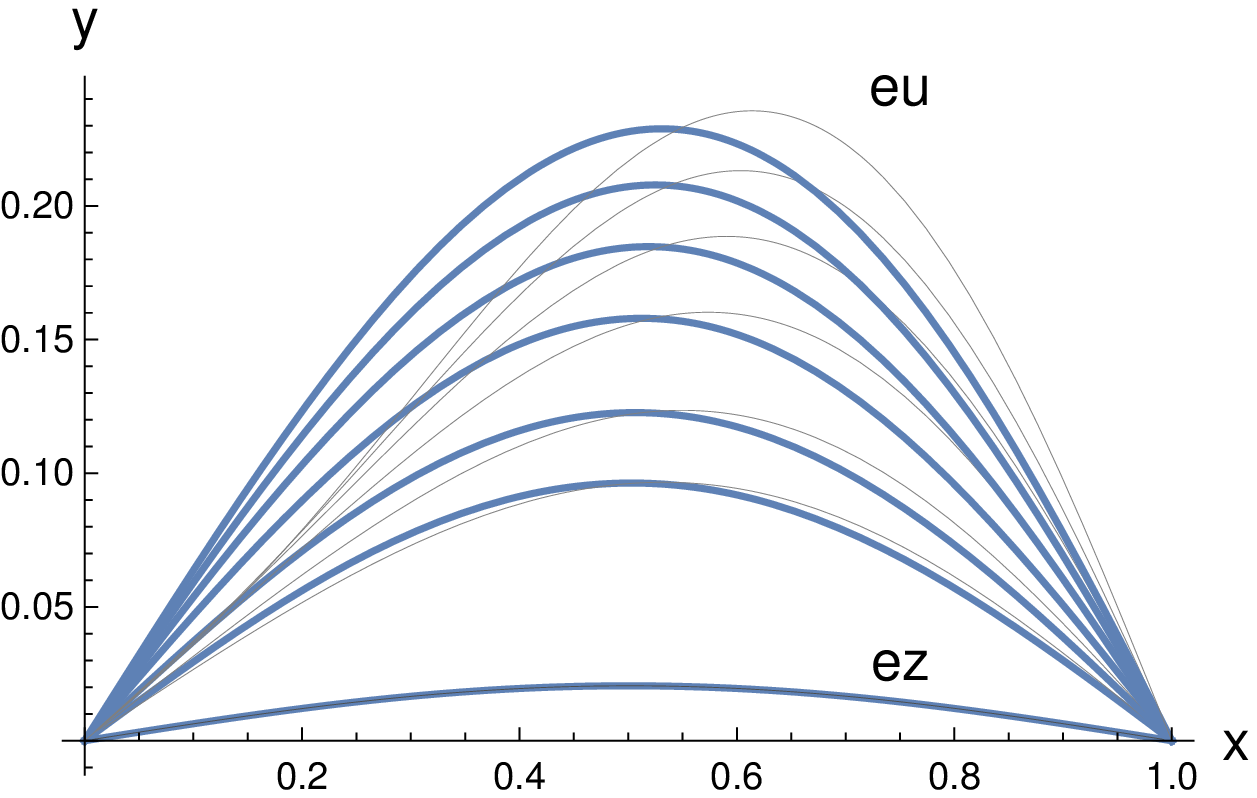}
 \caption{Buckling modes.}
 \label{SI:fig_Y1_profile_epsilon_DT}
 \end{subfigure}
 \begin{subfigure}[b]{0.45\textwidth} 
\psfrag{x}[ct][ct][1.]{$x$}
	 \psfrag{y}[ct][ct][1.]{$\Delta Y(x,\varepsilon)$}                                                                   
\psfrag{ez}[ct][ct][1.]{$\varepsilon=0.001$}                                                                   
\psfrag{eu}[ct][ct][1.]{$\varepsilon=1$}                                                                   
 \includegraphics[width=\linewidth]{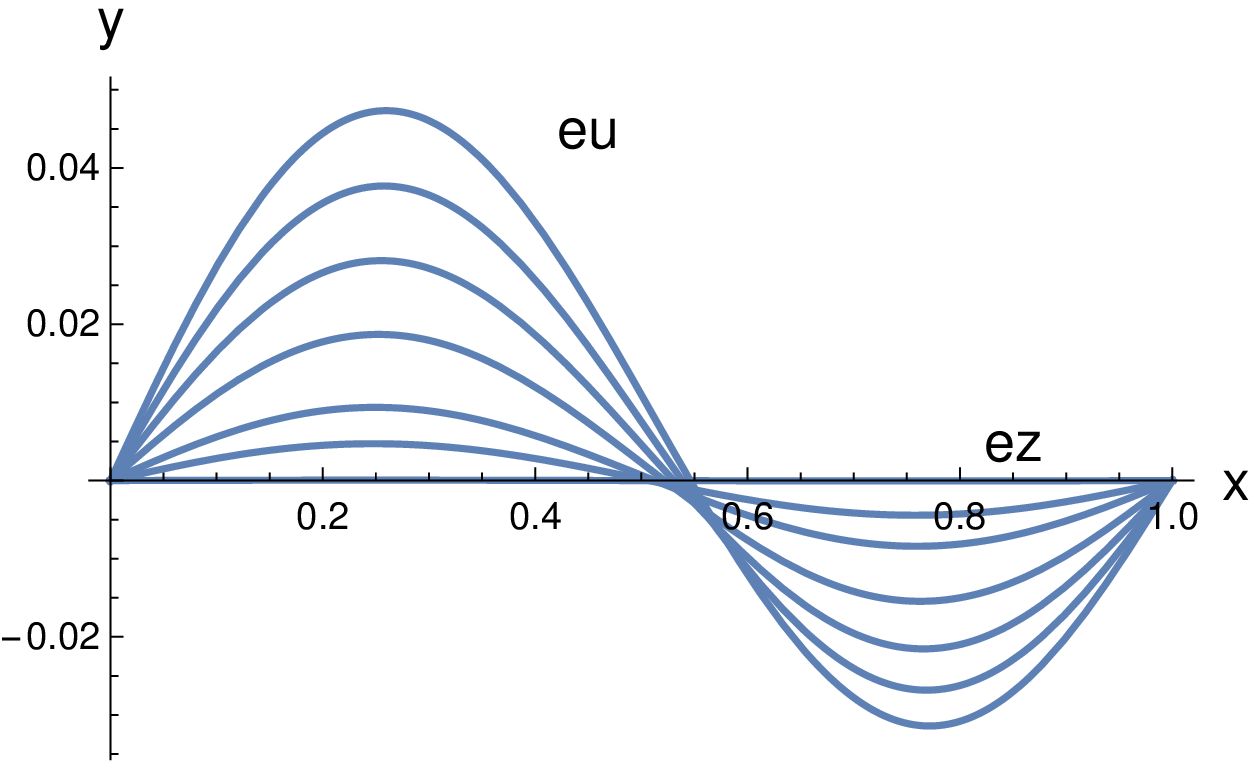}
	 \caption{Deviation between closed form Eq.(52) in the main text and the expansion with Eq.~\eqref{SI:eq_1st_ord_term_Y1}.}
 \label{fig_Y1_diff_epsilon_DT}
 \end{subfigure}                                                                                      
	\caption{First buckling mode in the DT model. (a) Blue lines correspond to the analytical expression Eq. (52) in the main text for $\phi_0=0.6$, $\gamma\prime=100$, and increasing $\varepsilon$, $\varepsilon=(0.001, 0.1, 0.2, 0.4, 0.6, 0.8, 1)$; the corresponding first order approximations Eq.~\eqref{SI:eq_perturb_1st_order_1st_mode} are plotted in gray lines.
	The constant $c_1$ in Eq.~\eqref{SI:eq_1st_ord_term_Y1} is fixed by imposing the same energy for the two configurations we compare at each $\varepsilon$.
	(b) Deviation between $Y(x,\varepsilon)$, Eq. (52) in the main text, and the corresponding first order approximation $Y_{\text{DT},0}^{(1)}(x)+Y_{\text{DT},1}^{(1)}(x)\,\varepsilon$, Eq.~\eqref{SI:eq_perturb_1st_order_1st_mode}, corresponding to the curves in (a).}
 \label{SI:fig_Y1_epsilon_DT}
\end{figure*}                                                                                          

In Fig.~\ref{SI:fig_Y1_profile_epsilon_DT} we plot the first buckling mode as a power series in $\varepsilon$, Eq.~\eqref{SI:eq_Y_epsilon},  truncated at first order,  with $Y_1^{(1)}$ given by Eq.~\eqref{SI:eq_1st_ord_term_Y1}.
Comparing these modes with the closed form solution in terms of Airy functions (Eq.~(52) in the main text) for different $\varepsilon$, we observe perfect agreement for $\varepsilon\to 0$, with the deviation $\Delta Y(x,\varepsilon)$ growing as $\varepsilon\to 1$ (see Fig.~\ref{fig_Y1_diff_epsilon_DT}). 
This is the expected behavior given that the validity of the expansions in the perturbation method break down as $\varepsilon$ grows.

\subsubsection{Second order correction to the first buckling mode: DT and MSW}
Proceeding along the same lines as above, the second order correction for $\Gamma_{\text{cr}}^{(1)}$ is
\begin{widetext}
\begin{align}
	\Gamma_{\text{cr},2}^{(1)}=-2\pi^2 \langle E_{\text{rel}}\rangle^2\left[E_{\text{rel},1}^{-1}\int_0^1\sin(\pi x)\left(\frac{1}{2}-x\right)\,Y_1^{(1)}(x)\,dx+E_{\text{rel},2}^{-1}\int_0^1\sin^2(\pi x)\left(\frac{1}{2}-x\right)^2\,dx\right].\label{SI:eq_2nd_ord_1st_mode}
\end{align}

Substituting the first order term in the expansion of $Y(x)$, Eq.~\eqref{SI:eq_1st_ord_term_Y1}, and the Young's modulus expansion in the DT model, Eq.~\eqref{SI:eq_Ec_DT_epsilon}, into Eq.~\eqref{SI:eq_2nd_ord_1st_mode}, gives
\begin{align}
	\Gamma_{\text{cr,DT},2}^{(1)}=-\frac{\pi^2}{2}\phi_0^2\frac{\left(1+\frac{5}{2}\gamma\prime\right)\left(\frac{5}{2}\gamma\prime-\frac{5}{3}\right)^2}{\left[1+\frac{5}{2}\gamma\prime-\frac{\phi_0}{2}\left(\frac{5}{2}\gamma\prime-\frac{5}{3}\right)\right]^3}\,\left(-\frac{1}{24}-\frac{1}{\pi}+\frac{1}{4\pi^2}+\frac{12}{\pi^3}\right).\label{SI:eq_gamma2_epsilon_DT}
\end{align}

Similarly, the second order correction for $\Gamma_{\text{cr}}^{(1)}$ in the MSW theory is given by
\begin{eqnarray}
	\Gamma_{\text{cr,MSW},2}^{(1)}&=&-\pi^2\phi_0^2\frac{10}{3}\frac{(2-3\gamma\prime)^2(2+5\gamma\prime)}{\left(2+\frac{2}{3}\phi_0+\gamma\prime(5-\phi_0)\right)^2\left(2-\phi_0+\gamma\prime\left(5+\frac{3}{2}\right)\right)}\nonumber\\
	&&\times\left[\frac{5}{12}\frac{(2+5\gamma\prime)}{2+\frac{2}{3}\phi_0+\gamma\prime(5-\phi_0)}\left(-\frac{1}{24}-\frac{1}{\pi}+\frac{1}{4\pi^2}+\frac{12}{\pi^3}\right)+\frac{1}{24}-\frac{1}{4\pi^2}\right].\label{SI:eq_gamma2_epsilon_MSW}
\end{eqnarray}
\end{widetext}

\begin{figure*}                                                                                        
\begin{subfigure}[b]{0.45\textwidth}
\psfrag{x}[ct][ct][1.]{$\varepsilon$}
	\psfrag{y}[ct][ct][1.]{$\Gamma_{\text{cr}}$}                                                                   
	\psfrag{e}[ct][ct][1.]{$\Gamma_{\text{cr}}(\varepsilon)$}                                                                   
	\psfrag{ez}[ct][ct][1.]{Leading--order approximation to $\Gamma_{\text{cr}}(\varepsilon)$}                                                                   
	\psfrag{eu}[ct][ct][1.]{Second--order approximation to $\Gamma_{\text{cr}}(\varepsilon)$}                                                                   
 \includegraphics[width=\linewidth]{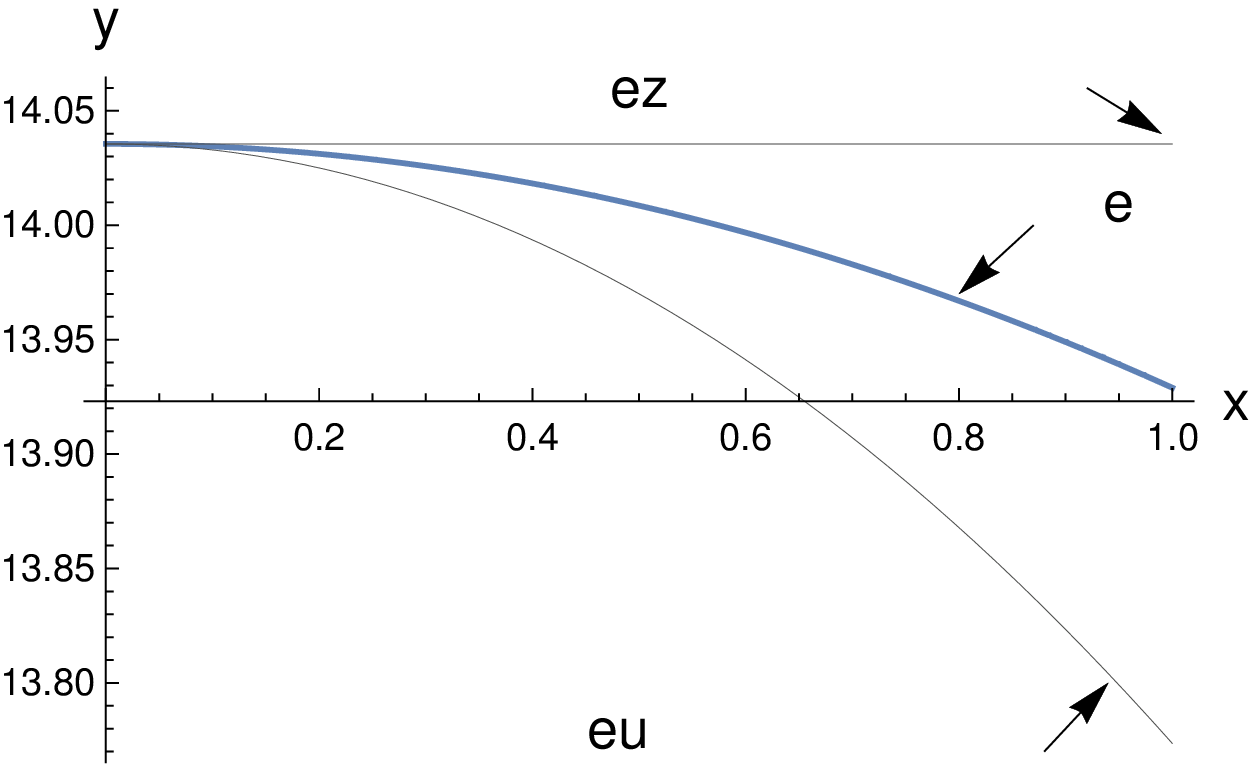}
 \caption{DT.}
 \label{SI:fig_gamma2_epsilon_DT}
 \end{subfigure}
 \begin{subfigure}[b]{0.45\textwidth} 
\psfrag{x}[ct][ct][1.]{$\varepsilon$}
	\psfrag{y}[ct][ct][1.]{$\Gamma_{\text{cr}}$}                                                                   
	\psfrag{e}[ct][ct][1.]{$\Gamma_{\text{cr}}(\varepsilon)$}                                                                   
	\psfrag{ez}[ct][ct][1.]{Leading--order approximation to $\Gamma_{\text{cr}}(\varepsilon)$}                                                                   
	\psfrag{eu}[ct][ct][1.]{Second--order approximation to $\Gamma_{\text{cr}}(\varepsilon)$}                                                                   
 \includegraphics[width=\linewidth]{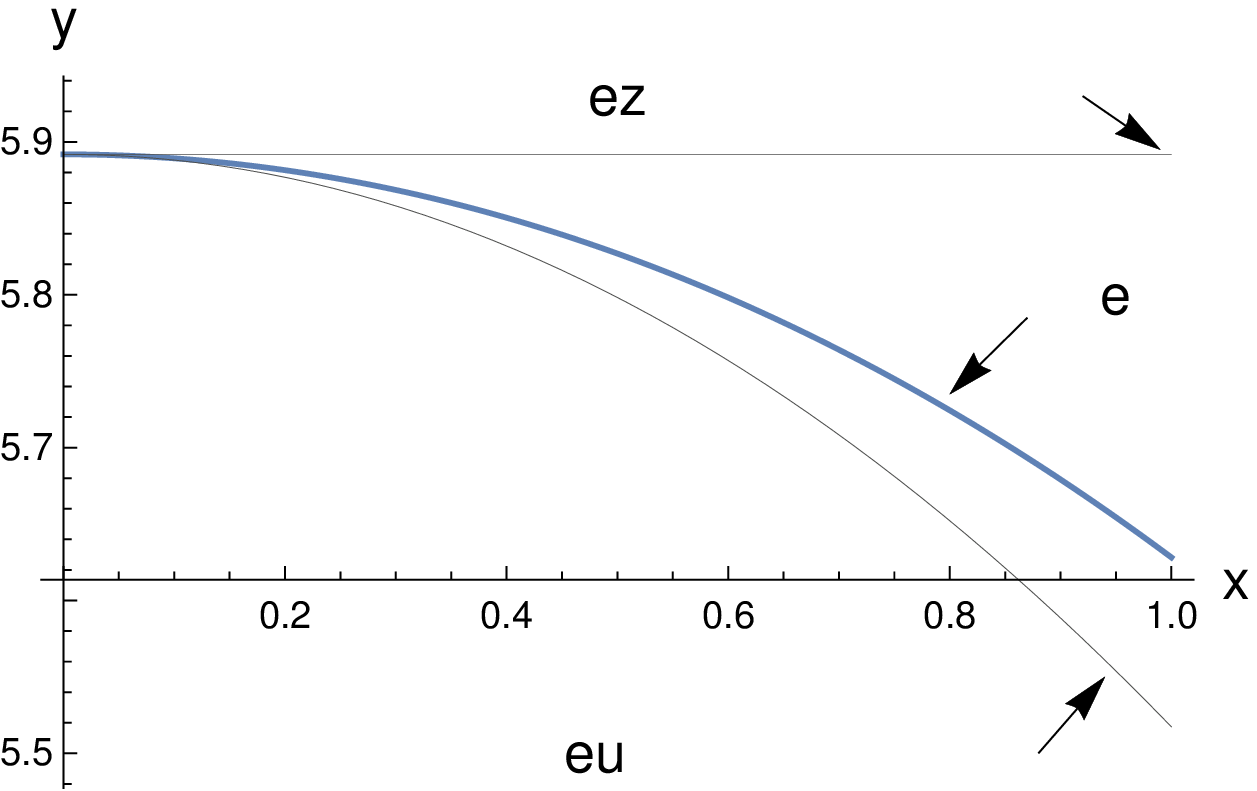}
	 \caption{MSW.}
 \label{SI:fig_gamma2_epsilon_MSW}
 \end{subfigure}                                                                                      
	\caption{A comparison of the two approximations to the critical compression force $\Gamma_{\text{cr}}$ by means of perturbation methods.
	The leading--order approximation in Eq.~\eqref{SI:eq_gamma_epsilon} is simply given by Eq.~\eqref{SI:eq_gamma_crit}.
	The second--order approximation is given by Eq.~\eqref{SI:eq_gamma2_epsilon_DT} for the DT ($\phi_0=0.6,\gamma\prime=100$), in panel (a), and by Eq.~\eqref{SI:eq_gamma2_epsilon_MSW} for the MSW ($\phi_0=0.6,\gamma\prime=0.01$), in panel (b).}
 \label{SI:fig_gamma2_epsilon}
\end{figure*}

We plot in Fig.~\ref{SI:fig_gamma2_epsilon} the approximations to the critical compressing force $\Gamma_{\text{cr}}$ up to second order, with the second--order terms of the DT and MSW models given by  Eqs.~\eqref{SI:eq_gamma2_epsilon_DT} and ~\eqref{SI:eq_gamma2_epsilon_MSW} respectively. 
The leading--order result, $\Gamma_{\text{cr}}(\varepsilon)\sim\Gamma_{\text{cr},0}\,(\varepsilon\to 0^+)$, is a good approximation to $\Gamma_{\text{cr}}$.
However, Fig.~\ref{SI:fig_gamma2_epsilon} clearly shows that the first non--vanishing correction to the leading order behavior (i.e., the second--order correction), is negative and hence the critical compression approximation tends to decrease below the leading--order approximation as $\varepsilon$ increases.
Importantly, recalling from Fig.~\ref{SI:fig_Erel_epsilon} that $\varepsilon$ controls the spatial \emph{inhomogeneity} of the properties of the rod (namely the homogeneous rod is recovered when $\varepsilon\to 0^+$)  and hence the zeroth-order approximation corresponds to the homogeneous rod.  Therefore, an \emph{inhomogeneous} column collapses under the influence of a lighter load.
(Note that the approximation of the critical compression for an \emph{inhomogeneous} rod Eq.~\eqref{SI:eq_gamma_epsilon} up to second order is plotted in Fig.~\ref{fig_critical_tension-DT-MSW} in the main text.)



%% file: Buckling_arXiv_v3.bbl
\begin{thebibliography}{58}%
\makeatletter
\providecommand \@ifxundefined [1]{%
 \@ifx{#1\undefined}
}%
\providecommand \@ifnum [1]{%
 \ifnum #1\expandafter \@firstoftwo
 \else \expandafter \@secondoftwo
 \fi
}%
\providecommand \@ifx [1]{%
 \ifx #1\expandafter \@firstoftwo
 \else \expandafter \@secondoftwo
 \fi
}%
\providecommand \natexlab [1]{#1}%
\providecommand \enquote  [1]{``#1''}%
\providecommand \bibnamefont  [1]{#1}%
\providecommand \bibfnamefont [1]{#1}%
\providecommand \citenamefont [1]{#1}%
\providecommand \href@noop [0]{\@secondoftwo}%
\providecommand \href [0]{\begingroup \@sanitize@url \@href}%
\providecommand \@href[1]{\@@startlink{#1}\@@href}%
\providecommand \@@href[1]{\endgroup#1\@@endlink}%
\providecommand \@sanitize@url [0]{\catcode `\\12\catcode `\$12\catcode
  `\&12\catcode `\#12\catcode `\^12\catcode `\_12\catcode `\%12\relax}%
\providecommand \@@startlink[1]{}%
\providecommand \@@endlink[0]{}%
\providecommand \url  [0]{\begingroup\@sanitize@url \@url }%
\providecommand \@url [1]{\endgroup\@href {#1}{\urlprefix }}%
\providecommand \urlprefix  [0]{URL }%
\providecommand \Eprint [0]{\href }%
\providecommand \doibase [0]{http://dx.doi.org/}%
\providecommand \selectlanguage [0]{\@gobble}%
\providecommand \bibinfo  [0]{\@secondoftwo}%
\providecommand \bibfield  [0]{\@secondoftwo}%
\providecommand \translation [1]{[#1]}%
\providecommand \BibitemOpen [0]{}%
\providecommand \bibitemStop [0]{}%
\providecommand \bibitemNoStop [0]{.\EOS\space}%
\providecommand \EOS [0]{\spacefactor3000\relax}%
\providecommand \BibitemShut  [1]{\csname bibitem#1\endcsname}%
\let\auto@bib@innerbib\@empty
\bibitem [{\citenamefont {Euler}(1744)}]{Euler1744}%
  \BibitemOpen
  \bibfield  {author} {\bibinfo {author} {\bibfnamefont {L.}~\bibnamefont
  {Euler}},\ }\href@noop {} {\emph {\bibinfo {title} {Methodus inveniendi
  lineas curvas maximi minimive proprietate gaudentes sivesolutio problematis
  isoperimetrici latissimo sensu accepti}}}\ (\bibinfo  {publisher} {Lausanne,
  Geneva: Marc-Michel Bousquet \& Co},\ \bibinfo {year} {1744})\BibitemShut
  {NoStop}%
\bibitem [{\citenamefont {Euler}(1759)}]{Euler1759}%
  \BibitemOpen
  \bibfield  {author} {\bibinfo {author} {\bibfnamefont {L.}~\bibnamefont
  {Euler}},\ }\href@noop {} {\bibfield  {journal} {\bibinfo  {journal} {Berlin,
  Germany: Memoires de l'Acad\'emie de Berlin}\ }\textbf {\bibinfo {volume}
  {13}} (\bibinfo {year} {1759})}\BibitemShut {NoStop}%
\bibitem [{\citenamefont {Lagrange}(1770)}]{Lagrange1770}%
  \BibitemOpen
  \bibfield  {author} {\bibinfo {author} {\bibfnamefont {J.~L.}\ \bibnamefont
  {Lagrange}},\ }\href@noop {} {\bibfield  {journal} {\bibinfo  {journal}
  {Miscellanea Taurinensia}\ }\textbf {\bibinfo {volume} {5}},\ \bibinfo
  {pages} {123} (\bibinfo {year} {1770})}\BibitemShut {NoStop}%
\bibitem [{\citenamefont {Erbil}\ \emph {et~al.}(2020)\citenamefont {Erbil},
  \citenamefont {Hatipoglu}, \citenamefont {Yanik}, \citenamefont {Ghavami},
  \citenamefont {Ari}, \citenamefont {Yuksel},\ and\ \citenamefont
  {Hanay}}]{ErbilHatipoglu20}%
  \BibitemOpen
  \bibfield  {author} {\bibinfo {author} {\bibfnamefont {S.~O.}\ \bibnamefont
  {Erbil}}, \bibinfo {author} {\bibfnamefont {U.}~\bibnamefont {Hatipoglu}},
  \bibinfo {author} {\bibfnamefont {C.}~\bibnamefont {Yanik}}, \bibinfo
  {author} {\bibfnamefont {M.}~\bibnamefont {Ghavami}}, \bibinfo {author}
  {\bibfnamefont {A.~B.}\ \bibnamefont {Ari}}, \bibinfo {author} {\bibfnamefont
  {M.}~\bibnamefont {Yuksel}}, \ and\ \bibinfo {author} {\bibfnamefont {M.~S.}\
  \bibnamefont {Hanay}},\ }\href {\doibase 10.1103/PhysRevLett.124.046101}
  {\bibfield  {journal} {\bibinfo  {journal} {Phys. Rev. Lett.}\ }\textbf
  {\bibinfo {volume} {124}},\ \bibinfo {pages} {046101} (\bibinfo {year}
  {2020})}\BibitemShut {NoStop}%
\bibitem [{\citenamefont {Timoshenko}\ and\ \citenamefont
  {Gere}(1963)}]{TimoshenkoGere63}%
  \BibitemOpen
  \bibfield  {author} {\bibinfo {author} {\bibfnamefont {S.}~\bibnamefont
  {Timoshenko}}\ and\ \bibinfo {author} {\bibfnamefont {J.}~\bibnamefont
  {Gere}},\ }\href@noop {} {\emph {\bibinfo {title} {Theory of Elastic
  Stability}}},\ \bibinfo {edition} {second edition}\ ed.\ (\bibinfo
  {publisher} {McGraw-Hill Book Company, Inc.},\ \bibinfo {year}
  {1963})\BibitemShut {NoStop}%
\bibitem [{\citenamefont {Nezamabadi}\ \emph {et~al.}(2009)\citenamefont
  {Nezamabadi}, \citenamefont {Yvonnet}, \citenamefont {Zahrouni},\ and\
  \citenamefont {Potier-Ferry}}]{NezambadiYvonnet09}%
  \BibitemOpen
  \bibfield  {author} {\bibinfo {author} {\bibfnamefont {S.}~\bibnamefont
  {Nezamabadi}}, \bibinfo {author} {\bibfnamefont {J.}~\bibnamefont {Yvonnet}},
  \bibinfo {author} {\bibfnamefont {H.}~\bibnamefont {Zahrouni}}, \ and\
  \bibinfo {author} {\bibfnamefont {M.}~\bibnamefont {Potier-Ferry}},\ }\href
  {\doibase https://doi.org/10.1016/j.cma.2009.02.026} {\bibfield  {journal}
  {\bibinfo  {journal} {Computer Methods in Applied Mechanics and Engineering}\
  }\textbf {\bibinfo {volume} {198}},\ \bibinfo {pages} {2099 } (\bibinfo
  {year} {2009})}\BibitemShut {NoStop}%
\bibitem [{\citenamefont {Gong}\ \emph {et~al.}(2005)\citenamefont {Gong},
  \citenamefont {Kyriakides},\ and\ \citenamefont
  {Triantafyllidis}}]{GongKyriakides05}%
  \BibitemOpen
  \bibfield  {author} {\bibinfo {author} {\bibfnamefont {L.}~\bibnamefont
  {Gong}}, \bibinfo {author} {\bibfnamefont {S.}~\bibnamefont {Kyriakides}}, \
  and\ \bibinfo {author} {\bibfnamefont {N.}~\bibnamefont {Triantafyllidis}},\
  }\href {\doibase https://doi.org/10.1016/j.jmps.2004.10.007} {\bibfield
  {journal} {\bibinfo  {journal} {Journal of the Mechanics and Physics of
  Solids}\ }\textbf {\bibinfo {volume} {53}},\ \bibinfo {pages} {771 }
  (\bibinfo {year} {2005})}\BibitemShut {NoStop}%
\bibitem [{\citenamefont {Triantafyllidis}\ \emph {et~al.}(2005)\citenamefont
  {Triantafyllidis}, \citenamefont {Nestorovi{\'c}},\ and\ \citenamefont
  {Schraad}}]{TriantafyllidisNestorovic05}%
  \BibitemOpen
  \bibfield  {author} {\bibinfo {author} {\bibfnamefont {N.}~\bibnamefont
  {Triantafyllidis}}, \bibinfo {author} {\bibfnamefont {M.~D.}\ \bibnamefont
  {Nestorovi{\'c}}}, \ and\ \bibinfo {author} {\bibfnamefont {M.~W.}\
  \bibnamefont {Schraad}},\ }\href {\doibase 10.1115/1.2126695} {\bibfield
  {journal} {\bibinfo  {journal} {Journal of Applied Mechanics}\ }\textbf
  {\bibinfo {volume} {73}},\ \bibinfo {pages} {505} (\bibinfo {year}
  {2005})}\BibitemShut {NoStop}%
\bibitem [{\citenamefont {deBotton}\ \emph {et~al.}(2006)\citenamefont
  {deBotton}, \citenamefont {Hariton},\ and\ \citenamefont
  {Socolsky}}]{deBottonHariton06}%
  \BibitemOpen
  \bibfield  {author} {\bibinfo {author} {\bibfnamefont {G.}~\bibnamefont
  {deBotton}}, \bibinfo {author} {\bibfnamefont {I.}~\bibnamefont {Hariton}}, \
  and\ \bibinfo {author} {\bibfnamefont {E.}~\bibnamefont {Socolsky}},\ }\href
  {\doibase https://doi.org/10.1016/j.jmps.2005.10.001} {\bibfield  {journal}
  {\bibinfo  {journal} {Journal of the Mechanics and Physics of Solids}\
  }\textbf {\bibinfo {volume} {54}},\ \bibinfo {pages} {533 } (\bibinfo {year}
  {2006})}\BibitemShut {NoStop}%
\bibitem [{\citenamefont {Jamal}\ \emph {et~al.}(1999)\citenamefont {Jamal},
  \citenamefont {Midani}, \citenamefont {Damil},\ and\ \citenamefont
  {Potier-Ferry}}]{JamalMidani99}%
  \BibitemOpen
  \bibfield  {author} {\bibinfo {author} {\bibfnamefont {M.}~\bibnamefont
  {Jamal}}, \bibinfo {author} {\bibfnamefont {M.}~\bibnamefont {Midani}},
  \bibinfo {author} {\bibfnamefont {N.}~\bibnamefont {Damil}}, \ and\ \bibinfo
  {author} {\bibfnamefont {M.}~\bibnamefont {Potier-Ferry}},\ }\href {\doibase
  https://doi.org/10.1016/S0020-7683(98)00028-6} {\bibfield  {journal}
  {\bibinfo  {journal} {International Journal of Solids and Structures}\
  }\textbf {\bibinfo {volume} {36}},\ \bibinfo {pages} {441 } (\bibinfo {year}
  {1999})}\BibitemShut {NoStop}%
\bibitem [{\citenamefont {Michel}\ \emph {et~al.}(2007)\citenamefont {Michel},
  \citenamefont {Lopez-Pamies}, \citenamefont {Casta{\~n}eda]},\ and\
  \citenamefont {Triantafyllidis}}]{MichelLopez-Pamies07}%
  \BibitemOpen
  \bibfield  {author} {\bibinfo {author} {\bibfnamefont {J.}~\bibnamefont
  {Michel}}, \bibinfo {author} {\bibfnamefont {O.}~\bibnamefont
  {Lopez-Pamies}}, \bibinfo {author} {\bibfnamefont {P.~P.}\ \bibnamefont
  {Casta{\~n}eda]}}, \ and\ \bibinfo {author} {\bibfnamefont {N.}~\bibnamefont
  {Triantafyllidis}},\ }\href {\doibase
  https://doi.org/10.1016/j.jmps.2006.11.006} {\bibfield  {journal} {\bibinfo
  {journal} {Journal of the Mechanics and Physics of Solids}\ }\textbf
  {\bibinfo {volume} {55}},\ \bibinfo {pages} {900 } (\bibinfo {year}
  {2007})}\BibitemShut {NoStop}%
\bibitem [{\citenamefont {Goriely}\ \emph {et~al.}(2008)\citenamefont
  {Goriely}, \citenamefont {Vandiver},\ and\ \citenamefont
  {Destrade}}]{GorielyVandiver08}%
  \BibitemOpen
  \bibfield  {author} {\bibinfo {author} {\bibfnamefont {A.}~\bibnamefont
  {Goriely}}, \bibinfo {author} {\bibfnamefont {R.}~\bibnamefont {Vandiver}}, \
  and\ \bibinfo {author} {\bibfnamefont {M.}~\bibnamefont {Destrade}},\ }\href
  {\doibase 10.1098/rspa.2008.0184} {\bibfield  {journal} {\bibinfo  {journal}
  {Proceedings of the Royal Society A: Mathematical, Physical and Engineering
  Sciences}\ }\textbf {\bibinfo {volume} {464}},\ \bibinfo {pages} {3003}
  (\bibinfo {year} {2008})}\BibitemShut {NoStop}%
\bibitem [{\citenamefont {Lifshitz}\ \emph {et~al.}(1986)\citenamefont
  {Lifshitz}, \citenamefont {Kosevich},\ and\ \citenamefont
  {Pitaevskii}}]{LandauLifshitz86}%
  \BibitemOpen
  \bibfield  {author} {\bibinfo {author} {\bibfnamefont {E.}~\bibnamefont
  {Lifshitz}}, \bibinfo {author} {\bibfnamefont {A.}~\bibnamefont {Kosevich}},
  \ and\ \bibinfo {author} {\bibfnamefont {L.}~\bibnamefont {Pitaevskii}},\
  }\href {\doibase https://doi.org/10.1016/B978-0-08-057069-3.50009-7} {\emph
  {\bibinfo {title} {Theory of Elasticity (Third Edition)}}},\ \bibinfo
  {edition} {third edition}\ ed.,\ edited by\ \bibinfo {editor} {\bibfnamefont
  {E.}~\bibnamefont {Lifshitz}}, \bibinfo {editor} {\bibfnamefont
  {A.}~\bibnamefont {Kosevich}}, \ and\ \bibinfo {editor} {\bibfnamefont
  {L.}~\bibnamefont {Pitaevskii}}\ (\bibinfo  {publisher}
  {Butterworth-Heinemann},\ \bibinfo {address} {Oxford},\ \bibinfo {year}
  {1986})\ pp.\ \bibinfo {pages} {38 -- 86}\BibitemShut {NoStop}%
\bibitem [{\citenamefont {Casta{\~n}eda}\ and\ \citenamefont
  {Suquet}(1997)}]{PonteCastanedaSuquet97}%
  \BibitemOpen
  \bibfield  {author} {\bibinfo {author} {\bibfnamefont {P.~P.}\ \bibnamefont
  {Casta{\~n}eda}}\ and\ \bibinfo {author} {\bibfnamefont {P.}~\bibnamefont
  {Suquet}}\ }(\bibinfo  {publisher} {Elsevier},\ \bibinfo {year} {1997})\ pp.\
  \bibinfo {pages} {171 -- 302}\BibitemShut {NoStop}%
\bibitem [{\citenamefont {Kirchhoff}(1859)}]{Kirchhoff59}%
  \BibitemOpen
  \bibfield  {author} {\bibinfo {author} {\bibfnamefont {G.}~\bibnamefont
  {Kirchhoff}},\ }\href@noop {} {\bibfield  {journal} {\bibinfo  {journal} {J.
  f. reine. angew. Math. (Crelle)}\ }\textbf {\bibinfo {volume} {56}},\
  \bibinfo {pages} {285} (\bibinfo {year} {1859})}\BibitemShut {NoStop}%
\bibitem [{\citenamefont {Kirchhoff}(1876)}]{Kirchhoff76}%
  \BibitemOpen
  \bibfield  {author} {\bibinfo {author} {\bibfnamefont {G.}~\bibnamefont
  {Kirchhoff}},\ }\href@noop {} {\emph {\bibinfo {title} {Vorlesungen \"uber
  mathematische Physik, Mechanik}}}\ (\bibinfo  {publisher} {B. G. Teubner,
  Leipzig},\ \bibinfo {year} {1876})\BibitemShut {NoStop}%
\bibitem [{\citenamefont {Clebsch}(1862)}]{Clebsch62}%
  \BibitemOpen
  \bibfield  {author} {\bibinfo {author} {\bibfnamefont {A.}~\bibnamefont
  {Clebsch}},\ }\href@noop {} {\emph {\bibinfo {title} {Theorie der
  Elasticit\"at Fester K\"orper}}}\ (\bibinfo  {publisher} {B. G. Teubner,
  Leipzig},\ \bibinfo {year} {1862})\BibitemShut {NoStop}%
\bibitem [{\citenamefont {Clebsch}(1883)}]{Clebsch83}%
  \BibitemOpen
  \bibfield  {author} {\bibinfo {author} {\bibfnamefont {A.}~\bibnamefont
  {Clebsch}},\ }\href@noop {} {\emph {\bibinfo {title} {Th\'eorie de
  l'Elasticit\'e des Corps Solides}}}\ (\bibinfo  {publisher} {Translation
  of~\cite{Clebsch62} by Saint-Venant \& Flamant, Dunod, Paris},\ \bibinfo
  {year} {1883})\BibitemShut {NoStop}%
\bibitem [{\citenamefont {Coleman}\ \emph {et~al.}(1993)\citenamefont
  {Coleman}, \citenamefont {Dill}, \citenamefont {Lembo}, \citenamefont {Lu},\
  and\ \citenamefont {Tobias}}]{ColemanDill93}%
  \BibitemOpen
  \bibfield  {author} {\bibinfo {author} {\bibfnamefont {B.~D.}\ \bibnamefont
  {Coleman}}, \bibinfo {author} {\bibfnamefont {E.~H.}\ \bibnamefont {Dill}},
  \bibinfo {author} {\bibfnamefont {M.}~\bibnamefont {Lembo}}, \bibinfo
  {author} {\bibfnamefont {Z.}~\bibnamefont {Lu}}, \ and\ \bibinfo {author}
  {\bibfnamefont {I.}~\bibnamefont {Tobias}},\ }\href {\doibase
  10.1007/BF00375625} {\bibfield  {journal} {\bibinfo  {journal} {Archive for
  Rational Mechanics and Analysis}\ }\textbf {\bibinfo {volume} {121}},\
  \bibinfo {pages} {339} (\bibinfo {year} {1993})}\BibitemShut {NoStop}%
\bibitem [{\citenamefont {Mora}\ \emph {et~al.}(2010)\citenamefont {Mora},
  \citenamefont {Phou}, \citenamefont {Fromental}, \citenamefont {Pismen},\
  and\ \citenamefont {Pomeau}}]{MoraPhou10}%
  \BibitemOpen
  \bibfield  {author} {\bibinfo {author} {\bibfnamefont {S.}~\bibnamefont
  {Mora}}, \bibinfo {author} {\bibfnamefont {T.}~\bibnamefont {Phou}}, \bibinfo
  {author} {\bibfnamefont {J.-M.}\ \bibnamefont {Fromental}}, \bibinfo {author}
  {\bibfnamefont {L.~M.}\ \bibnamefont {Pismen}}, \ and\ \bibinfo {author}
  {\bibfnamefont {Y.}~\bibnamefont {Pomeau}},\ }\href {\doibase
  10.1103/PhysRevLett.105.214301} {\bibfield  {journal} {\bibinfo  {journal}
  {Phys. Rev. Lett.}\ }\textbf {\bibinfo {volume} {105}},\ \bibinfo {pages}
  {214301} (\bibinfo {year} {2010})}\BibitemShut {NoStop}%
\bibitem [{\citenamefont {Duclou{\'e}}\ \emph {et~al.}(2014)\citenamefont
  {Duclou{\'e}}, \citenamefont {Pitois}, \citenamefont {Goyon}, \citenamefont
  {Chateau},\ and\ \citenamefont {Ovarlez}}]{DuclouePitois14}%
  \BibitemOpen
  \bibfield  {author} {\bibinfo {author} {\bibfnamefont {L.}~\bibnamefont
  {Duclou{\'e}}}, \bibinfo {author} {\bibfnamefont {O.}~\bibnamefont {Pitois}},
  \bibinfo {author} {\bibfnamefont {J.}~\bibnamefont {Goyon}}, \bibinfo
  {author} {\bibfnamefont {X.}~\bibnamefont {Chateau}}, \ and\ \bibinfo
  {author} {\bibfnamefont {G.}~\bibnamefont {Ovarlez}},\ }\href {\doibase
  10.1039/C4SM00200H} {\bibfield  {journal} {\bibinfo  {journal} {Soft Matter}\
  }\textbf {\bibinfo {volume} {10}},\ \bibinfo {pages} {5093} (\bibinfo {year}
  {2014})}\BibitemShut {NoStop}%
\bibitem [{\citenamefont {Style}\ \emph
  {et~al.}(2015{\natexlab{a}})\citenamefont {Style}, \citenamefont
  {Boltyanskiy}, \citenamefont {Allen}, \citenamefont {Jensen}, \citenamefont
  {Foote}, \citenamefont {Wettlaufer},\ and\ \citenamefont
  {Dufresne}}]{StyleBoltyanskiy15}%
  \BibitemOpen
  \bibfield  {author} {\bibinfo {author} {\bibfnamefont {R.~W.}\ \bibnamefont
  {Style}}, \bibinfo {author} {\bibfnamefont {R.}~\bibnamefont {Boltyanskiy}},
  \bibinfo {author} {\bibfnamefont {B.}~\bibnamefont {Allen}}, \bibinfo
  {author} {\bibfnamefont {K.~E.}\ \bibnamefont {Jensen}}, \bibinfo {author}
  {\bibfnamefont {H.~P.}\ \bibnamefont {Foote}}, \bibinfo {author}
  {\bibfnamefont {J.~S.}\ \bibnamefont {Wettlaufer}}, \ and\ \bibinfo {author}
  {\bibfnamefont {E.~R.}\ \bibnamefont {Dufresne}},\ }\href {\doibase
  10.1038/nphys3181} {\bibfield  {journal} {\bibinfo  {journal} {Nature Phys}\
  }\textbf {\bibinfo {volume} {11}},\ \bibinfo {pages} {82} (\bibinfo {year}
  {2015}{\natexlab{a}})}\BibitemShut {NoStop}%
\bibitem [{\citenamefont {Style}\ \emph
  {et~al.}(2015{\natexlab{b}})\citenamefont {Style}, \citenamefont
  {Wettlaufer},\ and\ \citenamefont {Dufresne}}]{StyleWettlaufer15}%
  \BibitemOpen
  \bibfield  {author} {\bibinfo {author} {\bibfnamefont {R.~W.}\ \bibnamefont
  {Style}}, \bibinfo {author} {\bibfnamefont {J.~S.}\ \bibnamefont
  {Wettlaufer}}, \ and\ \bibinfo {author} {\bibfnamefont {E.~R.}\ \bibnamefont
  {Dufresne}},\ }\href {\doibase 10.1039/C4SM02413C} {\bibfield  {journal}
  {\bibinfo  {journal} {Soft Matter}\ }\textbf {\bibinfo {volume} {11}},\
  \bibinfo {pages} {672} (\bibinfo {year} {2015}{\natexlab{b}})}\BibitemShut
  {NoStop}%
\bibitem [{\citenamefont {Mancarella}\ \emph
  {et~al.}(2016{\natexlab{a}})\citenamefont {Mancarella}, \citenamefont
  {Style},\ and\ \citenamefont {Wettlaufer}}]{MancarellaStyle16}%
  \BibitemOpen
  \bibfield  {author} {\bibinfo {author} {\bibfnamefont {F.}~\bibnamefont
  {Mancarella}}, \bibinfo {author} {\bibfnamefont {R.~W.}\ \bibnamefont
  {Style}}, \ and\ \bibinfo {author} {\bibfnamefont {J.~S.}\ \bibnamefont
  {Wettlaufer}},\ }\href {\doibase 10.1098/rspa.2015.0853} {\bibfield
  {journal} {\bibinfo  {journal} {Proceedings of the Royal Society A:
  Mathematical, Physical and Engineering Sciences}\ }\textbf {\bibinfo {volume}
  {472}},\ \bibinfo {pages} {20150853} (\bibinfo {year}
  {2016}{\natexlab{a}})}\BibitemShut {NoStop}%
\bibitem [{\citenamefont {Eshelby}\ and\ \citenamefont
  {Peierls}(1957)}]{Eshelby57}%
  \BibitemOpen
  \bibfield  {author} {\bibinfo {author} {\bibfnamefont {J.~D.}\ \bibnamefont
  {Eshelby}}\ and\ \bibinfo {author} {\bibfnamefont {R.~E.}\ \bibnamefont
  {Peierls}},\ }\href {\doibase 10.1098/rspa.1957.0133} {\bibfield  {journal}
  {\bibinfo  {journal} {Proceedings of the Royal Society of London. Series A.
  Mathematical and Physical Sciences}\ }\textbf {\bibinfo {volume} {241}},\
  \bibinfo {pages} {376} (\bibinfo {year} {1957})}\BibitemShut {NoStop}%
\bibitem [{\citenamefont {Peisker}\ \emph {et~al.}(2013)\citenamefont
  {Peisker}, \citenamefont {Michels},\ and\ \citenamefont
  {Gorb}}]{PeiskerMichels13}%
  \BibitemOpen
  \bibfield  {author} {\bibinfo {author} {\bibfnamefont {H.}~\bibnamefont
  {Peisker}}, \bibinfo {author} {\bibfnamefont {J.}~\bibnamefont {Michels}}, \
  and\ \bibinfo {author} {\bibfnamefont {S.~N.}\ \bibnamefont {Gorb}},\
  }\href@noop {} {\bibfield  {journal} {\bibinfo  {journal} {Nature
  Communications}\ } (\bibinfo {year} {2013})}\BibitemShut {NoStop}%
\bibitem [{\citenamefont {Schmitt}\ \emph {et~al.}(2018)\citenamefont
  {Schmitt}, \citenamefont {B{\"u}scher}, \citenamefont {Gorb},\ and\
  \citenamefont {Rajabi}}]{SchmittBuescher18}%
  \BibitemOpen
  \bibfield  {author} {\bibinfo {author} {\bibfnamefont {M.}~\bibnamefont
  {Schmitt}}, \bibinfo {author} {\bibfnamefont {T.~H.}\ \bibnamefont
  {B{\"u}scher}}, \bibinfo {author} {\bibfnamefont {S.~N.}\ \bibnamefont
  {Gorb}}, \ and\ \bibinfo {author} {\bibfnamefont {H.}~\bibnamefont
  {Rajabi}},\ }\href {\doibase 10.1242/jeb.173047} {\bibfield  {journal}
  {\bibinfo  {journal} {Journal of Experimental Biology}\ }\textbf {\bibinfo
  {volume} {221}} (\bibinfo {year} {2018}),\ 10.1242/jeb.173047}\BibitemShut
  {NoStop}%
\bibitem [{\citenamefont {R{\"u}ggeberg}\ \emph {et~al.}(2008)\citenamefont
  {R{\"u}ggeberg}, \citenamefont {Speck}, \citenamefont {Paris}, \citenamefont
  {Lapierre}, \citenamefont {Pollet}, \citenamefont {Koch},\ and\ \citenamefont
  {Burgert}}]{RuggebergSpeck08}%
  \BibitemOpen
  \bibfield  {author} {\bibinfo {author} {\bibfnamefont {M.}~\bibnamefont
  {R{\"u}ggeberg}}, \bibinfo {author} {\bibfnamefont {T.}~\bibnamefont
  {Speck}}, \bibinfo {author} {\bibfnamefont {O.}~\bibnamefont {Paris}},
  \bibinfo {author} {\bibfnamefont {C.}~\bibnamefont {Lapierre}}, \bibinfo
  {author} {\bibfnamefont {B.}~\bibnamefont {Pollet}}, \bibinfo {author}
  {\bibfnamefont {G.}~\bibnamefont {Koch}}, \ and\ \bibinfo {author}
  {\bibfnamefont {I.}~\bibnamefont {Burgert}},\ }\href {\doibase
  10.1098/rspb.2008.0531} {\bibfield  {journal} {\bibinfo  {journal}
  {Proceedings of the Royal Society B: Biological Sciences}\ }\textbf {\bibinfo
  {volume} {275}},\ \bibinfo {pages} {2221} (\bibinfo {year}
  {2008})}\BibitemShut {NoStop}%
\bibitem [{\citenamefont {Speck}\ and\ \citenamefont
  {Burgert}(2011)}]{SpeckBurgert11}%
  \BibitemOpen
  \bibfield  {author} {\bibinfo {author} {\bibfnamefont {T.}~\bibnamefont
  {Speck}}\ and\ \bibinfo {author} {\bibfnamefont {I.}~\bibnamefont
  {Burgert}},\ }\href {\doibase 10.1146/annurev-matsci-062910-100425}
  {\bibfield  {journal} {\bibinfo  {journal} {Annual Review of Materials
  Research}\ }\textbf {\bibinfo {volume} {41}},\ \bibinfo {pages} {169}
  (\bibinfo {year} {2011})}\BibitemShut {NoStop}%
\bibitem [{\citenamefont {Fritsch}\ and\ \citenamefont
  {Hellmich}(2007)}]{FritschHellmich07}%
  \BibitemOpen
  \bibfield  {author} {\bibinfo {author} {\bibfnamefont {A.}~\bibnamefont
  {Fritsch}}\ and\ \bibinfo {author} {\bibfnamefont {C.}~\bibnamefont
  {Hellmich}},\ }\href {\doibase https://doi.org/10.1016/j.jtbi.2006.09.013}
  {\bibfield  {journal} {\bibinfo  {journal} {Journal of Theoretical Biology}\
  }\textbf {\bibinfo {volume} {244}},\ \bibinfo {pages} {597 } (\bibinfo {year}
  {2007})}\BibitemShut {NoStop}%
\bibitem [{\citenamefont {Mendelson}(1982)}]{Mendelson82}%
  \BibitemOpen
  \bibfield  {author} {\bibinfo {author} {\bibfnamefont {N.~H.}\ \bibnamefont
  {Mendelson}},\ }\href {https://mmbr.asm.org/content/46/3/341} {\bibfield
  {journal} {\bibinfo  {journal} {Microbiology and Molecular Biology Reviews}\
  }\textbf {\bibinfo {volume} {46}},\ \bibinfo {pages} {341} (\bibinfo {year}
  {1982})}\BibitemShut {NoStop}%
\bibitem [{\citenamefont {Jaffe}\ and\ \citenamefont
  {Galston}(1968)}]{JaffeGalston68}%
  \BibitemOpen
  \bibfield  {author} {\bibinfo {author} {\bibfnamefont {M.~J.}\ \bibnamefont
  {Jaffe}}\ and\ \bibinfo {author} {\bibfnamefont {A.~W.}\ \bibnamefont
  {Galston}},\ }\href {\doibase 10.1146/annurev.pp.19.060168.002221} {\bibfield
   {journal} {\bibinfo  {journal} {Annual Review of Plant Physiology}\ }\textbf
  {\bibinfo {volume} {19}},\ \bibinfo {pages} {417} (\bibinfo {year}
  {1968})}\BibitemShut {NoStop}%
\bibitem [{\citenamefont {Goriely}(2017{\natexlab{a}})}]{Goriely2017-4}%
  \BibitemOpen
  \bibfield  {author} {\bibinfo {author} {\bibfnamefont {A.}~\bibnamefont
  {Goriely}},\ }\enquote {\bibinfo {title} {Growing on a line},}\ in\ \href
  {\doibase 10.1007/978-0-387-87710-5_4} {\emph {\bibinfo {booktitle} {The
  Mathematics and Mechanics of Biological Growth}}}\ (\bibinfo  {publisher}
  {Springer New York},\ \bibinfo {address} {New York, NY},\ \bibinfo {year}
  {2017})\ pp.\ \bibinfo {pages} {63--96}\BibitemShut {NoStop}%
\bibitem [{\citenamefont {Goriely}(2017{\natexlab{b}})}]{Goriely2017-5}%
  \BibitemOpen
  \bibfield  {author} {\bibinfo {author} {\bibfnamefont {A.}~\bibnamefont
  {Goriely}},\ }\enquote {\bibinfo {title} {Elastic rods},}\ in\ \href
  {\doibase 10.1007/978-0-387-87710-5_5} {\emph {\bibinfo {booktitle} {The
  Mathematics and Mechanics of Biological Growth}}}\ (\bibinfo  {publisher}
  {Springer New York},\ \bibinfo {address} {New York, NY},\ \bibinfo {year}
  {2017})\ pp.\ \bibinfo {pages} {97--123}\BibitemShut {NoStop}%
\bibitem [{\citenamefont {Goldstein}\ and\ \citenamefont
  {Goriely}(2006)}]{GoldsteinGoriely06}%
  \BibitemOpen
  \bibfield  {author} {\bibinfo {author} {\bibfnamefont {R.~E.}\ \bibnamefont
  {Goldstein}}\ and\ \bibinfo {author} {\bibfnamefont {A.}~\bibnamefont
  {Goriely}},\ }\href {\doibase 10.1103/PhysRevE.74.010901} {\bibfield
  {journal} {\bibinfo  {journal} {Phys. Rev. E}\ }\textbf {\bibinfo {volume}
  {74}},\ \bibinfo {pages} {010901(R)} (\bibinfo {year} {2006})}\BibitemShut
  {NoStop}%
\bibitem [{\citenamefont {Goriely}(2017{\natexlab{c}})}]{Goriely2017-6}%
  \BibitemOpen
  \bibfield  {author} {\bibinfo {author} {\bibfnamefont {A.}~\bibnamefont
  {Goriely}},\ }\enquote {\bibinfo {title} {Morphoelastic rods},}\ in\ \href
  {\doibase 10.1007/978-0-387-87710-5_6} {\emph {\bibinfo {booktitle} {The
  Mathematics and Mechanics of Biological Growth}}}\ (\bibinfo  {publisher}
  {Springer New York},\ \bibinfo {address} {New York, NY},\ \bibinfo {year}
  {2017})\ pp.\ \bibinfo {pages} {125--172}\BibitemShut {NoStop}%
\bibitem [{\citenamefont {Caflisch}\ and\ \citenamefont
  {Maddocks}(1984)}]{CaflischMaddocks84}%
  \BibitemOpen
  \bibfield  {author} {\bibinfo {author} {\bibfnamefont {R.~E.}\ \bibnamefont
  {Caflisch}}\ and\ \bibinfo {author} {\bibfnamefont {J.~H.}\ \bibnamefont
  {Maddocks}},\ }\href {\doibase 10.1017/S0308210500025920} {\bibfield
  {journal} {\bibinfo  {journal} {Proceedings of the Royal Society of
  Edinburgh: Section A Mathematics}\ }\textbf {\bibinfo {volume} {99}},\
  \bibinfo {pages} {1–23} (\bibinfo {year} {1984})}\BibitemShut {NoStop}%
\bibitem [{\citenamefont {Maddocks}(1984)}]{Maddocks84}%
  \BibitemOpen
  \bibfield  {author} {\bibinfo {author} {\bibfnamefont {J.~H.}\ \bibnamefont
  {Maddocks}},\ }\href {\doibase 10.1007/BF00275737} {\bibfield  {journal}
  {\bibinfo  {journal} {Archive for Rational Mechanics and Analysis}\ }\textbf
  {\bibinfo {volume} {85}},\ \bibinfo {pages} {311–354} (\bibinfo {year}
  {1984})}\BibitemShut {NoStop}%
\bibitem [{\citenamefont {Antman}\ and\ \citenamefont
  {Rosenfeld}(1978)}]{AntmanRosenfeld78}%
  \BibitemOpen
  \bibfield  {author} {\bibinfo {author} {\bibfnamefont {S.~S.}\ \bibnamefont
  {Antman}}\ and\ \bibinfo {author} {\bibfnamefont {G.}~\bibnamefont
  {Rosenfeld}},\ }\href {http://www.jstor.org/stable/2030353} {\bibfield
  {journal} {\bibinfo  {journal} {SIAM Review}\ }\textbf {\bibinfo {volume}
  {20}},\ \bibinfo {pages} {513} (\bibinfo {year} {1978})}\BibitemShut
  {NoStop}%
\bibitem [{\citenamefont {Hashin}(1962)}]{Hashin62}%
  \BibitemOpen
  \bibfield  {author} {\bibinfo {author} {\bibfnamefont {Z.}~\bibnamefont
  {Hashin}},\ }\href {\doibase 10.1115/1.3636446} {\bibfield  {journal}
  {\bibinfo  {journal} {Journal of Applied Mechanics}\ }\textbf {\bibinfo
  {volume} {29}},\ \bibinfo {pages} {143} (\bibinfo {year} {1962})}\BibitemShut
  {NoStop}%
\bibitem [{\citenamefont {Hashin}\ and\ \citenamefont
  {Shtrikman}(1963)}]{HashinShtrikman63}%
  \BibitemOpen
  \bibfield  {author} {\bibinfo {author} {\bibfnamefont {Z.}~\bibnamefont
  {Hashin}}\ and\ \bibinfo {author} {\bibfnamefont {S.}~\bibnamefont
  {Shtrikman}},\ }\href {\doibase https://doi.org/10.1016/0022-5096(63)90060-7}
  {\bibfield  {journal} {\bibinfo  {journal} {Journal of the Mechanics and
  Physics of Solids}\ }\textbf {\bibinfo {volume} {11}},\ \bibinfo {pages} {127
  } (\bibinfo {year} {1963})}\BibitemShut {NoStop}%
\bibitem [{\citenamefont {Christensen}\ and\ \citenamefont
  {Lo}(1979)}]{ChristensenLo79}%
  \BibitemOpen
  \bibfield  {author} {\bibinfo {author} {\bibfnamefont {R.}~\bibnamefont
  {Christensen}}\ and\ \bibinfo {author} {\bibfnamefont {K.}~\bibnamefont
  {Lo}},\ }\href {\doibase https://doi.org/10.1016/0022-5096(79)90032-2}
  {\bibfield  {journal} {\bibinfo  {journal} {Journal of the Mechanics and
  Physics of Solids}\ }\textbf {\bibinfo {volume} {27}},\ \bibinfo {pages} {315
  } (\bibinfo {year} {1979})}\BibitemShut {NoStop}%
\bibitem [{\citenamefont {Mora}\ \emph {et~al.}(2013)\citenamefont {Mora},
  \citenamefont {Maurini}, \citenamefont {Phou}, \citenamefont {Fromental},
  \citenamefont {Audoly},\ and\ \citenamefont {Pomeau}}]{MoraMaurini13}%
  \BibitemOpen
  \bibfield  {author} {\bibinfo {author} {\bibfnamefont {S.}~\bibnamefont
  {Mora}}, \bibinfo {author} {\bibfnamefont {C.}~\bibnamefont {Maurini}},
  \bibinfo {author} {\bibfnamefont {T.}~\bibnamefont {Phou}}, \bibinfo {author}
  {\bibfnamefont {J.-M.}\ \bibnamefont {Fromental}}, \bibinfo {author}
  {\bibfnamefont {B.}~\bibnamefont {Audoly}}, \ and\ \bibinfo {author}
  {\bibfnamefont {Y.}~\bibnamefont {Pomeau}},\ }\href {\doibase
  10.1103/PhysRevLett.111.114301} {\bibfield  {journal} {\bibinfo  {journal}
  {Phys. Rev. Lett.}\ }\textbf {\bibinfo {volume} {111}},\ \bibinfo {pages}
  {114301} (\bibinfo {year} {2013})}\BibitemShut {NoStop}%
\bibitem [{\citenamefont {Style}\ \emph {et~al.}(2013)\citenamefont {Style},
  \citenamefont {Boltyanskiy}, \citenamefont {Che}, \citenamefont {Wettlaufer},
  \citenamefont {Wilen},\ and\ \citenamefont {Dufresne}}]{StyleBoltyanskiy13}%
  \BibitemOpen
  \bibfield  {author} {\bibinfo {author} {\bibfnamefont {R.~W.}\ \bibnamefont
  {Style}}, \bibinfo {author} {\bibfnamefont {R.}~\bibnamefont {Boltyanskiy}},
  \bibinfo {author} {\bibfnamefont {Y.}~\bibnamefont {Che}}, \bibinfo {author}
  {\bibfnamefont {J.~S.}\ \bibnamefont {Wettlaufer}}, \bibinfo {author}
  {\bibfnamefont {L.~A.}\ \bibnamefont {Wilen}}, \ and\ \bibinfo {author}
  {\bibfnamefont {E.~R.}\ \bibnamefont {Dufresne}},\ }\href {\doibase
  10.1103/PhysRevLett.110.066103} {\bibfield  {journal} {\bibinfo  {journal}
  {Phys. Rev. Lett.}\ }\textbf {\bibinfo {volume} {110}},\ \bibinfo {pages}
  {066103} (\bibinfo {year} {2013})}\BibitemShut {NoStop}%
\bibitem [{\citenamefont {Nadermann}\ \emph {et~al.}(2013)\citenamefont
  {Nadermann}, \citenamefont {Hui},\ and\ \citenamefont
  {Jagota}}]{NadermannHui13}%
  \BibitemOpen
  \bibfield  {author} {\bibinfo {author} {\bibfnamefont {N.}~\bibnamefont
  {Nadermann}}, \bibinfo {author} {\bibfnamefont {C.-Y.}\ \bibnamefont {Hui}},
  \ and\ \bibinfo {author} {\bibfnamefont {A.}~\bibnamefont {Jagota}},\ }\href
  {\doibase 10.1073/pnas.1304587110} {\bibfield  {journal} {\bibinfo  {journal}
  {Proceedings of the National Academy of Sciences}\ }\textbf {\bibinfo
  {volume} {110}},\ \bibinfo {pages} {10541} (\bibinfo {year}
  {2013})}\BibitemShut {NoStop}%
\bibitem [{\citenamefont {Xu}\ \emph {et~al.}(2014)\citenamefont {Xu},
  \citenamefont {Jagota},\ and\ \citenamefont {Hui}}]{XuJagota14}%
  \BibitemOpen
  \bibfield  {author} {\bibinfo {author} {\bibfnamefont {X.}~\bibnamefont
  {Xu}}, \bibinfo {author} {\bibfnamefont {A.}~\bibnamefont {Jagota}}, \ and\
  \bibinfo {author} {\bibfnamefont {C.-Y.}\ \bibnamefont {Hui}},\ }\href
  {\doibase 10.1039/C4SM00216D} {\bibfield  {journal} {\bibinfo  {journal}
  {Soft Matter}\ }\textbf {\bibinfo {volume} {10}},\ \bibinfo {pages} {4625}
  (\bibinfo {year} {2014})}\BibitemShut {NoStop}%
\bibitem [{\citenamefont {Mora}\ \emph {et~al.}(2011)\citenamefont {Mora},
  \citenamefont {Abkarian}, \citenamefont {Tabuteau},\ and\ \citenamefont
  {Pomeau}}]{MoraAbkarian11}%
  \BibitemOpen
  \bibfield  {author} {\bibinfo {author} {\bibfnamefont {S.}~\bibnamefont
  {Mora}}, \bibinfo {author} {\bibfnamefont {M.}~\bibnamefont {Abkarian}},
  \bibinfo {author} {\bibfnamefont {H.}~\bibnamefont {Tabuteau}}, \ and\
  \bibinfo {author} {\bibfnamefont {Y.}~\bibnamefont {Pomeau}},\ }\href
  {\doibase 10.1039/C1SM06051A} {\bibfield  {journal} {\bibinfo  {journal}
  {Soft Matter}\ }\textbf {\bibinfo {volume} {7}},\ \bibinfo {pages} {10612}
  (\bibinfo {year} {2011})}\BibitemShut {NoStop}%
\bibitem [{\citenamefont {Style}\ and\ \citenamefont
  {Dufresne}(2012)}]{StyleDufresne12}%
  \BibitemOpen
  \bibfield  {author} {\bibinfo {author} {\bibfnamefont {R.~W.}\ \bibnamefont
  {Style}}\ and\ \bibinfo {author} {\bibfnamefont {E.~R.}\ \bibnamefont
  {Dufresne}},\ }\href {\doibase 10.1039/C2SM25540E} {\bibfield  {journal}
  {\bibinfo  {journal} {Soft Matter}\ }\textbf {\bibinfo {volume} {8}},\
  \bibinfo {pages} {7177} (\bibinfo {year} {2012})}\BibitemShut {NoStop}%
\bibitem [{\citenamefont {Mancarella}\ \emph
  {et~al.}(2016{\natexlab{b}})\citenamefont {Mancarella}, \citenamefont
  {Style},\ and\ \citenamefont {Wettlaufer}}]{MancarellaStyle16b}%
  \BibitemOpen
  \bibfield  {author} {\bibinfo {author} {\bibfnamefont {F.}~\bibnamefont
  {Mancarella}}, \bibinfo {author} {\bibfnamefont {R.~W.}\ \bibnamefont
  {Style}}, \ and\ \bibinfo {author} {\bibfnamefont {J.~S.}\ \bibnamefont
  {Wettlaufer}},\ }\href {\doibase 10.1039/C5SM03029C} {\bibfield  {journal}
  {\bibinfo  {journal} {Soft Matter}\ }\textbf {\bibinfo {volume} {12}},\
  \bibinfo {pages} {2744} (\bibinfo {year} {2016}{\natexlab{b}})}\BibitemShut
  {NoStop}%
\bibitem [{\citenamefont {Mancarella}\ and\ \citenamefont
  {Wettlaufer}(2017)}]{MW2017}%
  \BibitemOpen
  \bibfield  {author} {\bibinfo {author} {\bibfnamefont {F.}~\bibnamefont
  {Mancarella}}\ and\ \bibinfo {author} {\bibfnamefont {J.~S.}\ \bibnamefont
  {Wettlaufer}},\ }\href@noop {} {\bibfield  {journal} {\bibinfo  {journal}
  {Soft Matter}\ }\textbf {\bibinfo {volume} {13}},\ \bibinfo {pages} {945}
  (\bibinfo {year} {2017})}\BibitemShut {NoStop}%
\bibitem [{\citenamefont {Mori}\ and\ \citenamefont
  {Tanaka}(1973)}]{MoriTanaka73}%
  \BibitemOpen
  \bibfield  {author} {\bibinfo {author} {\bibfnamefont {T.}~\bibnamefont
  {Mori}}\ and\ \bibinfo {author} {\bibfnamefont {K.}~\bibnamefont {Tanaka}},\
  }\href {\doibase https://doi.org/10.1016/0001-6160(73)90064-3} {\bibfield
  {journal} {\bibinfo  {journal} {Acta Metallurgica}\ }\textbf {\bibinfo
  {volume} {21}},\ \bibinfo {pages} {571 } (\bibinfo {year}
  {1973})}\BibitemShut {NoStop}%
\bibitem [{Note1()}]{Note1}%
  \BibitemOpen
  \bibinfo {note} {In order to benchmark our results for heterogeneous elastic
  rods against the constant Young's modulus case, we have to assume an
  additional constraint associated with the fact that the bending shapes are
  defined up to a constant. Here we impose the condition that the reference
  case Eqs.~\protect \textup {\hbox {\mathsurround \z@ \protect \normalfont
  (\ignorespaces \ref {eq_reference_modes}\unskip \@@italiccorr )}}, and the
  heterogeneous elastic rod Eqs.~\protect \textup {\hbox {\mathsurround \z@
  \protect \normalfont (\ignorespaces \ref {eq_sol_bc}\unskip \@@italiccorr
  )}},\protect \textup {\hbox {\mathsurround \z@ \protect \normalfont
  (\ignorespaces \ref {eq_sol_bc-Mori-Tanaka}\unskip \@@italiccorr )}} have the
  same potential energy as given by Eq.~\protect \textup {\hbox {\mathsurround
  \z@ \protect \normalfont (\ignorespaces \ref {eq_energy}\unskip \@@italiccorr
  )}}.}\BibitemShut {Stop}%
\bibitem [{Note2()}]{Note2}%
  \BibitemOpen
  \bibinfo {note} {The critical loads for the buckling modes in Fig.~\ref
  {fig_buckling_MSW_DT}, $\gamma \prime =0.1$, are---MSW in first place, DT
  second: $\Gamma _{\protect \text {cr}}^{(1)}=6.785(4),7.344(9)$; $\Gamma
  _{\protect \text {cr}}^{(2)}=27.104(2),29.557(6)$, for $\phi _0=0.6$; and
  $\Gamma _{\protect \text {cr}}^{(1)}=8.265(4),8.427(8)$; $\Gamma _{\protect
  \text {cr}}^{(2)}=33.059(4),33.778(8)$, for $\phi _0=0.3$.}\BibitemShut
  {Stop}%
\bibitem [{\citenamefont {Bender}\ and\ \citenamefont
  {Orszag}(1999)}]{BenderOrszag}%
  \BibitemOpen
  \bibfield  {author} {\bibinfo {author} {\bibfnamefont {C.}~\bibnamefont
  {Bender}}\ and\ \bibinfo {author} {\bibfnamefont {S.}~\bibnamefont
  {Orszag}},\ }\href@noop {} {\emph {\bibinfo {title} {Advanced Mathematical
  Methods for Scientists and Engineers I}}}\ (\bibinfo  {publisher}
  {Springer-Verlag New York},\ \bibinfo {year} {1999})\BibitemShut {NoStop}%
\bibitem [{Note3()}]{Note3}%
  \BibitemOpen
  \bibinfo {note} {The numerical results for the critical forces corresponding
  to the first two buckling modes of a polar rod with \protect \{$\gamma \prime
  _0=2,\protect \tmspace +\thinmuskip {.1667em}\gamma \prime _1=0.01,\protect
  \tmspace +\thinmuskip {.1667em}\phi =0.4$\protect \} are $\Gamma _{\protect
  \text {cr}}^{(1)}=10.601(6)$ and $\Gamma _{\protect \text
  {cr}}^{(2)}=41.265(1)$ (DT), and $\Gamma _{\protect \text
  {cr}}^{(1)}=10.502(6)$ and $\Gamma _{\protect \text {cr}}^{(2)}=40.549(2)$
  (MSW)\label {foot_polar}}\BibitemShut {NoStop}%
\bibitem [{\citenamefont {Hofhuis}\ \emph {et~al.}(2016)\citenamefont
  {Hofhuis}, \citenamefont {Moulton}, \citenamefont {Lessinnes}, \citenamefont
  {Routier-Kierzkowska}, \citenamefont {Bomphrey}, \citenamefont {Mosca},
  \citenamefont {Reinhardt}, \citenamefont {Sarchet}, \citenamefont {Gan},
  \citenamefont {Tsiantis}, \citenamefont {Ventikos}, \citenamefont {Walker},
  \citenamefont {Goriely}, \citenamefont {Smith},\ and\ \citenamefont
  {Hay}}]{Hofhuis2016}%
  \BibitemOpen
  \bibfield  {author} {\bibinfo {author} {\bibfnamefont {H.}~\bibnamefont
  {Hofhuis}}, \bibinfo {author} {\bibfnamefont {D.}~\bibnamefont {Moulton}},
  \bibinfo {author} {\bibfnamefont {T.}~\bibnamefont {Lessinnes}}, \bibinfo
  {author} {\bibfnamefont {A.-L.}\ \bibnamefont {Routier-Kierzkowska}},
  \bibinfo {author} {\bibfnamefont {R.~J.}\ \bibnamefont {Bomphrey}}, \bibinfo
  {author} {\bibfnamefont {G.}~\bibnamefont {Mosca}}, \bibinfo {author}
  {\bibfnamefont {H.}~\bibnamefont {Reinhardt}}, \bibinfo {author}
  {\bibfnamefont {P.}~\bibnamefont {Sarchet}}, \bibinfo {author} {\bibfnamefont
  {X.}~\bibnamefont {Gan}}, \bibinfo {author} {\bibfnamefont {M.}~\bibnamefont
  {Tsiantis}}, \bibinfo {author} {\bibfnamefont {Y.}~\bibnamefont {Ventikos}},
  \bibinfo {author} {\bibfnamefont {S.}~\bibnamefont {Walker}}, \bibinfo
  {author} {\bibfnamefont {A.}~\bibnamefont {Goriely}}, \bibinfo {author}
  {\bibfnamefont {R.}~\bibnamefont {Smith}}, \ and\ \bibinfo {author}
  {\bibfnamefont {A.}~\bibnamefont {Hay}},\ }\href@noop {} {\bibfield
  {journal} {\bibinfo  {journal} {Cell}\ }\textbf {\bibinfo {volume} {166}},\
  \bibinfo {pages} {222} (\bibinfo {year} {2016})}\BibitemShut {NoStop}%
\bibitem [{\citenamefont {Kim}\ \emph {et~al.}(2020)\citenamefont {Kim},
  \citenamefont {Liu}, \citenamefont {Weon}, \citenamefont {Cohen},
  \citenamefont {Hui}, \citenamefont {Dufresne},\ and\ \citenamefont
  {Style}}]{KimLiu20}%
  \BibitemOpen
  \bibfield  {author} {\bibinfo {author} {\bibfnamefont {J.~Y.}\ \bibnamefont
  {Kim}}, \bibinfo {author} {\bibfnamefont {Z.}~\bibnamefont {Liu}}, \bibinfo
  {author} {\bibfnamefont {B.~M.}\ \bibnamefont {Weon}}, \bibinfo {author}
  {\bibfnamefont {T.}~\bibnamefont {Cohen}}, \bibinfo {author} {\bibfnamefont
  {C.-Y.}\ \bibnamefont {Hui}}, \bibinfo {author} {\bibfnamefont {E.~R.}\
  \bibnamefont {Dufresne}}, \ and\ \bibinfo {author} {\bibfnamefont {R.~W.}\
  \bibnamefont {Style}},\ }\href {\doibase 10.1126/sciadv.aaz0418} {\bibfield
  {journal} {\bibinfo  {journal} {Science Advances}\ }\textbf {\bibinfo
  {volume} {6}} (\bibinfo {year} {2020}),\ 10.1126/sciadv.aaz0418}\BibitemShut
  {NoStop}%
\bibitem [{\citenamefont {Riva}\ \emph {et~al.}(2020)\citenamefont {Riva},
  \citenamefont {Rosa},\ and\ \citenamefont {Ruzzene}}]{RivaRosa20}%
  \BibitemOpen
  \bibfield  {author} {\bibinfo {author} {\bibfnamefont {E.}~\bibnamefont
  {Riva}}, \bibinfo {author} {\bibfnamefont {M.~I.~N.}\ \bibnamefont {Rosa}}, \
  and\ \bibinfo {author} {\bibfnamefont {M.}~\bibnamefont {Ruzzene}},\ }\href
  {\doibase 10.1103/PhysRevB.101.094307} {\bibfield  {journal} {\bibinfo
  {journal} {Phys. Rev. B}\ }\textbf {\bibinfo {volume} {101}},\ \bibinfo
  {pages} {094307} (\bibinfo {year} {2020})}\BibitemShut {NoStop}%
\end{thebibliography}%
